\documentclass[12pt]{article}
\usepackage{amsmath,amssymb,graphicx} 
\usepackage{epsfig}
\usepackage{feynarts}
\usepackage{slashed}
\usepackage{pstricks}
\usepackage{xspace}
\newcommand{\sstrike}[1]{\ensuremath{\not\!#1}\xspace}

\newcommand{\beq}{\begin{eqnarray}}
\newcommand{\eeq}{\end{eqnarray}}

\newcommand{\centeron}[2]{{\setbox0=\hbox{#1}\setbox1=\hbox{#2}\ifdim
                                        
\wd1>\wd0\kern.5\wd1\kern-.5\wd0\fi
\copy0

\kern-.5\wd0\kern-.5\wd1\copy1\ifdim\wd0>\wd1
                                       \kern.5\wd0\kern-.5\wd1\fi}}
\newcommand{\ltap}{\>\centeron{\raise.35ex\hbox{$<$}}
                               {\lower.65ex\hbox{$\sim$}}\>}
\newcommand{\gtap}{\>\centeron{\raise.35ex\hbox{$>$}}
                               {\lower.65ex\hbox{$\sim$}}\>}

\newcommand\ZZ{\hbox{\zfont Z\kern-.4emZ}}
\font\zfont = cmss10 
\newcommand{\sfrac}[2]{{\textstyle\frac{#1}{#2}}}

\begin{document}
\begin{titlepage}
\begin{flushright}
{\tt LYCEN 2011-03}
\end{flushright}

\vskip.5cm
\begin{center}
{\huge \bf 
The Universal Real Projective Plane: LHC phenomenology at one Loop
}

\vskip.1cm
\end{center}
\vskip0.2cm

\begin{center}
{\bf
Giacomo Cacciapaglia$^{a,b}$, Aldo Deandrea$^{a}$ {\rm and}
J\'er\'emie Llodra-Perez$^{a}$}
\end{center}
\vskip 8pt

\begin{center}
$^{a}$ {\it Universit\'e de Lyon, F-69622 Lyon, France; Universit\'e Lyon 1, Villeurbanne;\\
CNRS/IN2P3, UMR5822, Institut de Physique Nucl\'eaire de Lyon\\
F-69622 Villeurbanne Cedex, France } \\

$^{b}$ {\it King's College London, Department of Physics, Strand, London WC2R 2LS, UK}\\

\vspace{0.2cm}

{\tt  g.cacciapaglia@ipnl.in2p3.fr,\\ deandrea@ipnl.in2p3.fr, \\jllodra@ipnl.in2p3.fr} 
\end{center}

\vglue 0.3truecm
\begin{abstract}
\vskip 3pt
\noindent
The Real Projective Plane is the lowest dimensional orbifold which, when combined with the usual Minkowski space-time, 
gives rise to a unique model in six flat dimensions possessing an exact Kaluza Klein (KK) parity as a relic symmetry of the 
broken six dimensional Lorentz group. As a consequence of this property, any model 
formulated on this background will include a stable Dark Matter candidate. 
Loop corrections play a crucial role because they remove mass degeneracy in the tiers of KK modes and induce new couplings 
which mediate decays. We study the full one loop structure of 
the corrections by means of counter-terms localised on the two singular points.
As an application, the LHC phenomenology of the $(2,0)$ and $(0,2)$ tiers is discussed. We identify promising signatures with single and di-lepton, top anti-top and 4 tops: in the di-lepton channel, 
present data from CMS and ATLAS may already exclude KK masses up to $250$ GeV, while by next year they may cover 
the whole mass range preferred by WMAP data.
\end{abstract}


\end{titlepage}
\newpage
\tableofcontents
\newpage
\section{Introduction}
\label{sec:intro}
\setcounter{equation}{0}
\setcounter{footnote}{0}

The presence of Dark Matter in the Universe has been suggested by observations both in cosmology and in astrophysics.
A reasonable, but not unique, candidate is a massive neutral and stable particle: this candidate is absent in the Standard Model (SM) 
because the relic abundance of neutrinos is too low.
Having a Dark Matter candidate is therefore a desirable feature for any model of New Physics beyond the Standard Model.
The usual lore is to impose a discrete parity that renders one or more of the new particles stable, and fit the parameter space to 
reproduce the relic abundance mainly extracted from the Cosmic Microwave Background data, under the assumption of a 
standard cosmology.
In many models, the presence of a parity is also required for other phenomenological reasons: for instance, in supersymmetric 
models, R-parity \cite{rp} forbids terms that violate baryon number and therefore ensures the stability of the proton; in Little Higgs 
models, T-parity \cite{tpar} ensures that corrections to electroweak precision measurements arise at loop level.

In recent years, extra dimensions have offered new mechanisms to solve the problems of the standard model, 
in particular the stability of the Higgs mass and the hierarchy in the Yukawa sector.
It has also been proposed that a parity, called Kaluza-Klein parity (KK-parity), can arise after the compactification of the extra 
space co-ordinates as a relic of the extended Lorentz symmetries~\cite{UED5,chiralsquare1} (in particular, 
from the conservation of momentum along the 
extra directions) and that the resulting stable particle is a good candidate for the Dark Matter~\cite{LKP,LKP2}. 
However, models present in the literature are tainted by the presence of counter-terms localised on fixed 
points of the orbifold compactification~\cite{Cheng:2002iz,Ponton:2005kx}: they in general violate the extra 
Lorentz invariance, including the KK-parity.
Therefore, the KK-parity is no more a consequence of the compactification, but it is imposed by hand on the general structure of the models and on its Ultra-violet completion which is the ultimate origin of the localised counter-terms. 

In~\cite{Cacciapaglia:2009pa,Cacciapaglia:2009zz}, it has been realised that a KK-parity can arise naturally in compactifications 
where fixed points are absent. With two flat extra dimensions, there is a unique orbifold without fixed points where a 
realistic model can be constructed: the Real Projective Plane, which is defined in terms of a 180 degree rotation and a glide, that form a discrete 
symmetry group of the flat plane\footnote{In~\cite{Dohi:2010vc} a model based on a real projective plane constructed starting from 
the two-sphere with its antipodal points being identified has been considered. The spectrum of this model in curved 
space is different from the one we consider.}.
Localised counter-terms are not avoided, however the geometry of the space ensures that they do respect the KK-parity.
In this scenario, the Dark Matter candidate is relatively light with a mass of few hundred GeV, and it is accompanied by a large 
number of states of similar masses, corresponding to the extra dimensional fields. 
Therefore, this scenario offers interesting phenomenology, possibly accessible to the early stage of the Large Hadron Collider 
(LHC) experiments (with low luminosity and low energy).
Studying the phenomenology is not trivial: a common feature of models with extra flat dimensions is that, at tree level, all the extra 
particles are degenerate and possible decays are exactly on threshold. 
To have a detailed idea of the phenomenology of the real projective plane, therefore, one needs to compute radiative corrections. 
In general loop corrections are dominated by the logarithmically divergent contribution due to the rotation symmetry, therefore it is 
in principle possible to use the localised counter-terms to estimate them.
In this paper we study the structure of the counter-terms of the Standard Model on the Real Projective plane in detail: the goal is 
to predict all the mass corrections and effective couplings relevant for the phenomenology of all levels of KK resonance by means 
of a limited number of localised counter-terms.
If we were to write down all lower dimension operators on the singular points, imposing gauge invariance, we would be left with a 
handful of parameters, and such parameters would rule the phenomenology of all the levels so that, calculating the divergences 
for a specific level would allow us to predict the masses and couplings for all the others. The main complication is that loop 
corrections are generically not gauge invariant. In the SM this behaviour is well-known and it is not a limitation. In our case, 
one-loop calculations in the six-dimensional (6D) model are more involved as one increases the number of fields propagating 
in the loop. In order to predict masses and couplings for phenomenology studies, it is necessary to predict the 6D divergences.

This problem has been already observed in other extra-dimension scenarios~\cite{Ponton:2005kx} by using calculations in generic $\xi$ gauge. 
The authors remarked that, for a peculiar gauge choice $\xi=-3$, one can see that some cancellations occur and the structure of the mass corrections respect the gauge invariant counter-terms.
As conjectured in~\cite{Ponton:2005kx}, this feature is also true for couplings: we will stress here that the peculiarity of the 
$\xi=-3$ gauge is also present in the Standard Model~\cite{4Dmagic}, and therefore it is not a feature of extra dimensional models.

In this paper, we study loop corrections in the case of the Real Projective Plane, but our results can be generalised to any 
compact space. To simplify the calculation, we limit ourselves to the one-loop induced interactions between the $(2 k, 2 l)$ level, with 
$k$ and $l$ generic non-zero integers, with $(0,0)$ modes, where zero modes correspond to standard model fields.
We calculate mass mixings and also trilinear couplings with two standard model particles.
The reason for this choice is that, due to the conservation of extra momentum on the tree level vertices, these loops can only be 
mediated by a particle in the level $(k, l)$: therefore, a complicated 6D calculation reduces to a standard four-dimensional (4D) 
loop calculation, without any sum over an infinite tower of KK modes involved.
This greatly simplifies the calculation of loop integrals.
We will exhaustively calculate loop corrections in generic $\xi$ gauge, and show that ``gauge invariance'' is recovered 
in the ``magic'' $\xi=-3$ gauge.
Furthermore, we show that the decays of level $(2k, 2l)$ particles into standard model ones are gauge invariant, and use the 
results in the magic gauge to calculate the full set of counter-terms for the standard model on the Real Projective Plane.
These results can be used to predict the phenomenology of any level, both mass spectra and decay rates.

As an example, we study the phenomenology at the LHC for the levels $(2,0)$ (and $(0,2)$).
We show that a generic particle in these tiers decays with comparable rates into a pair of SM particles, a pair of particles in 
tier $(1,0$) (or $(0,1)$) and a same-tier particle plus a standard model one.
From the counter-term structure, we can deduce that fermion decays into SM pair are suppressed, while gauge vectors 
decay significantly into a pair of SM fermions. Due to the many decay chains leading to a massive vector, the total 
production cross section is large even for electroweak bosons.
As a consequence, the effective inclusive cross section in di-leptons or top pairs can be rather large.
We added the complete interactions of tiers $(1,0)$ and $(2,0)$ in CalcHEP 2.5.5~\cite{calchep}, using our FeynRules~\cite{feynrules} implementation of the 6D Lagrangian~\cite{jeremie}.
We estimated such cross sections over the relevant mass range, and show that present data from CMS and ATLAS with integrated luminosity of 40 pb$^{-1}$ already 
pose serious bounds on the size of the extra dimensions.
A more detailed study, which is beyond the scope of this work, is nevertheless necessary to determine the excluded 
masse range and the prospect for next year results.
Our estimates show that the LHC with an integrated luminosity of 1 fb$^{-1}$ should be able to probe the mass range 
preferred by Dark Matter data.

In Section~\ref{sec:model} we introduce the model and in Sections~\ref{sec:gaugeb}, \ref{sec:scalar} and \ref{sec:fermion} we discuss the loop corrections and counter-terms for gauge boson, scalar and fermion fields respectively.
In Section~\ref{sec:physobs} we calculate the effective gauge-invariant couplings and in Section~\ref{sec:6Dloop} we show a simple 6D calculation for particles in tier $(2 n,0)$.
Finally, in Section~\ref{sec:tier2} we discuss the LHC phenomenology of tier $(2,0)$, before concluding in Section~\ref{sec:concl}.

\section{The model}
\label{sec:model}
\setcounter{equation}{0}
\setcounter{footnote}{0}

In this paper we consider the simplest model which is an extension of the Standard Model on the Real Projective 
Plane~\cite{Cacciapaglia:2009pa}: to each field in the SM we associate a 6D field that propagates in the bulk.
Each field is characterised by its parities $(p_r, p_g)$ under the two symmetries of the orbifold, the rotation $r$ and the glide $g$, 
with $p_{r,g} = \pm$. The gauge bosons have parities $(+,+)$ so that to each generator corresponds a zero mode and an 
unbroken 4D gauge symmetry.
The 6D scalar Higgs boson has also $(+,+)$ parities so that its (massless) zero mode can play the role of the SM Higgs boson and 
trigger electroweak symmetry breaking with a tree level potential.
To each chiral fermion there corresponds a vector 6D fermion (a full 4 Weyl component spinor), because the two 6D chiralities 
are interchanged by the glide symmetry: the parity $p_r$ determines if the 4D chirality of the zero mode is left or right-handed.
Under the usual reduction to Kaluza-Klein modes, the 6D fields are replaced by an infinite tower of massive states and each tier is 
labelled by 2 integer numbers $(k,l)$: physically they correspond to the discrete momenta carried by the state along the two extra 
directions in units of the two radii $R_5$ and $R_6$.
Neglecting the contribution of the Higgs vacuum expectation value, the spectrum is uniquely determined by these integers, 
therefore regardless of the type of particle the mass is
\beq
M_{k,l}^2 = \frac{k^2}{R_5^2} + \frac{l^2}{R_6^2} = \frac{k^2+l^2}{R^2} = (k^2 + l^2) m_{KK}^2\,,
\eeq
where we have assumed equal radii for the two directions and defined $m_{KK} = 1/R$.
The couplings are also determined by the parity of the fields via the wave functions of the KK modes.
As a general rule, the couplings do respect sum rules on the integers $(k,l)$ which are a relic of the momentum conservation 
along the extra co-ordinates.
For example, a trilinear coupling between three modes from tiers $(k_1,l_1)$, $(k_2,l_2)$ and $(k_3,l_3)$ is present only if there 
exists a choice of signs such that the sum of the three $k$'s with such signs vanish, and independently the same is possible for 
the $l$'s:
$$
k_1 \pm k_2 \pm k_3 = 0\, \quad \mbox{and} \quad l_1 \pm l_2 \pm l_3 = 0\,.
$$
These rules impose non trivial constraints on the loop contributions: in particular, if a loop-induced coupling violates the sum rules 
on both $k$'s and $l$'s, only one tier can run into the loop and the calculation reduces to a 4D calculation.
As an example, we can look to the mixing between a mode $(2k,2l)$ with both $k$ and $l$ non zero and a zero mode: in Figure~
\ref{fig:genloops} we show that only the modes $(k,l)$ can run in the loop.
The same is true if we calculate the loop induced couplings of $(2k,2l)$ to any number of zero modes.
This case is in fact the simplest calculation in terms of number of diagrams, and we will use it as a tool to calculate the counter-terms in a generic model. Another example is the mixing of modes $(2k+1, 2l)$ with one $(1,0)$ and arbitrary zero modes: the 
presence of the $(1,0)$ is due to the fact that $(2k+1,2l)$ is odd under the KK parity and therefore cannot couple to zero modes 
only. In Fig.~\ref{fig:genloops}, one can see that the unit of momentum running in the loop doubles the number of diagrams.
Note also that some couplings are not allowed at loop level even though they do not violate the KK parity: an example is the couplings of $(1,1)$ states to SM ones.
Such couplings will be generated by the localised higher dimensional counter-terms, and their properties are strongly UV dependent.

\setlength{\unitlength}{1cm}
\begin{figure}[tb]
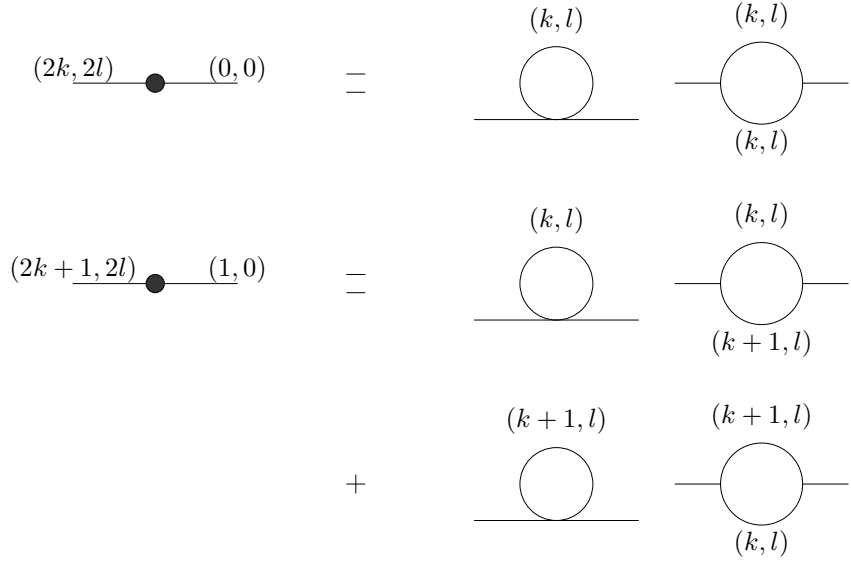

\begin{center}
\begin{feynartspicture}(15,8)(4,3)
\FADiagram{}
\FAProp(2.,12.)(20.,12.)(0.,){/Straight}{0}
\FALabel(2.,15.)[t]{\footnotesize{$(2 k,2 l)$}}
\FALabel(20.,15)[t]{\footnotesize{$(0,0)$}}
\FAVert(11.,12.){-3}
\FADiagram{}
\FAProp(10.,13.)(12.,13.)(0.,){/Straight}{0}
\FAProp(10.,11.)(12.,11.)(0.,){/Straight}{0}
\FADiagram{}
\FAProp(2.,8.)(11.,8.)(0.,){/Straight}{0}
\FAProp(11.,8.)(20.,8.)(0.,){/Straight}{0}
\FAProp(11.,8.)(11.,8.)(11.,16.){/Straight}{0}
\FALabel(11.,17.5)[b]{\footnotesize{$(k,l)$}}
\FADiagram{}
\FAProp(2.,12.)(7.,12.)(0.,){/Straight}{0}
\FAProp(7.,12.)(7.,12.)(16.,12.){/Straight}{0}
\FALabel(11.5,18)[b]{\footnotesize{$(k,l)$}}
\FALabel(11.5,4)[b]{\footnotesize{$(k,l)$}}
\FAProp(16.,12.)(21.,12.)(0.,){/Straight}{0}
\FADiagram{}
\FAProp(2.,12.)(20.,12.)(0.,){/Straight}{0}
\FALabel(2.,15.)[t]{\footnotesize{$(2 k+1,2 l)$}}
\FALabel(20.,15)[t]{\footnotesize{$(1,0)$}}
\FAVert(11.,12.){-3}
\FADiagram{}
\FAProp(10.,13.)(12.,13.)(0.,){/Straight}{0}
\FAProp(10.,11.)(12.,11.)(0.,){/Straight}{0}
\FADiagram{}
\FAProp(2.,8.)(11.,8.)(0.,){/Straight}{0}
\FAProp(11.,8.)(20.,8.)(0.,){/Straight}{0}
\FAProp(11.,8.)(11.,8.)(11.,16.){/Straight}{0}
\FALabel(11.,17.5)[b]{\footnotesize{$(k,l)$}}
\FADiagram{}
\FAProp(2.,12.)(7.,12.)(0.,){/Straight}{0}
\FAProp(7.,12.)(7.,12.)(16.,12.){/Straight}{0}
\FALabel(11.5,18)[b]{\footnotesize{$(k,l)$}}
\FALabel(11.5,4)[b]{\footnotesize{$(k+1,l)$}}
\FAProp(16.,12.)(21.,12.)(0.,){/Straight}{0}
\FADiagram{}
\FADiagram{}
\FAProp(10.,12.)(12.,12.)(0.,){/Straight}{0}
\FAProp(11.,11.)(11.,13.)(0.,){/Straight}{0}
\FADiagram{}
\FAProp(2.,8.)(11.,8.)(0.,){/Straight}{0}
\FAProp(11.,8.)(20.,8.)(0.,){/Straight}{0}
\FAProp(11.,8.)(11.,8.)(11.,16.){/Straight}{0}
\FALabel(11.,17.5)[b]{\footnotesize{$(k+1,l)$}}
\FADiagram{}
\FAProp(2.,12.)(7.,12.)(0.,){/Straight}{0}
\FAProp(7.,12.)(7.,12.)(16.,12.){/Straight}{0}
\FALabel(11.5,18)[b]{\footnotesize{$(k+1,l)$}}
\FALabel(11.5,4)[b]{\footnotesize{$(k,l)$}}
\FAProp(16.,12.)(21.,12.)(0.,){/Straight}{0}
\end{feynartspicture}
\end{center}
\caption{\footnotesize Generic loops for the mixing of level $(2k, 2l)$ with zero modes and $(2k+1,2l)$ with $(1,0)$. Note that any number of zero modes can be attached to vertices or loops to obtain three- and four-point functions without changing existing labels.}
\label{fig:genloops}
\end{figure}
In the following, we will summarise the spectrum of the relevant tiers.
The couplings that enter the loops have been calculated with FeynRules.
For simplicity we will study an SU(N) gauge theory with fundamental fermions and scalars; when necessary we will quote the 
analogous results for a U(1) and generalise the final results to the SM case. Note also that the spectrum outlined in this section is 
affected by the Higgs vacuum expectation value (VEV) at three level and by loop effects.
Generically
\beq
M_{k,l}^2 = (k^2 + l^2) m_{KK}^2 + m_0^2 + \delta_{\rm 1-loop}\,,
\eeq
where $m_0$ is the corresponding mass of the zero mode generated by the Higgs VEV and $\delta$ is the one loop correction.
In the following we will consider the contribution of electroweak symmetry breaking of the same order as the 1-loop correction (or 
smaller), which is numerically true for $m_{KK}$ of few hundred GeV. Therefore, we can consistently neglect the contribution of 
the Higgs VEV in the loop calculations.

\subsection*{Scalar field}

For a complex scalar field, the 6D action can be written as
\begin{align}
\mathcal{L}_{\rm s} =  \int_0^{2 \pi} d x_5\, d x_6\; \left\{(D^M\phi)^\dagger(D_M\phi)-m_\phi^2\ \phi^\dagger\phi\right\}
\end{align}
where the index $M = (\mu, 5, 6)$ and $D_M=\partial_M-igA^a_Mt_r^a$.
Minimising the variation of the action and Fourier transforming in the usual 4D, we obtain the following equation of motion for the 6D field:
\beq
(\partial_5^2 + \partial_6^2 + p^2 - m_\phi^2)\, \phi (p^2, x_5, x_6) = 0\,;
\eeq
 after the KK expansion of the 6D field
 \beq
  \phi (p^2, x_5, x_6) = \sum_{k,l} f_{k,l} (x_5, x_6) \phi_{k,l} (p^2)\,,
  \eeq
  we use the equation of motion to determine wave functions $f_{k,l}$ and masses
  \beq
  p^2 = M_{k,l}^2 + m_{\phi}^2\,.
  \eeq 
The detailed spectrum is determined by the parity assignment of the 6D field: in the case of the SM Higgs $(+,+)$, there exists a 
mode in the tiers $(0,0)$, $(2k, 0)$, $(0, 2l)$ and $(k,l)$, with $k$ and $l$ positive integers.
The propagator of each mode is given by a 4D scalar propagator with the appropriate mass.
Note also that in the Higgs case, $m_\phi$ is of the same order as the VEV, therefore we will consider this mass at the same 
order as a one loop corrections and neglect it in the loop calculations.

\subsection*{Gauge fields}

For a SU(N) gauge group, the 6D Lagrangian reads:
\beq
\mathcal{L}_{\mbox{gauge}} = \int_0^{2 \pi} d x_5\, d x_6\; \left\{- \frac{1}{4} F^a_{M N} F^{a\, M N} - \frac{1}{2 \xi} \left( \partial_\mu A^{a,\mu} - \xi (\partial_5 A^a_5 + \partial_6 A^a_6) \right)^2 \right\}\,,
\eeq
where $F^a_{M N} = \partial_M A^a_N - \partial_N A^a_M + g f^{abc} A^b_M A^c_N$, the indices $M$ and $N = (\mu, 5, 6)$, and 
the $\xi$-gauge fixing term is added to eliminate the mixing between $A_\mu$ and the scalar components $A_5$ and $A_6$.

For the 4D vector component, once the usual 4D equations of motion in $\xi$ gauge are applied, the wave functions satisfy an 
equation similar to the scalar one (with $m_\phi=0$), so that the spectrum and the wave functions are the same as described 
above. Regarding the two extra polarisations, which correspond to scalar adjoints in the 4D reduction, the situation is more 
complicated due to the presence of a $\xi$ dependent mixing between the two components: one can see that $A_5$ and $A_6$ 
are not the eigenvectors of the bilinear term
\beq
\mathcal{L}_{A_5,A_6} = \int_0^{2 \pi} d x_5\, d x_6\; \frac{1}{2} (A_5,A_6)\cdotp
\left(\begin{array}{cc}
-\partial_{\mu}^2+\partial^2_6+\xi \partial_5^2 &(\xi-1) \partial_5\partial_6\\
   (\xi-1) \partial_5\partial_6 &-\partial_{\mu}^2+\partial^2_5+\xi \partial_6^2\\
\end{array}\right)\cdotp
\left(\begin{array}{c}
A_5\\
A_6\\
\end{array} \right)
\,.
\eeq
Note that the off-diagonal terms vanish only in the Feynman gauge $\xi=1$.
One can nevertheless expand $A_5$ and $A_6$ in the same 4D fields, and diagonalise the bilinear action mode by mode: we 
then see that one of the mass eigenstates, $A_\pi^{k,l}$, has $\xi$ dependent mass ($M^2_{A_\pi^{k,l}} = \xi M_{k,l}^2$) and 
corresponds to the Goldstone boson eaten by the corresponding massive gauge boson, while the other eigenstate 
$A_\varphi^{k,l}$ is a physical scalar degree of freedom. Note also that they are both present for modes $(k,l)$ with
\beq
A_5^{k,l} &=& \sfrac{l}{M_{k,l} R}\ A_\varphi^{k,l} + \sfrac{k}{M_{k,l} R}\ A_\pi^{k,l}\,,\\
A_6^{k,l} &=& -\sfrac{k}{M_{k,l} R}\ A_\varphi^{k,l}+ \sfrac{l}{M_{k,l} R}\ A_\pi^{k,l}\,.
\eeq
For $(k,0)$ and $(0,k)$ modes, the physical scalar is present for $k$ odd, when there is no vector state, while for $k$ even only 
the Goldstone boson is in the spectrum.
Finally, for a zero mode $(0,0)$, only a massless vector is present in the spectrum.

Concerning the ghost part of the Lagrangian, we obtain:
\beq
\mathcal{L}_{\mbox{ghost}} = \int_0^{2 \pi} d x_5\, d x_6\; \bar{c}^a\left\{\left(-\partial_{\mu} (D^{\mu})^{ac}+\xi\partial_{5} (D_{5})^{ac} +\xi\partial_{6} (D_{6})^{ac}\right)\right\}c^c\,,
\eeq
where $(D_{M})^{ac}=\partial_M \delta^{ac} -i g (t_G^b)^{ac} =\partial_M \delta^{ac} + g f^{abc}A_M^b$.
After KK expansion, we see that the ghost behaves like a scalar field with the same parity as the vector component and mass 
proportional to $\xi$.

\subsection*{Fermions}

The action for a 6D fermion can be written as
\beq \label{eq:fermaction}
S_{\rm ferm} = \int_0^{2 \pi} d x_5\, d x_6\; \frac{i}{2} \Big\{ \bar \Psi \Gamma^M D_M \Psi - \left( D_M \bar \Psi_\pm \right) \Gamma^M \Psi_\pm \Big\}
\eeq
where $D_M=\partial_M-igA^a_Mt_r^a$. 
Here $\Gamma_M$ are 8$\times$8 6D Dirac matrices and $\Psi = (\chi_+, \bar{\eta}_-, \chi_-, \bar{\eta}_+)^T$ is an 8-component 
6D fermion (and $\chi$ and $\eta$ are 2-component Weyl spinors). Note that mass terms are not allowed by the rotation and that the two 
6D chiralities, labelled with a $\pm$ and in general independent, are exchanged by the glide.
A detailed description of the spectrum can be found in \cite{Cacciapaglia:2009pa}: here we will limit ourselves to summarise 
the main features. The parity under the rotation determines if the zero mode is left- or right-handed in 4D, while the parity under 
the glide affects the wave functions but not the spectrum.
In each $(k,0)$ and $(0,l)$ there is one massive state, while levels $(k,l)$ contain two degenerate states with mass $M_{k,l}$.
The wave functions of the two degenerate states are parameterised by an arbitrary mixing parameter $\theta$:
as it will be clear in the following sections, loop corrections will lift the degeneracy and determine the value of $\theta$ that 
describes the two mass eigenstates.
It will be useful later to chose $\theta = 0$, and for short notation we will label $a$ and $b$ the two states.
The wave functions of the two degenerate states (for a left-handed zero mode) are given by
\beq
\Psi_a^{(+ p_g)} =   \frac{1}{\sqrt{2} \pi R} \left( \begin{array}{c}
\left( \cos k x_5\, \cos l x_6  \right) \chi_a^{k,l} \\
p_g (-1)^{k+l} \left( -\frac{k}{\sqrt{k^2+l^2}} \sin k x_5\, \cos l x_6 + \frac{i l}{\sqrt{k^2+l^2}} \cos k x_5\, \sin l x_6 \right) \bar{\eta}_a^{k,l}\\
p_g (-1)^{k+l} \left(\cos k x_5\, \cos l x_6\right) \chi_a^{k,l} \\
\left(  -\frac{k}{\sqrt{k^2+l^2}}  \sin k x_5\, \cos l x_6 - \frac{i l}{\sqrt{k^2+l^2}} \cos k x_5\, \sin l x_6 \right) \bar{\eta}_a^{k,l} 
\end{array} \right)\,,
\eeq
and
\beq
\Psi_b^{(+ p_g)} =  \frac{1}{\sqrt{2} \pi R} \left( \begin{array}{c}
\left( \sin k x_5\, \sin l x_6 \right) \chi_b^{k,l} \\
 p_g (-1)^{k+l} \left( \frac{i l}{\sqrt{k^2+l^2}} \sin k x_5\, \cos l x_6 - \frac{k}{\sqrt{k^2+l^2}}  \cos k x_5\, \sin l x_6 \right) \bar{\eta}_b^{k,l}\\
p_g (-1)^{k+l} \left(  -  \sin k x_5\, \sin l x_6 \right) \chi_b^{k,l} \\
\left( \frac{i l}{\sqrt{k^2+l^2}} \sin k x_5\, \cos l x_6 +  \frac{k}{\sqrt{k^2+l^2}} \cos k x_5\, \sin l x_6 \right) \bar{\eta}_b^{k,l} 
\end{array} \right)\,,
\eeq
where $\chi^{k,l}_{a/b}$ and $\eta^{k,l}_{a/b}$ are the left- and right-handed components respectively of the 4D Dirac fermions.

\section{Gauge bosons}
\label{sec:gaugeb}
\setcounter{equation}{0}
\setcounter{footnote}{0}

In this section we discuss the loops generating mixing in the gauge sector between the $(2k, 2l)$ mode and zero modes.
We will consider only corrections to the two and three point functions and show that in the $\xi=-3$ gauge the two terms respect 
gauge invariance in the sense that the effective couplings are related to each other and correspond to a gauge invariant 
counter-term localised on the two singular points.
It is interesting to notice that fermion loops do not contribute, because of the presence of the two chiralities.

\subsection{Bilinear mixing terms}

Here we will consider mixing terms between the vector and two scalar modes in the level $(2k,2l)$ with the vector zero mode, 
which corresponds to a SM gauge boson.
The results have been computed in the usual dimensional regularisation in 4D, and they show the expected logarithmic sensitivity 
to the cut-off of the theory.

\subsubsection*{Vector mixing: $A_\mu^{2k,2l} - A_\nu^{0,0}$}
\setlength{\unitlength}{1cm}
\begin{figure}[!tb]
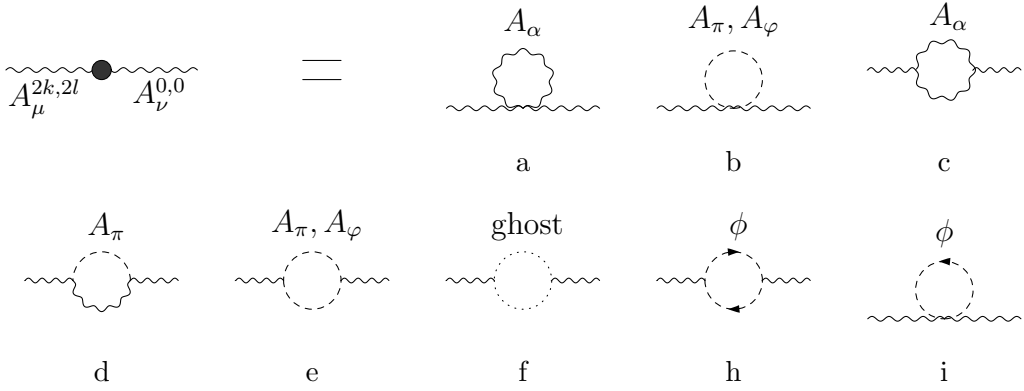

\begin{center}
\begin{feynartspicture}(14,7)(5,2)

\FADiagram{}
\FAProp(-0.,10.)(10.,10.)(0.,){/Sine}{0}
\FALabel(4.,8.98)[t]{$A_{\mu}^{2k,2l}$}
\FAProp(10.,10.)(20.,10.)(0.,){/Sine}{0}
\FALabel(16.,9.18)[t]{$A_{\nu}^{0,0}$}
\FAVert(10.,10.){-3}
\FADiagram{}
\FAProp(9.,11.)(13.,11.)(0.,){/Straight}{0}
\FAProp(9,9.)(13.,9.)(0.,){/Straight}{0}

\FADiagram{a}
\FAProp(2.,6.)(10.,6.)(0.,){/Sine}{0}
\FAProp(10.,6.)(18.,6.)(0.,){/Sine}{0}
\FAProp(10.,6.)(10.,6.)(10.,12.){/Sine}{0}
\FALabel(10.,13.52)[b]{$A_{\alpha}$}

\FADiagram{b}
\FAProp(2.,6.)(10.,6.)(0.,){/Sine}{0}
\FAProp(10.,6.)(18.,6.)(0.,){/Sine}{0}
\FAProp(10.,6.)(10.,6.)(10.,12.){/ScalarDash}{0}
\FALabel(10.,13.52)[b]{$A_\pi, A_\varphi$}

\FADiagram{c}
\FAProp(2.,10.)(7.,10.)(0.,){/Sine}{0}
\FAProp(7.,10.)(7.,10.)(13.,10.){/Sine}{0}
\FALabel(10.48,14.2)[b]{$A_\alpha$}
\FAProp(13.,10.)(18.,10.)(0.,){/Sine}{0}
\FALabel(10.48,4)[b]{}

\FADiagram{d}
\FAProp(2.,10.)(7.,10.)(0.,){/Sine}{0}
\FAProp(7.,10.)(13.,10.)(1,){/Sine}{0}
\FAProp(13.,10.)(7.,10.)(1,){/ScalarDash}{0}
\FALabel(10.48,14.2)[b]{$A_\pi$}
\FAProp(13.,10.)(18.,10.)(0.,){/Sine}{0}
\FALabel(10.48,4)[b]{$ $}

\FADiagram{e}
\FAProp(2.,10.)(7.,10.)(0.,){/Sine}{0}
\FAProp(7.,10.)(7.,10.)(13.,10.){/ScalarDash}{0}
\FALabel(10.48,14.2)[b]{$A_\pi,A_\varphi$}
\FAProp(13.,10.)(18.,10.)(0.,){/Sine}{0}

\FADiagram{f}
\FAProp(2.,10.)(7.,10.)(0.,){/Sine}{0}
\FAProp(7.,10.)(7.,10.)(13.,10.){/GhostDash}{0}
\FALabel(10.48,14.2)[b]{ghost}
\FAProp(13.,10.)(18.,10.)(0.,){/Sine}{0}

\FADiagram{h}
\FAProp(2.,10.)(7.,10.)(0.,){/Sine}{0}
\FAProp(7.,10.)(13.,10.)(1,){/ScalarDash}{-1}
\FAProp(13.,10.)(7.,10.)(1,){/ScalarDash}{-1}
\FALabel(10.48,14.2)[b]{$\phi$}
\FAProp(13.,10.)(18.,10.)(0.,){/Sine}{0}

\FADiagram{i}
\FAProp(2.,6.)(10.,6.)(0.,){/Sine}{0}
\FAProp(10.,6.)(18.,6.)(0.,){/Sine}{0}
\FAProp(10.,6.)(10.,6.)(10.,12.){/ScalarDash}{1}
\FALabel(10.,13.52)[b]{$\phi$}
\end{feynartspicture}
\end{center}
\caption{\footnotesize $A_\mu^{2k,2l} - A_\nu^{0,0} $ mixing: loops a--f are gauge corrections while h and i are loops of a scalar field.}
\label{fig:gaugemixing}
\end{figure}

For gauge loops, the mixing can be parameterised as~\footnote{The group theory factor is defined as $f^{acd} f^{bcd} = C_2 (G) \delta^{ab}$. For SU(N), $C_2(G) = N$.}:
\begin{eqnarray}
i \Pi^{\mu \nu} =\frac{g^2 C_2(G) \delta^{ab}}{16 \pi^4} \left[\left(\lambda_{1} \frac{i \pi^2}{\epsilon}+ \kappa_1 
\frac{ i \pi^3}{\eta}\right) g^{\mu \nu} + \lambda_{2} \frac{i \pi^2}{\epsilon} q^{\mu }q^{\nu}\right] \,, \label{eq:gaugemixing}
\end{eqnarray}
where $1/\epsilon = \log \Lambda R$ and $1/\eta = 1/4\, \Lambda^2$. We have computed the log divergent parts in the usual 4D 
dimensional regularisation scheme in $d = 4 - \epsilon$ and the quadratic divergence in 2D dimensional regularisation with 
$d=2-\eta$.
The same results can be obtained with a cut-off regularisation, with $\Lambda$ equal to the UV cut-off of the integrals.
For the scalar loop, the same parameterisation is valid: the only difference is that the group factor in front is different.
For a non Abelian group and a scalar in the representation $r_h$, the coefficient $C_2(G)$ is replaced by 
$C(r_h)$~\footnote{$\mbox{Tr} (t_r^a t_r^b) = C (r) \delta^{ab}$. For a fundamental of SU(N), $C(N) = \frac{1}{2}$.}; while for a U(1),  it is replaced by the square of the charge of the scalar field $Y_h^2$ times the 
number of complex components of the scalar.

The contribution of each loop to the coefficients $\lambda_1$, $\lambda_2$ and $\kappa_1$ are listed in table~\ref{tab:gauge2} 
(for simplicity we define $M = M_{k,l} = \sqrt{k^2+l^2} \, m_{KK}$).
\begin{table}[tb]
\begin{center} 
\begin{tabular}{|lc|lc|lc|}
\hline
$\lambda_{1a}$ & $\frac{1}{4}\left(-3M^2(\xi^2+3)\right)$ & $\kappa_{1a}$ & $2(\xi+1)$ & $\lambda_{2a}$& $0 $ \\
\hline
$\lambda_{1b}$ & $ M^2(\xi+1)$ &  $\kappa_{1b}$ & $-8 $ & $\lambda_{2b}$& $0 $\\
\hline
$\lambda_{1c}$ & $\frac{1}{12}\left( 9M^2(\xi^2+\xi+4)+q^2(25-6\xi)\right)$ & $\kappa_{1c}$ &  $ -2(\xi+2)$ & $\lambda_{2c}$& $\frac{1}{6}\left(3\xi-14\right) $\\
\hline
$\lambda_{1d}$ & $\frac{1}{4}\left(3M^2(\xi+3) \right)$ & $\kappa_{1d}$ &  $ 0$ & $\lambda_{2d}$& $0 $\\
\hline
$\lambda_{1e}$ &   $\frac{1}{6}\left(-6M^2(\xi+1)+2q^2 \right)$ & $\kappa_{1e}$ &  $ 8$ & $\lambda_{2e}$& $-\frac{1}{3}$\\
\hline
$\lambda_{1f}$ & $ \frac{1}{12}\left(-6M^2\xi+q^2\right)$ & $\kappa_{1f}$ &  $ 2$ & $\lambda_{2f}$& $\frac{1}{6} $\\
\hline
\hline
$\lambda_{1gauge}$ & $\frac{1}{2}\left(2M^2(\xi+3)-q^2(\xi-5)\right)$ & $\kappa_{1gauge}$ &  $0$& $\lambda_{2gauge}$& $\frac{1}{2}\left(\xi-5\right)$ \\
\hline
\hline
$\lambda_{1h}$ &   $\frac{1}{3}\left(6M^2-q^2 \right)$ & $\kappa_{1h}$ &  $ -8$ & $\lambda_{2h}$& $\frac{1}{3}$\\
\hline
$\lambda_{1i}$ &   $-2M^2$ & $\kappa_{1i}$ &  $ +8$ & $\lambda_{2i}$& $0$\\
\hline
\hline
$\lambda_{1scalar}$ & $-\frac{1}{3} q^2$ & $\kappa_{1scalar}$ &  $0$& $\lambda_{2scalar}$& $\frac{1}{3}$ \\
\hline
\end{tabular} 
\end{center}
\caption{\footnotesize Contribution of the loops in Figure~\ref{fig:gaugemixing} to the parameters in Eq.~\ref{eq:gaugemixing}. In the table, $M = M_{k,l} = \sqrt{k^2+l^2} \, m_{KK}$.} \label{tab:gauge2}
\end{table} 
Note that, as expected, the quadratic divergence vanishes.
The total contribution of gauge loops is therefore given by:
\begin{eqnarray} \label{eq:gauge2gauge}
i \Pi^{\mu \nu}_{\rm gauge}=\frac{g^2 C_2(G) \delta^{ab}}{16 \pi^4} \frac{i \pi^2}{\epsilon} \left[M^2(\xi+3)g^{\mu \nu}-\frac{(\xi-5)}{2} (q^2g^{\mu \nu}-q^{\mu }q^{\nu})\right]\,. 
\end{eqnarray}
The total contribution of scalar loops:
\begin{eqnarray}
i \Pi^{\mu \nu}_{\rm scalar}=\frac{g^2 C(r_h) \delta^{ab}}{16 \pi^4} \frac{i \pi^2}{\epsilon} \left[- \frac{1}{3} (q^2g^{\mu \nu}-q^{\mu }q^{\nu})\right]. \label{eq:gauge2scalar}
\end{eqnarray}

\subsubsection*{Physical scalar: $A_\varphi^{2k,2l} - A_\mu^{0,0} $}

This mixing is absent at 1-loop, because the trilinear coupling between a physical scalar and a gauge boson of level $(k,l)$ vanishes, 
and similarly for the ghosts.

\subsubsection*{Goldstone-vector mixing: $A_\pi^{2k,2l} - A_\mu^{0,0}$}

\setlength{\unitlength}{1cm}
\begin{figure}[!tb]
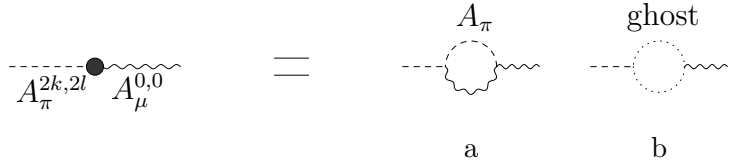

\begin{center}
\begin{feynartspicture}(10,4)(4,1)
\FADiagram{}

\FAProp(-0.,10.)(10.,10.)(0.,){/ScalarDash}{0}
\FALabel(5.,8.98)[t]{$A_\pi^{2k,2l}$}
\FAProp(10.,10.)(20.,10.)(0.,){/Sine}{0}
\FALabel(15.,9.18)[t]{$A_{\mu}^{0,0}$}
\FAVert(10.,10.){-3}

\FADiagram{}
\FAProp(9.,11.)(13.,11.)(0.,){/Straight}{0}
\FAProp(9,9.)(13.,9.)(0.,){/Straight}{0}

\FADiagram{a}
\FAProp(2.,10.)(7.,10.)(0.,){/ScalarDash}{0}
\FAProp(7.,10.)(13.,10.)(1,){/Sine}{0}
\FAProp(13.,10.)(7.,10.)(1,){/ScalarDash}{0}
\FALabel(10.48,14.2)[b]{$A_\pi$}
\FAProp(13.,10.)(18.,10.)(0.,){/Sine}{0}
\FALabel(10.48,4)[b]{$ $}

\FADiagram{b}
\FAProp(2.,10.)(7.,10.)(0.,){/ScalarDash}{0}
\FAProp(7.,10.)(7.,10.)(13.,10.){/GhostDash}{0}
\FALabel(10.48,14.2)[b]{ghost}
\FAProp(13.,10.)(18.,10.)(0.,){/Sine}{0}
\end{feynartspicture}
\end{center}
\caption{\footnotesize Diagrams generating the $ A_\pi^{2k,2l} - A_\mu^{0,0} $ mixing. For the physical scalar $A_\varphi^{2k,2l}$ the mixing is absent.}
\label{fig:S2Ammixing}
\end{figure}

The mixing between the Goldstone boson and the zero mode is given by the diagrams in Figure~\ref{fig:S2Ammixing}: the scalar 
field does not contribute because the coupling of the Goldstone to two scalars in tier $(k,l)$ vanishes.
The general structure of the mixing is as follows
\begin{eqnarray}
i \Pi^{\mu}=\frac{ g^2 C_2(G) \delta^{ab}}{16 \pi^4} \frac{i \pi^2}{\epsilon}\ \lambda\ (-i M) q^{\mu}\,;
\end{eqnarray}
the contributions of the two diagrams in Figure~\ref{fig:S2Ammixing} to the coefficient $\lambda$ are given by
\beq
\lambda_a = - \frac{\xi}{2}\,, \qquad \lambda_b = - \frac{3}{2}\,.
\eeq
All in all, the mixing is given by:
\begin{eqnarray}
i \Pi^{\mu}=\frac{g^2 C_2(G) \delta^{ab}}{16 \pi^4} \frac{i \pi^2}{2\epsilon} \left[iM(\xi+3)q^{\mu }\right]\,; \label{eq:goldstone}
\end{eqnarray}
the presence of this new mixing between Goldstone and vector would require a redefinition of the gauge fixing parameter $\xi$ at 
one loop. Note however that the definition of the Goldstone field is not modified because there is no mixing generated with the physical 
scalar $A_\varphi$, and also that the mixing vanishes in the $\xi=-3$ gauge.

\subsection{Trilinear couplings}

In this section we list the results for the three point functions with one gauge state from level $(2k,2l)$ and two vector zero modes.
\subsubsection*{Gauge vector trilinear coupling}

\setlength{\unitlength}{1cm}
\begin{figure}[!tb]
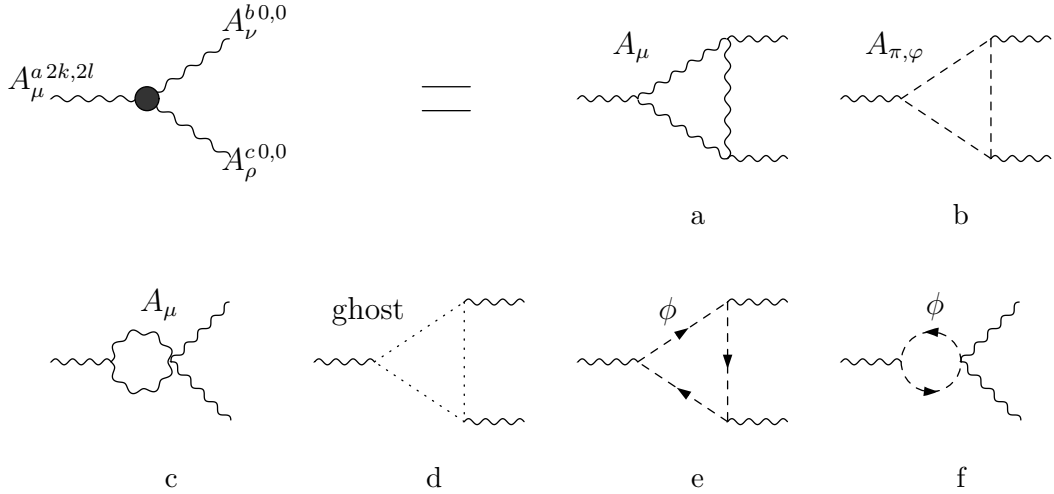

\begin{center}
\begin{feynartspicture}(16,7)(4,2)

\FADiagram{}
\FAProp(-0.,10.)(8.,10.)(0.,){/Sine}{0}
\FAProp(8.,10.)(15.,15.)(0.,){/Sine}{0}
\FAProp(8.,10.)(15.,5.)(0.,){/Sine}{0}
\FAVert(8.,10.){-3}
\FALabel(17.,15.02)[b]{$A_\nu^{b\, 0,0}$}
\FALabel(17.,6.18)[t]{$A_\rho^{c\, 0,0}$}
\FALabel(-0,13)[t]{$A_\mu^{a\, 2k,2l}$}

\FADiagram{}
\FAProp(9.,11.)(13.,11.)(0.,){/Straight}{0}
\FAProp(9,9.)(13.,9.)(0.,){/Straight}{0}

\FADiagram{a}
\FAProp(0.,10.)(5.,10.)(0.,){/Sine}{0}
\FAProp(5.,10.)(12.5,15.)(0.,){/Sine}{0}
\FAProp(12.5,15.)(12.5,5.)(0.,){/Sine}{0}
\FALabel(13.52,10.)[l]{}
\FAProp(12.5,5.)(5.,10.)(0.,){/Sine}{0}
\FAProp(12.5,15.)(17.5,15.)(0.,){/Sine}{0}
\FAProp(12.5,5.)(17.5,5.)(0.,){/Sine}{0}
\FALabel(4.48,13.02)[b]{$A_\mu$}
\FALabel(7.48,6.98)[t]{}

\FADiagram{b}
\FAProp(0.,10.)(5.,10.)(0.,){/Sine}{0}
\FAProp(5.,10.)(12.5,15.)(0.,){/ScalarDash}{0}
\FAProp(12.5,15.)(12.5,5.)(0.,){/ScalarDash}{0}
\FALabel(13.52,10.)[l]{}
\FAProp(12.5,5.)(5.,10.)(0.,){/ScalarDash}{0}
\FAProp(12.5,15.)(17.5,15.)(0.,){/Sine}{0}
\FAProp(12.5,5.)(17.5,5.)(0.,){/Sine}{0}
\FALabel(4.48,13.02)[b]{$A_{\pi, \varphi}$}
\FALabel(7.48,6.98)[t]{}

\FADiagram{c}
\FAProp(0.,10.)(5.,10.)(0.,){/Sine}{0}
\FAProp(10.,10.)(15.,15.)(0.,){/Sine}{0}
\FAProp(10.,10.)(15.,5.)(0.,){/Sine}{0}
\FAProp(5.,10.)(5.,10.)(10.,10.){/Sine}{0}
\FALabel(10.6071,13.5171)[br]{$A_\mu$}
\FALabel(11,7.0607)[tr]{}

\FADiagram{d}
\FAProp(0.,10.)(5.,10.)(0.,){/Sine}{0}
\FAProp(5.,10.)(12.5,15.)(0.,){/GhostDash}{0}
\FALabel(4.48,13.02)[b]{ghost}
\FAProp(12.5,15.)(12.5,5.)(0.,){/GhostDash}{0}
\FAProp(12.5,5.)(5.,10.)(0.,){/GhostDash}{0}
\FALabel(7.48,6.98)[t]{}
\FAProp(12.5,15.)(17.5,15.)(0.,){/Sine}{0}
\FAProp(12.5,5.)(17.5,5.)(0.,){/Sine}{0}

\FADiagram{e}
\FAProp(0.,10.)(5.,10.)(0.,){/Sine}{0}
\FAProp(5.,10.)(12.5,15.)(0.,){/ScalarDash}{1}
\FALabel(7.48,13.02)[b]{$\phi$}
\FAProp(12.5,15.)(12.5,5.)(0.,){/ScalarDash}{1}
\FAProp(12.5,5.)(5.,10.)(0.,){/ScalarDash}{1}
\FALabel(7.48,6.98)[t]{}
\FAProp(12.5,15.)(17.5,15.)(0.,){/Sine}{0}
\FAProp(12.5,5.)(17.5,5.)(0.,){/Sine}{0}

\FADiagram{f}
\FAProp(0.,10.)(5.,10.)(0.,){/Sine}{0}
\FAProp(10.,10.)(15.,15.)(0.,){/Sine}{0}
\FAProp(10.,10.)(15.,5.)(0.,){/Sine}{0}
\FAProp(5.,10.)(10.,10.)(1.,){/ScalarDash}{1}
\FAProp(10.,10.)(5.,10.)(1.,){/ScalarDash}{1}
\FALabel(8.6071,13.5171)[br]{$\phi$}
\FALabel(11,7.0607)[tr]{}
\end{feynartspicture}
\end{center}
\caption{\footnotesize Vector $A_\mu^{2k,2l} $ coupling to two SM vectors. In graphs (c) and (f), the graphs with loops on 
the zero mode legs are included.}
\label{fig:gaugetriangle}
\end{figure}

The effective coupling can be written as:
\beq
i V^{\mu\nu\rho}=\frac{ig^3 f^{abc}C_2(G)}{16 \pi^4}\frac{i \pi^2}{4\epsilon}\ \tau \  \mathcal{O}^{\mu\nu\rho}\,,
\eeq
where $\mathcal{O}^{\mu\nu\rho}= (q_2^\mu-q_1^\mu)g^{\nu\rho} -(q_1^\nu+2q_2^\nu) g^{\mu\rho}+(2q_1^\rho+q_2^\rho)
g^{\mu\nu}$  is the usual tensor from tree level trilinear couplings, $q_1^\nu$ and $q_2^\rho$ being the momenta of the zero 
modes. For the scalar loops, the gauge factor $C_2(G)$ is replaced by $C(r_h)$ for an SU(N) representation $r_h$.
Note that the vertex vanishes for a U(1) gauge group, because the tensor is antisymmetric.
The contributing loops are in Figure~\ref{fig:gaugetriangle}, and their individual contribution is listed in table~\ref{tab:gauge3}.
\begin{table}[tb]
\begin{center} \begin{tabular}{|l|c||l|c|}
\hline
$\tau_{a}$ & $-\frac{1}{2}(9\xi+4)$ & $\tau_{e} $ & $-\frac{4}{3}$ \\
\hline
$\tau_{b}(A_\pi + A_\varphi)$ & $\frac{4}{3}$ & $\tau_{f} $ & $0$ \\
\hline
$\tau_{c}$ & $\frac{3}{2}(\xi+5)$ & & \\
\hline
$\tau_{d}$ & $\frac{1}{6}$ & & \\
\hline
\hline
$\tau_{gauge}$ & $(-3\xi+7)$  & $\tau_{scalar} $ & $-\frac{4}{3}$ \\
\hline
\end{tabular} \end{center}
\caption{\footnotesize Contributions to the trilinear vector coupling of diagrams in Figure~\ref{fig:gaugetriangle}} \label{tab:gauge3}
\end{table}
The total gauge contribution is therefore given by:
\begin{align}
i V^{\mu\nu\rho}_{\rm gauge}=\frac{i g^3 f^{abc}C_2(G)}{16 \pi^4}\frac{i \pi^2}{4\epsilon}\ (-3\xi+7) \  \mathcal{O}^{\mu\nu\rho} \label{eq:gauge3gauge}
\end{align}
For the scalar contribution:
\begin{align}
i V^{\mu\nu\rho}_{\rm scalar} = \frac{i g^3 f^{abc}C(r_h)}{16 \pi^4}\left(\frac{-i \pi^2}{3\epsilon}\right) \  \mathcal{O}^{\mu\nu\rho}\,. \label{eq:gauge3scalar}
\end{align}

\subsubsection*{Goldstone $A_\pi^{2k,2l}$ to two vectors}

\begin{figure}[!tb]
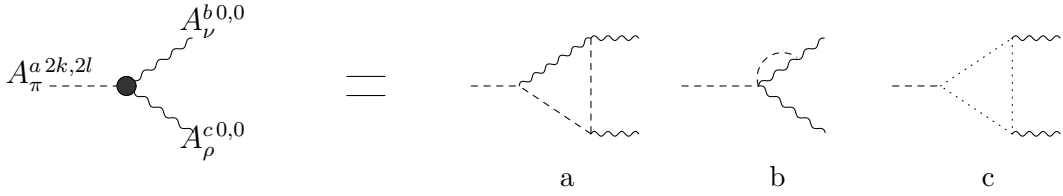

\begin{center}
\begin{feynartspicture}(14,4)(5,1)

\FADiagram{}
\FAProp(-0.,10.)(8.,10.)(0.,){/ScalarDash}{0}
\FAProp(8.,10.)(15.,15.)(0.,){/Sine}{0}
\FAProp(8.,10.)(15.,5.)(0.,){/Sine}{0}
\FAVert(8.,10.){-3}
\FALabel(17.,15.02)[b]{$A_\nu^{b\, 0,0}$}
\FALabel(17.,6.18)[t]{$A_\rho^{c\, 0,0}$}
\FALabel(-0,13)[t]{$A_\pi^{a\, 2k,2l}$}

\FADiagram{}
\FAProp(9.,11.)(13.,11.)(0.,){/Straight}{0}
\FAProp(9,9.)(13.,9.)(0.,){/Straight}{0}

\FADiagram{a}
\FAProp(0.,10.)(5.,10.)(0.,){/ScalarDash}{0}
\FAProp(5.,10.)(12.5,15.)(0.,){/Sine}{0}
\FAProp(12.5,15.)(12.5,5.)(0.,){/ScalarDash}{0}
\FAProp(12.5,5.)(5.,10.)(0.,){/ScalarDash}{0}
\FAProp(12.5,15.)(17.5,15.)(0.,){/Sine}{0}
\FAProp(12.5,5.)(17.5,5.)(0.,){/Sine}{0}

\FADiagram{b}
\FAProp(0.,10.)(8.,10.)(0.,){/ScalarDash}{0}
\FAProp(8.,10.)(15.,15.)(0.,){/Sine}{0}
\FAProp(8.,10.)(15.,5.)(0.,){/Sine}{0}
\FAProp(8.,10.)(12.,13.)(-0.7285,){/ScalarDash}{0}

\FADiagram{c}
\FAProp(0.,10.)(5.,10.)(0.,){/ScalarDash}{0}
\FAProp(5.,10.)(12.5,15.)(0.,){/GhostDash}{0}
\FAProp(12.5,15.)(12.5,5.)(0.,){/GhostDash}{0}
\FAProp(12.5,5.)(5.,10.)(0.,){/GhostDash}{0}
\FAProp(12.5,15.)(17.5,15.)(0.,){/Sine}{0}
\FAProp(12.5,5.)(17.5,5.)(0.,){/Sine}{0}
\end{feynartspicture}
\end{center}
\caption{\footnotesize $A_\pi^{2k,2l}$ coupling to SM vectors}
\label{fig:scalartriangle}
\end{figure}

The coupling of the Goldstone with two zero mode vectors is given by the loops in  Fig.~\ref{fig:scalartriangle}. 
However, we can show that the loops give a finite result: for loops $a$ and $b$, the divergent part cancels out after symmetrising 
in the vector legs, while the ghost loop is finite.

\subsubsection*{Scalar $A_\varphi^{2k,2l}$ to SM vectors}

We can show that there is no way to close the loop by using the new extra physical scalar. So in this case, this field does not 
couple to SM model gauge bosons.

\subsection{Gauge counter-terms}

The divergences in these loops correspond to the projection under the rotation and they require the presence of counter-terms 
localised on the singular points of the geometry: the rotation has four fixed points (the vertices of the fundamental square), while 
the glide identifies them in pairs.
At the end, the flat Real Projective Plane has two singular points: $(0, 0) \sim (\pi R, \pi R)$ and $(0, \pi R) \sim (\pi R, 0)$.
For convenience and to keep the glide invariance explicit, we define two localisation operators~\cite{Cacciapaglia:2009pa}:
\beq
\delta_0 &=& \frac{1}{2} \left( \delta (x_5) \delta (x_6) +  \delta (x_5 - \pi R) \delta (x_6 - \pi R)  \right)\,, \\
\delta_\pi &=& \frac{1}{2} \left( \delta (x_5) \delta (x_6-\pi R) +  \delta (x_5 - \pi R) \delta (x_6)  \right)\,,
\eeq
and label the two singular points with a subscript ``$0$'' for the point $(0,0) = (\pi R, \pi R)$, and ``$\pi$'' for $(0,\pi R) = (\pi R,0)$.
In general the two counter-term Lagrangians can be different: however, the bulk is invariant under a rotation by 90 degrees around 
the centre, symmetry that exchange the two singular points.
Therefore, as our purpose is to study the divergence structure of the bulk loops, we can consider identical counter-terms on the 
two fixed points.
The generic counter-term Lagrangian will therefore be
\beq
\mathcal{\delta L} = \frac{\delta_0 + \delta_\pi}{\Lambda^2} \mathcal{L}_{\rm c-t}\,,
\eeq
where the cut-off $\Lambda$ shows that we are interested in operators of dimension 6 in $\mathcal{L}_{\rm c-t}$ (note that the 
localisation operators have dimension 2 in mass, while $\mathcal{\delta L}$ has dimension 6).
Operators of dimension 6 in the 6D fields are, however, kinetic terms in the bulk Lagrangian!

In general the localised interactions must only respect 4-dimensional Lorentz invariance.
Furthermore, they do depend on the gauge choice, thus they are generically not gauge invariant operators.
This is true in any number of dimensions.
An important observation is that the counter-terms assume a gauge invariant form in a special gauge $\xi = -3$: this fact is also true 
in 4 dimensions~\cite{4Dmagic}.
It is crucial to use this property here: in fact, due to the presence of extra gauge scalars, the number of operators that one can write down is very large!
Imposing gauge invariance, allows us to reduce the number to a manageable level.
In the following, therefore, we will restrict ourselves to gauge invariant operators and verify that the $\xi=-3$ gauge does work in 
our model too.
One may naively expect that only 4D gauge invariance can be imposed, because only the vector part of the 6D vector $A_M$ is 
non-zero on the fixed points (both $A_5$ and $A_6$ are odd under the rotation symmetry, therefore their wave functions vanish 
on the singular points).
The 6D vector transforms in general as $A_M \to A_M + \partial_M \theta (x^\mu, x_5, x_6)$, where the gauge parameter $\theta$ 
has the same parities of the $A_\mu$ component. While $A_5$, for instance, vanishes on the singular points, $\partial_5 A_5$ 
does not. However, the residual gauge transformation $\partial_5 A_5 \to \partial_5 A_5 + \partial_5^2 \theta$ constraints the 
possible couplings containing $\partial_5$.
In order to preserve the full gauge invariance, we therefore construct the operators using the energy-stress tensor components 
$F_{\mu \nu}$, $F_{5 \mu}$, $F_{6\mu}$ and $F_{56}$.
The most general counter-term Lagrangian is~\cite{Ponton:2005kx}:
\beq
\mathcal{L}_{\rm c-t} = - \frac{r_{1}}{4} F_{\mu \nu}^2 - \frac{r_{2}}{2} F_{56}^2 + \frac{r_{5}}{2} 
F_{5 \mu}^2 +  \frac{r_{6}}{2} F_{6 \mu}^2 +  \frac{r_{56}}{2} F_{5 \mu} F_6^\mu\,;
\eeq
for a standard model gauge boson, with parities $(+,+)$, both $A_{5,6}$ and 
$\partial_{5,6} A_\mu$ vanish on the singular points, therefore $F_{5 \mu} = F_{6 \mu} =  0$ and 
$F_{56} = \partial_5 A_6 - \partial_6 A_5$.
The counter-term Lagrangian reduces to two operators:
\beq
\mathcal{L}_{\rm c-t} = - \frac{r_{1}}{4} F_{\mu \nu}^2 - \frac{r_{2}}{2} ( \partial_5 A_6 - \partial_6 A_5)^2\,.
\eeq
Note that gauge-invariance forbids any new mixing between the gauge scalars and the vector, which is in agreement with Eq.~\ref{eq:goldstone} in the magic gauge, and that scalars only receive a 
correction to the masses but not to couplings, as we found in the previous section.
For a zero mode, the wave function is constant and equal~\footnote{Here we integrate the 6D Lagrangian on the torus 
fundamental space, $x_5 \subset [0, 2 \pi R)$ and $x_6 \subset [0, 2 \pi R)$. This is however equivalent to integrating over the 
fundamental square $x_5, x_6 \subset [0, \pi R)$.} to $\frac{1}{2 \pi R}$, therefore the counter-term Lagrangian for zero modes is
\beq
\left. \delta \mathcal{L} \right|_{(0,0)} = \frac{2 r_{1}}{4 \pi^2 \Lambda^2 R^2} \left\{ -  \frac{1} {4} (F_{\mu \nu}^{0,0})^2 \right\}\,,
\eeq
where the 4D gauge coupling is defined as $g_4 = g_6/(2 \pi R)$.
For the mixing of a level $(2k,2l)$ vector and a zero mode, we have an extra factor of 2 from the different normalisation of the 
wave function of the massive state.
Therefore, the mixing and trilinear couplings from the counter-term are
\beq
A_\mu^{2k,2l} A_\mu^{0,0} &\rightarrow& i \Pi_{\mu \nu} = i \frac{r_1}{\pi^2 \Lambda^2 R^2} 
\left( q^2 g_{\mu \nu} - q_\mu q_\nu \right) \,, \\
A_\mu^{2k,2l} A_\mu^{0,0} A_\rho^{0,0} &\rightarrow& i V_{\mu \nu \rho} = i \frac{r_1}{\pi^2 \Lambda^2 R^2} g 
f^{abc} \mathcal{O}_{\mu \nu \rho} \,.
\eeq
This result implies that, in the magic gauge, the coefficient of the vector mixing and trilinear couplings should be the same: this is in fact true for Eq.s~\ref{eq:gauge2gauge} and \ref{eq:gauge3gauge} for $\xi=-3$, while the scalar contributions in Eq.s ~\ref{eq:gauge2scalar} and \ref{eq:gauge3scalar} also agree.
Therefore, in the magic gauge we can express the counter-term coefficients as
\beq
\frac{r_1}{\pi^2 \Lambda^2 R^2} = \frac{g^2 C_2 (G)}{4 \pi^2 \epsilon} = \frac{\alpha C_2(G)}{\pi} \log \Lambda R\,.
\eeq
Finally, putting together the contributions of the gauge bosons and the Higgs in the Standard Model, for the three SM gauge 
groups U(1) (with gauge coupling $g_1$), SU(2) ($g_2$) and SU(3) ($g_3$), we obtain the three counter-terms:
\beq
\frac{r_1^{(1)}}{\pi^2 \Lambda^2 R^2} & = & - \frac{2 Y_h^2}{12} \frac{\alpha_1}{\pi} \log \Lambda R = - \frac{1}{24}  
\frac{\alpha_1}{\pi} \log \Lambda R\,; \\
\frac{r_1^{(2)}}{\pi^2 \Lambda^2 R^2} & = & \left( C_2 (G) - \frac{1}{12} C (r_h) \right)\frac{\alpha_2}{\pi} 
\log \Lambda R = \frac{47}{24}  \frac{\alpha_2}{\pi} \log \Lambda R\,; \\
\frac{r_1^{(3)}}{\pi^2 \Lambda^2 R^2} & = & C_2 (G) \frac{\alpha_3}{\pi} \log \Lambda R = 3  \frac{\alpha_3}{\pi} \log \Lambda R\,.
\eeq
For completeness, the value of $r_2$ should be calculated from the corrections to odd tiers, that contain a physical scalar in the 
spectrum. As an example, in~\cite{Cacciapaglia:2009pa} we computed the mass corrections to the gauge scalars in the levels $(n,0)$ and $(0,n)$ with 
$n$ odd: comparing the mass corrections with the counter-term we obtain:
\beq
\frac{r_2^{(1)}}{\pi^2 \Lambda^2 R^2} & = &  \frac{2 Y_h^2}{4} \frac{\alpha_1}{\pi} \log \Lambda R =  \frac{1}{8}  
\frac{\alpha_1}{\pi} \log \Lambda R\,; \\
\frac{r_2^{(2)}}{\pi^2 \Lambda^2 R^2} & = & \left( C_2 (G) + \frac{1}{4} C (r_h) \right)\frac{\alpha_2}{\pi} \log \Lambda R = 
\frac{17}{8}  \frac{\alpha_2}{\pi} \log \Lambda R\,; \\
\frac{r_2^{(3)}}{\pi^2 \Lambda^2 R^2} & = & C_2 (G) \frac{\alpha_3}{\pi} \log \Lambda R = 3  \frac{\alpha_3}{\pi} \log \Lambda R\,.
\eeq

\section{Scalar fields}
\label{sec:scalar}
\setcounter{equation}{0}
\setcounter{footnote}{0}

Let us now turn to loop corrections to a scalar field.
For simplicity, we will quote the results for a representation $r$ of a SU(N) group, and how to modify them for U(1) gauge fields.
We will ignore the potentially large contribution of the scalar quartic coupling due to the poor understanding of the Higgs sector in universal extra dimension models and also because of our main interest in the gauge dependence of the result. As already mentioned, we also drop the dependency on the scalar bulk mass because we assume that it is of the order of the VEV.
Finally, fermion loops do not contribute.

\subsection{Mixing terms}
\label{sec:scalarMixingterms}
\setlength{\unitlength}{1cm}

The mixing between the massive scalar and the zero mode can be parametreised in general as~\footnote{$\sum_a t^a_r t^a_r = C_2 (r) \cdot 1$. For a fundamental of SU(N), $C_2 (N) = \frac{N^2-1}{2N}$.}
\begin{eqnarray}
i \Pi= - \frac{g^2 C_2(r_h)}{16 \pi^4}\left(\lambda \frac{i \pi^2}{\epsilon} + \kappa \frac{4 i \pi^3}{\eta}\right)\,. \label{eq:scalarmixing}
\end{eqnarray} 
In the case of a U(1), it would be enough to replace $C_2 (r_h) \to Y_h^2$.

\begin{figure}[!tb]
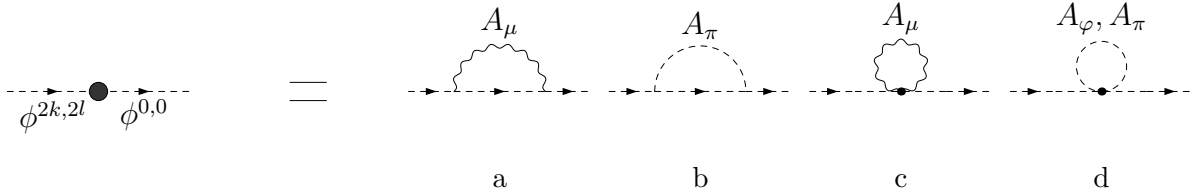

\begin{center}
\begin{feynartspicture}(16,4)(6,1)

\FADiagram{}
\FAProp(-0.,10.)(10.,10.)(0.,){/ScalarDash}{1}
\FALabel(5.,8.98)[t]{$\phi^{2k,2l}$}
\FAProp(10.,10.)(20.,10.)(0.,){/ScalarDash}{1}
\FALabel(15.,9.18)[t]{$\phi^{0,0}$}
\FAVert(10.,10.){-3}

\FADiagram{}
\FAProp(9.,11.)(13.,11.)(0.,){/Straight}{0}
\FAProp(9,9.)(13.,9.)(0.,){/Straight}{0}

\FADiagram{a}
\FAProp(-0.,10.)(5.,10.)(0.,){/ScalarDash}{1}
\FAProp(5.,10.)(15.,10.)(0.,){/ScalarDash}{1}
\FAProp(15.,10.)(20.,10.)(0.,){/ScalarDash}{1}
\FAProp(5.,10.)(15.,10.)(-1.,){/Sine}{0}
\FALabel(10.,15.82)[b]{$A_{\mu}$}

\FADiagram{b}
\FAProp(-0.,10.)(5.,10.)(0.,){/ScalarDash}{1}
\FAProp(5.,10.)(15.,10.)(0.,){/ScalarDash}{1}
\FAProp(15.,10.)(20.,10.)(0.,){/ScalarDash}{1}
\FAProp(5.,10.)(15.,10.)(-1.,){/ScalarDash}{0}
\FALabel(10.,15.82)[b]{$A_\pi$}

\FADiagram{c}
\FAProp(-0.,10.)(5.,10.)(0.,){/ScalarDash}{1}
\FAProp(5.,10.)(15.,10.)(0.,){/ScalarDash}{0}
\FAProp(15.,10.)(20.,10.)(0.,){/ScalarDash}{1}
\FAProp(10.,10.)(10.,10.)(10.,15.5){/Sine}{0}
\FALabel(10.,16.32)[b]{$A_{\mu}$}
\FAVert(10.,10.){0}

\FADiagram{d}
\FAProp(-0.,10.)(5.,10.)(0.,){/ScalarDash}{1}
\FAProp(5.,10.)(15.,10.)(0.,){/ScalarDash}{0}
\FAProp(15.,10.)(20.,10.)(0.,){/ScalarDash}{1}
\FAProp(10.,10.)(10.,10.)(10.,15.5){/ScalarDash}{0}
\FALabel(10.,16.32)[b]{$A_\varphi, A_\pi$}
\FAVert(10.,10.){0}
\end{feynartspicture}
\end{center}
\caption{\footnotesize Scalar mixings generated by gauge loops.} \label{fig:scalarmixing}
\end{figure}
The contribution of the diagrams in Fig.~\ref{fig:scalarmixing} are listed in table~\ref{tab:scalar2}.
Summing all the gauge contributions we obtain
\begin{eqnarray}
i \Pi= - \frac{g^2 C_2(r_h)}{16 \pi^4}\left(  (M^2 (2 \xi+1)-q^2 (\xi-3)) \frac{i \pi^2}{\epsilon}- \frac{4 i \pi^3}{\eta}\right)\,. \label{eq:scalar2}
\end{eqnarray}
Note that in this case the quadratic divergence remains, signalling the presence of a localised counter-term for the mass.
This localised mass will add up to the bulk mass for the scalar, in particular for the zero mode which we want to identify with 
the Higgs boson.
As we want the mass of the zero mode Higgs to be small, in particular smaller than the KK mass, we can consider two cases: 
either there is a cancellation between the bulk and localised masses, or the localised mass is small.
In both cases, this requires an additional fine tuning with respect to the 4D standard model case.
In the following, we will take the second point of view and assume that the localised mass is small, of the order of the electroweak scale.
Note also that a large localised mass would significantly distort the wave function of the Higgs boson, and therefore the distorted 
Higgs VEV would induce large mixing between different KK tiers.
\begin{table}[tb]
\begin{center} \begin{tabular}{|l|c|c|c|}
\hline
$\lambda_a$ & $(M^2 \xi (\xi+1)-q^2 (\xi-3))$ & $\kappa_a$ & $-\xi$ \\
\hline
$\lambda_b$ & $3 M^2$ &  $\kappa_b$ & $0$\\
\hline
$\lambda_c$ & $- M^2 (\xi^2+3)$ & $\kappa_c$ &  $\xi+1$ \\
\hline
$\lambda_d$ & $ M^2 (\xi+1)$ & $\kappa_d$ &  $-2$ \\
\hline
\hline
$\lambda_{\rm gauge}$ & $  (M^2 (2 \xi+1)-q^2 (\xi-3))$ & $\kappa_{\rm gauge}$ &  $-1$ \\
\hline
\end{tabular} \end{center}
\caption{\footnotesize Contributions to the scalar mixing of the diagrams in Figure~\ref{fig:scalarmixing}.} \label{tab:scalar2}
\end{table}

\subsection{Gauge couplings}

In this section, we calculate the loop contribution to trilinear couplings involving two scalars and a gauge field.
There are 2 kinds of couplings: a massive gauge field coupling to two zero mode scalars, and a massive scalar coupling to a zero 
mode scalar and a massless gauge boson.
\setlength{\unitlength}{1cm}

\subsubsection*{Vector $A_\mu^{2k,2l}$ to SM scalars}

\begin{figure}[!tb]
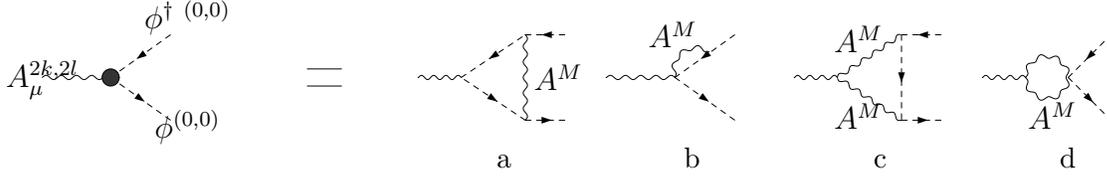

\begin{center}
\begin{feynartspicture}(15,4)(6,1)

\FADiagram{}
\FAProp(-0.,10.)(8.,10.)(0.,){/Sine}{0}
\FAProp(8.,10.)(15.,15.)(0.,){/ScalarDash}{-1}
\FAProp(8.,10.)(15.,5.)(0.,){/ScalarDash}{1}
\FAVert(8.,10.){-3}
\FALabel(17.,15.02)[b]{$\phi^{\dagger\ (0,0)}$}
\FALabel(17.,6.18)[t]{$\phi^{(0,0)}$}
\FALabel(-0,12)[t]{$A_\mu^{2k,2l}$}

\FADiagram{}
\FAProp(9.,11.)(13.,11.)(0.,){/Straight}{0}
\FAProp(9,9.)(13.,9.)(0.,){/Straight}{0}

\FADiagram{a}
\FAProp(0.,10.)(5.,10.)(0.,){/Sine}{0}
\FAProp(5.,10.)(12.5,15.)(0.,){/ScalarDash}{-1}
\FAProp(12.5,15.)(12.5,5.)(0.,){/Sine}{0}
\FALabel(13.52,10.)[l]{$A^{M}$}
\FAProp(12.5,5.)(5.,10.)(0.,){/ScalarDash}{-1}
\FAProp(12.5,15.)(17.5,15.)(0.,){/ScalarDash}{-1}
\FAProp(12.5,5.)(17.5,5.)(0.,){/ScalarDash}{1}

\FADiagram{b}
\FAProp(0.,10.)(8.,10.)(0.,){/Sine}{0}
\FAProp(8.,10.)(15.,15.)(0.,){/ScalarDash}{-1}
\FAProp(8.,10.)(15.,5.)(0.,){/ScalarDash}{1}
\FAProp(8.,10.)(12.,13.)(-0.7285,){/Sine}{0}
\FALabel(10.6071,13.5171)[br]{$A^{M}$}

\FADiagram{c}
\FAProp(0.,10.)(5.,10.)(0.,){/Sine}{0}
\FAProp(5.,10.)(12.5,15.)(0.,){/Sine}{0}
\FALabel(7.48,13.02)[b]{$A^{M}$}
\FAProp(12.5,15.)(12.5,5.)(0.,){/ScalarDash}{1}
\FAProp(12.5,5.)(5.,10.)(0.,){/Sine}{0}
\FALabel(7.48,6.98)[t]{$A^{M}$}
\FAProp(12.5,15.)(17.5,15.)(0.,){/ScalarDash}{-1}
\FAProp(12.5,5.)(17.5,5.)(0.,){/ScalarDash}{1}

\FADiagram{d}
\FAProp(0.,10.)(5.,10.)(0.,){/Sine}{0}
\FAProp(10.,10.)(15.,15.)(0.,){/ScalarDash}{-1}
\FAProp(10.,10.)(15.,5.)(0.,){/ScalarDash}{1}
\FALabel(11,7.0607)[tr]{$A^{M}$}
\FAProp(5.,10.)(5.,10.)(10.,10.){/Sine}{0}
\end{feynartspicture}
\end{center}
\caption{\footnotesize Coupling of $A_\mu^{2k,2l}$ with two SM scalars}
\label{fig:scalartriangle1}
\end{figure}

This coupling can be parameterised as:
\begin{eqnarray}
i V^\mu_A = \frac{g^3}{16 \pi^4}\frac{i \pi^2}{\epsilon}\  \tau_A \  (q_1^\mu-q_2^\mu) t^a
\end{eqnarray}
where $q_1$ and $q_2$ are the momenta of the two massless scalars. 
The loop computations are also done using dimensional regularisation, and the diagrams are given in Fig.~\ref{fig:scalartriangle1}.

For SU(N) gauge fields, the contributions of individual diagrams are listed in Table~\ref{tab:scalar3amu}.
\begin{table}[tb]
\begin{center} \begin{tabular}{|l|c|}
\hline
$\tau_{Aa}$ & $-(C_2(r_h)-1/2C_2(G)) \xi$  \\
\hline
$\tau_{Ab}$ & $\frac{3}{2} (2C_2(r_h)-1/2C_2(G))$\\
\hline
$\tau_{Ac}$ & $-\frac{3}{4} C_2(G)\xi$  \\
\hline
$\tau_{Ad}$ & $0$ \\
\hline
\hline
$\tau_{A tot}$ & $-\frac{1}{4}(4(\xi-3) C_2(r_h)+(\xi+3) C_2(G))$ \\
\hline
\end{tabular} \end{center}
\caption{\footnotesize Contributions to the coupling of a massive vector to massless scalars from diagrams in Figure~\ref{fig:scalartriangle1}.} \label{tab:scalar3amu}
\end{table}
Summing all the contributions, we obtain:
\begin{eqnarray}
i V^\mu_A = -\frac{g^3}{16 \pi^4}\frac{i \pi^2}{4\epsilon}\ (4(\xi-3) C_2(r_h)+(\xi+3) C_2(G)) \  (q_1^\mu-q_2^\mu) t^a\,. \label{eq:scalar3amu}
\end{eqnarray}
This result is valid if both the external and internal gauge bosons belong to the same SU(N) gauge group.
For U(1) gauge groups, it suffices to replace $C_2 (r_h) \to Y_\phi^2$, $C_2 (G) \to 0$ and $t^a \to Y_\phi$.
If the gauge bosons in the external and internal lines belong to different gauge groups, then only the diagrams a and b in 
Figure~\ref{fig:scalartriangle1} contribute: the result is the same as in the table with $C_2 (r_h) \to C_2 (r'_h)$ and $C_2 (G) \to 0$, 
and the total result is
\begin{eqnarray}
i V^\mu_{A'} = -\frac{g {g'}^2}{16 \pi^4}\frac{i \pi^2}{\epsilon}\ (\xi-3) C_2(r'_h) \  (q_1^\mu-q_2^\mu) t^a\,.
\end{eqnarray}
If the external gauge boson is a U(1), $t^a \to Y_\phi$; if the internal one is a U(1), $C_2(r'_h) \to Y_\phi^2$.

\subsubsection*{Gauge scalar $A_{\varphi}^{2k,2l}$ and Goldstone $A_{\pi}^{2k,2l}$ to SM scalars}

\begin{figure}[!tb]
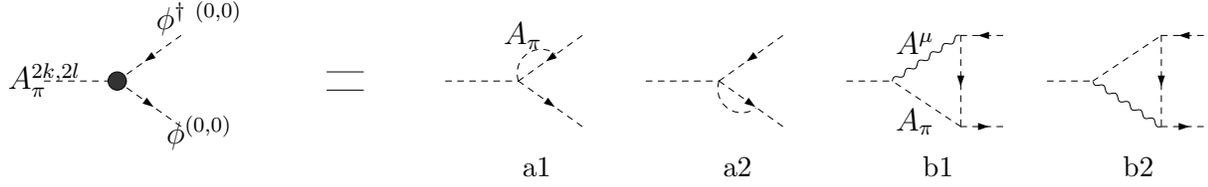

\begin{center}
\begin{feynartspicture}(16,5)(6,1)

\FADiagram{}
\FAProp(-0.,10.)(8.,10.)(0.,){/ScalarDash}{0}
\FAProp(8.,10.)(15.,15.)(0.,){/ScalarDash}{-1}
\FAProp(8.,10.)(15.,5.)(0.,){/ScalarDash}{1}
\FAVert(8.,10.){-3}
\FALabel(17.,15.02)[b]{$\phi^{\dagger\ (0,0)}$}
\FALabel(17.,6.18)[t]{$\phi^{(0,0)}$}
\FALabel(-0,12)[t]{$A_\pi^{2k,2l}$}

\FADiagram{}
\FAProp(9.,11.)(13.,11.)(0.,){/Straight}{0}
\FAProp(9,9.)(13.,9.)(0.,){/Straight}{0}

\FADiagram{a1}
\FAProp(0.,10.)(8.,10.)(0.,){/ScalarDash}{0}
\FAProp(8.,10.)(15.,15.)(0.,){/ScalarDash}{-1}
\FAProp(8.,10.)(15.,5.)(0.,){/ScalarDash}{1}
\FAProp(8.,10.)(12.,13.)(-0.7285,){/ScalarDash}{0}
\FALabel(10.6071,13.5171)[br]{$A_\pi$}

\FADiagram{a2}
\FAProp(0.,10.)(8.,10.)(0.,){/ScalarDash}{0}
\FAProp(8.,10.)(15.,15.)(0.,){/ScalarDash}{-1}
\FAProp(8.,10.)(15.,5.)(0.,){/ScalarDash}{1}
\FAProp(8.,10.)(12.,7.)(+0.7285,){/ScalarDash}{0}

\FADiagram{b1}
\FAProp(0.,10.)(5.,10.)(0.,){/ScalarDash}{0}
\FAProp(5.,10.)(12.5,15.)(0.,){/Sine}{0}
\FALabel(7.48,13.02)[b]{$A^{\mu}$}
\FAProp(12.5,15.)(12.5,5.)(0.,){/ScalarDash}{1}
\FAProp(12.5,5.)(5.,10.)(0.,){/ScalarDash}{0}
\FALabel(7.48,6.98)[t]{$A_\pi$}
\FAProp(12.5,15.)(17.5,15.)(0.,){/ScalarDash}{-1}
\FAProp(12.5,5.)(17.5,5.)(0.,){/ScalarDash}{1}

\FADiagram{b2}
\FAProp(0.,10.)(5.,10.)(0.,){/ScalarDash}{0}
\FAProp(5.,10.)(12.5,15.)(0.,){/ScalarDash}{0}
\FAProp(12.5,15.)(12.5,5.)(0.,){/ScalarDash}{1}
\FAProp(12.5,5.)(5.,10.)(0.,){/Sine}{0}
\FAProp(12.5,15.)(17.5,15.)(0.,){/ScalarDash}{-1}
\FAProp(12.5,5.)(17.5,5.)(0.,){/ScalarDash}{1}

\end{feynartspicture}
\end{center}
\caption{\footnotesize $A_\pi^{2k,2l}$ coupling to SM scalars}
\label{fig:scalartriangle3}
\end{figure}

Concerning the coupling of gauge scalars to two SM scalars, we see that for the physical scalar $A_\varphi^{2k,2l}$ no 
coupling at one loop is possible.
For the Goldstone $A_\pi^{2k,2l}$, the only loops that can contribute are in Figure~\ref{fig:scalartriangle3}.
The couplings between $A_\pi$ and two scalars with the same KK-masses are indeed zero.
The divergent contribution of these loops is vanishing because of a cancellation coming from symmetric diagrams 
($a1/a2$  and $b_1/b_2$). 

\subsubsection*{Scalar $\phi^{2k,2l}$ to SM gauge boson and massless scalar}
\begin{figure}[!tb]
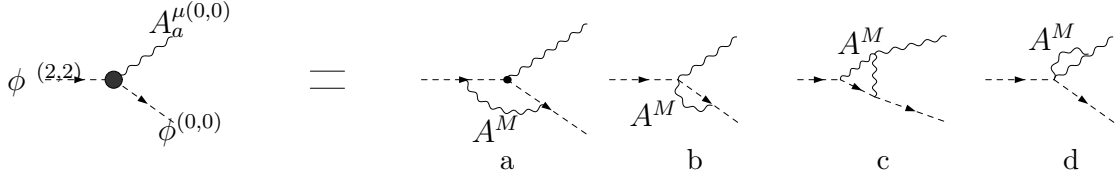

\begin{center}
\begin{feynartspicture}(15,4)(6,1)

\FADiagram{}
\FAProp(-0.,10.)(8.,10.)(0.,){/ScalarDash}{1}
\FAProp(8.,10.)(15.,15.)(0.,){/Sine}{0}
\FAProp(8.,10.)(15.,5.)(0.,){/ScalarDash}{1}
\FAVert(8.,10.){-3}
\FALabel(17.,15.02)[b]{$A_a^{\mu(0,0)}$}
\FALabel(17.,6.18)[t]{$\phi^{(0,0)}$}
\FALabel(-0,12)[t]{$\phi^{\ (2,2)}$}

\FADiagram{}
\FAProp(9.,11.)(13.,11.)(0.,){/Straight}{0}
\FAProp(9,9.)(13.,9.)(0.,){/Straight}{0}

\FADiagram{a}
\FAProp(-0.,10.)(10.,10.)(0.,){/ScalarDash}{1}
\FAProp(10.,10.)(19.5,16.5)(0.,){/Sine}{0}
\FAProp(10.,10.)(19.5,3.5)(0.,){/ScalarDash}{1}
\FAProp(5.,10.)(14.5,7.)(0.4222,){/Sine}{0}
\FALabel(8.725,5.7344)[t]{$A^{M}$}
\FAVert(10.,10.){0}

\FADiagram{b}
\FAProp(0.,10.)(8.,10.)(0.,){/ScalarDash}{1}
\FAProp(8.,10.)(15.,15.)(0.,){/Sine}{0}
\FAProp(8.,10.)(15.,5.)(0.,){/ScalarDash}{1}
\FAProp(8.,10.)(12.,7.)(+0.7285,){/Sine}{0}
\FALabel(8.,4.5171)[br]{$A^{M}$}

\FADiagram{c}
\FAProp(0.,10.)(5.,10.)(0.,){/ScalarDash}{1}
\FAProp(5.,10.)(9,13.)(0.,){/Sine}{0}
\FALabel(7.48,13.02)[b]{$A^{M}$}
\FAProp(9,13.)(9,8.)(0.,){/Sine}{0}
\FAProp(9,8.)(5.,10.)(0.,){/ScalarDash}{-1}
\FAProp(9,13)(17.5,15.)(0.,){/Sine}{0}
\FAProp(9,8)(17.5,5.)(0.,){/ScalarDash}{1}

\FADiagram{d}
\FAProp(0.,10.)(8.,10.)(0.,){/ScalarDash}{1}
\FAProp(8.,10.)(15.,15.)(0.,){/Sine}{0}
\FAProp(8.,10.)(15.,5.)(0.,){/ScalarDash}{1}
\FAProp(8.,10.)(12.,13.)(-0.7285,){/Sine}{0}
\FALabel(10.6071,13.5171)[br]{$A^{M}$}
\end{feynartspicture}
\end{center}
\caption{\footnotesize $\phi^{2k,2l}$ decay in SM scalar and gauge boson}
\label{fig:scalartriangle2}
\end{figure}
We also computed the decay of a heavy scalar into a SM one and a SM gauge boson. It is expected to give similar results as the previous one and that is what we are going to check.
The coupling can be written as:
\begin{eqnarray}
i V^\mu_\phi=\frac{g^3}{16 \pi^4}\frac{i \pi^2}{\epsilon}\  \tau_\phi (-q_1^\mu-2q_2^\mu) t^a\,,
\end{eqnarray} 
where $q_1$ is the momentum of the gauge boson and $q_2$ of the massless scalar.
The diagrams in Figure~\ref{fig:scalartriangle2} are the same as in Figure~\ref{fig:scalartriangle1} with reversed external legs, 
and, because the coefficient $\tau$ does not depend on the external masses or momenta, we expect the same results to apply 
here. The $q$-dependent factor comes from the replacement $q_1\longrightarrow -q=-q_1-q_2$.
We calculated the diagrams and checked explicitly that the contributions are the same as in the previous table.
For U(1) and different gauge groups, the same considerations as above apply.

\subsection{Scalar counter-terms}
After imposing gauge symmetry, many counter-terms are allowed for a scalar field:
\begin{multline}
\mathcal{L}_{\rm c-t} = c_{1} \left( D_\mu \phi\right)^\dagger D^\mu \phi - c_5  \left( D_5 \phi\right)^\dagger D_5 \phi  
- c_6  \left( D_6 \phi\right)^\dagger D_6 \phi - c_{56} \left( \left( D_5 \phi\right)^\dagger D_6 \phi + \left( D_6 \phi\right)^\dagger 
D_5 \phi \right) + \\
c'_5 \left( \left( D_5 D_5 \phi\right)^\dagger \phi + \phi^\dagger \left(D_5 D_5 \phi\right) \right)  + c'_6 
\left( \left( D_6 D_6 \phi\right)^\dagger \phi + \phi^\dagger \left(D_6 D_6 \phi\right) \right)  + \\
c'_{56} \left( \left( D_5 D_6 \phi\right)^\dagger \phi + \phi^\dagger \left(D_5 D_6 \phi\right) \right)   + c'_{65} 
\left( \left( D_6 D_5 \phi\right)^\dagger \phi + \phi^\dagger \left(D_6 D_5 \phi\right) \right) - \delta m^2 \phi^\dagger \phi\,.
\end{multline}
For a SM Higgs, with parities $(+,+)$, $\partial_5 \phi = \partial_6 \phi = 0$, while terms like $\partial_5 \partial_6 \phi = 0$ 
as they are odd under the glide. Therefore, we are left with a much simpler Lagrangian
\beq
\mathcal{L}_{\rm c-t} = c_{1} \left( D_\mu \phi\right)^\dagger D^\mu \phi  + c'_5 \left( \partial_5^2 \phi^\dagger \phi + \phi^\dagger 
\partial_5^2 \phi \right) + c'_6 \left( \partial_6^2 \phi^\dagger \phi + \phi^\dagger \partial_6^2 \phi \right)  - \delta m^2 \phi^\dagger 
\phi\,.
\eeq
Only 4 counter-terms survive, and no gauge couplings involving gauge scalars are left, confirming the result of our loop 
calculation.
From this Lagrangian, we can extract a prediction for mixing and gauge couplings:
\beq
\phi^{2k,2l} \phi^{0,0} &\rightarrow& i \Pi = \frac{i c_1}{\pi^2 \Lambda^2 R^2} q^2 - \frac{i c'_5}{\pi^2 \Lambda^2 R^2} \frac{(2k)^2}
{R^2} - \frac{i c'_6}{\pi^2 \Lambda^2 R^2} \frac{(2l)^2}{R^2} - i \frac{\delta m^2}{\pi^2 \Lambda^2 R^2}\,, \\
\phi^{2k,2l} \phi^{0,0} A^{0,0}_\mu &\rightarrow& - i V_\mu = \frac{c_1}{\pi^2 \Lambda^2 R^2} g (-q_1^\mu -2 q_2^\mu) t^a\,, \\
A^{2k,2l}_\mu \phi^{0,0} \phi^{0,0} &\rightarrow& - i V_\mu = \frac{c_1}{\pi^2 \Lambda^2 R^2} g (q_1^\mu - q_2^\mu) t^a\,.
\eeq
Once more our loop calculations respect the predictions in the $\xi=-3$ gauge with
\beq
\frac{c_1}{\pi^2 \Lambda^2 R^2} &=& - \frac{3}{2} C_2(r_h) \frac{\alpha}{\pi} \log \Lambda R\,, \\
\frac{c'_5}{\pi^2 \Lambda^2 R^2} = \frac{c'_6}{\pi^2 \Lambda^2 R^2} &=& - \frac{5}{16} C_2 (r_h)  \frac{\alpha}{\pi} \log 
\Lambda R\,, \\
\frac{\delta m^2}{\pi^2 \Lambda^2 R^2} &=& C_2 (r_h) \frac{\alpha}{\pi} \frac{1}{\eta}\,.
\eeq
For the SM Higgs doublet, both U(1) and SU(2) gauge bosons contribute (we neglect here the contribution of the quartic Higgs coupling):
\beq
\frac{c_1}{\pi^2 \Lambda^2 R^2} &=&- \frac{3}{2} \left( Y_h^2 \frac{\alpha_1}{\pi} + C_2(r_h) \frac{\alpha_2}{\pi} \right) \log \Lambda 
R = \nonumber \\
 &=& -\frac{3}{2} \left( \frac{1}{4} \frac{\alpha_1}{\pi} + \frac{3}{4} \frac{\alpha_2}{\pi} \right) \log \Lambda R\,, \\
\frac{c'_5}{\pi^2 \Lambda^2 R^2} = \frac{c'_6}{\pi^2 \Lambda^2 R^2} &=& - \frac{5}{16} \left( Y_h^2 \frac{\alpha_1}{\pi} + C_2(r_h) 
\frac{\alpha_2}{\pi} \right)  \log \Lambda R\,, \\
\frac{\delta m^2}{\pi^2 \Lambda^2 R^2} &=& \left( Y_h^2 \frac{\alpha_1}{\pi} + C_2(r_h) \frac{\alpha_2}{\pi} \right)  \frac{1}{\eta}\,.
\eeq

\section{Fermion fields}
\label{sec:fermion}
\setcounter{equation}{0}
\setcounter{footnote}{0}

\subsection{Mixing terms}

In a generic level $(m,n)$ with non-zero integers, there are two degenerate fermions at tree level, labelled 
$\psi_a^{m,n}$ and $\psi_b^{m,n}$.
With our choice of basis, the couplings of the two states are very different: for instance, $\psi_a$ couples to the Goldstone 
while $\psi_b$ couples to the physical gauge scalar.
The main consequence is that only $\psi_a$ can mix with a zero mode fermion, see Figure~\ref{fig:fermionmixing}, while no loop 
can be closed for $\psi_b$.
This means that, in general, the states $\psi_a$ couple to the counter-terms and receive divergent contributions from the loops, 
while corrections for the $\psi_b$ fermions are finite (at one-loop order).
\setlength{\unitlength}{1cm}
\begin{figure}[!tb]
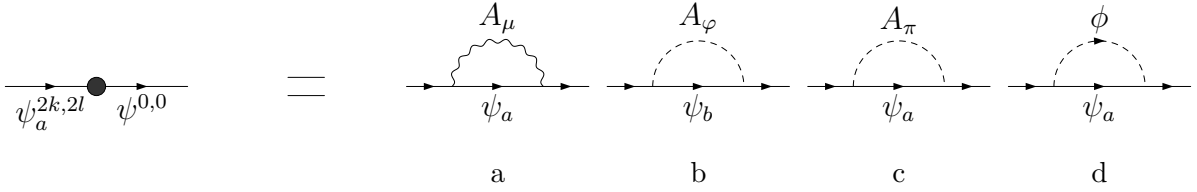

\begin{center}
\begin{feynartspicture}(16,4)(6,1)

\FADiagram{}
\FAProp(-0.,10.)(10.,10.)(0.,){/Straight}{1}
\FALabel(5.,8.98)[t]{$\psi_a^{2k,2l}$}
\FAProp(10.,10.)(20.,10.)(0.,){/Straight}{1}
\FALabel(15.,9.18)[t]{$\psi^{0,0}$}
\FAVert(10.,10.){-3}

\FADiagram{}
\FAProp(9.,11.)(13.,11.)(0.,){/Straight}{0}
\FAProp(9,9.)(13.,9.)(0.,){/Straight}{0}

\FADiagram{a}
\FAProp(0.,10.)(5.,10.)(0.,){/Straight}{1}
\FAProp(5.,10.)(15.,10.)(0.,){/Straight}{1}
\FAProp(15.,10.)(20.,10.)(0.,){/Straight}{1}
\FALabel(10.,9.18)[t]{$\psi_a$}
\FAProp(5.,10.)(15.,10.)(-1.,){/Sine}{0}
\FALabel(10.,15.82)[b]{$A_{\mu}$}

\FADiagram{b}
\FAProp(-0.,10.)(5.,10.)(0.,){/Straight}{1}
\FAProp(5.,10.)(15.,10.)(0.,){/Straight}{1}
\FAProp(15.,10.)(20.,10.)(0.,){/Straight}{1}
\FALabel(2.,8.98)[t]{$ $}
\FALabel(10.,9.18)[t]{$\psi_b$}
\FALabel(18,9.18)[t]{$ $}
\FAProp(5.,10.)(15.,10.)(-1.,){/ScalarDash}{0}
\FALabel(10.,15.82)[b]{$A_\varphi$}

\FADiagram{c}
\FAProp(-0.,10.)(5.,10.)(0.,){/Straight}{1}
\FAProp(5.,10.)(15.,10.)(0.,){/Straight}{1}
\FAProp(15.,10.)(20.,10.)(0.,){/Straight}{1}
\FALabel(2.,8.98)[t]{$ $}
\FALabel(10.,9.18)[t]{$\psi_a$}
\FALabel(18,9.18)[t]{$ $}
\FAProp(5.,10.)(15.,10.)(-1.,){/ScalarDash}{0}
\FALabel(10.,15.82)[b]{$A_\pi$}

\FADiagram{d}
\FAProp(-0.,10.)(5.,10.)(0.,){/Straight}{1}
\FAProp(5.,10.)(15.,10.)(0.,){/Straight}{1}
\FAProp(15.,10.)(20.,10.)(0.,){/Straight}{1}
\FALabel(2.,8.98)[t]{$ $}
\FALabel(10.,9.18)[t]{$\psi_a$}
\FALabel(18,9.18)[t]{$ $}
\FAProp(5.,10.)(15.,10.)(-1.,){/ScalarDash}{1}
\FALabel(10.,15.82)[b]{$\phi$}
\end{feynartspicture}
\end{center}
\caption{\footnotesize Fermion mixing: no one-loop diagram exists for $\psi_b$.}
\label{fig:fermionmixing}
\end{figure}
In the following, we will focus on a fermion with left-handed zero modes: results for the right-handed zero modes can be easily 
obtained from this calculation.
The mixing term can be written as
\beq
\mathcal{L}_{\rm mix} = i \bar{\psi}_a^{2k,2l}\; i \Sigma\, P_L \; \psi^{0,0} + \mbox{h.c.}\,,
\eeq
where $P_L$ is the chirality projector on left-handed zero mode and 
\begin{eqnarray}
i \Sigma = \frac{g^2 C_2(r_f)}{16 \pi^4} \frac{i \pi^2}{\epsilon} \left( \lambda_1 \sstrike{q} + \lambda_2 M \right)\,.
\end{eqnarray}
 and the contribution of each diagram is given table~\ref{tab:fermion2}.
\begin{table}[tb]
\begin{center} \begin{tabular}{|l|c|l|c|}
\hline
$\lambda_{1a}$ & $\xi$ & $\lambda_{2a}$ & $-(\xi+3)$  \\
\hline
$\lambda_{1b}$ & $-\frac{1}{2}$  & $\lambda_{2b}$ & $1$ \\
\hline
$\lambda_{1c}$ & $-\frac{1}{2}$ &$\lambda_{2c}$ & $-1 $ \\
\hline
\hline
$\lambda_{1\, tot}$ & $(\xi-1)$ & $l_{2\, tot}$ & $- (\xi+3)$\\
\hline
\end{tabular} \end{center}
\caption{\footnotesize Contribution of gauge loops in Figure~\ref{fig:fermionmixing} to the fermion mixing.} \label{tab:fermion2}
\end{table}
The total contribution is therefore
\begin{eqnarray}
i \Sigma_{\rm gauge}=\frac{g^2 C_2(r_f)}{16 \pi^4} \frac{i \pi^2}{\epsilon}\left(-M(\xi+3)+(\xi-1) \sstrike{q} \right)\,. \label{eq:fermionmixing}
\end{eqnarray}
For U(1) gauge bosons, it is enough to replace $C_2 (r_f) \to Y_f^2$.
For fields with right-handed zero modes, it is enough to replace the chirality projector.

\subsubsection*{Yukawa couplings}

In the standard model, fermions also couple to the scalar Higgs field via Yukawa interactions.
The Higgs loops will only be numerically relevant for the top Yukawa, therefore they affect the third generation doublet 
(left-handed top and bottom) and the top singlet (right-handed top). 
This contribution comes through the loop (d) and, for a doublet, is given by: 
\begin{eqnarray}
i \Sigma_{\rm scalar}^{D}=\frac{y_f^2 }{16 \pi^4} \frac{i \pi^2}{\epsilon}\left(\frac{1}{2}\sstrike{q}-M \right)\,,
\end{eqnarray}
where $y_f$ is the 4D Yukawa of the fermions.
Note that the result is the same as the gauge scalar $A_\varphi$ loop, up to a sign which is the different parity under the rotation symmetry.
For the top singlet,  the result is the same up to a factor of two that counts the contribution of the up and down components of the Higgs and fermion doublet and to a change in the chirality projector.

\subsection{Fermion gauge couplings}

\subsubsection*{Gauge vector $A_\mu^{2k,2l}$ to two SM fermions}
\setlength{\unitlength}{1cm}

\begin{figure}[!tb]
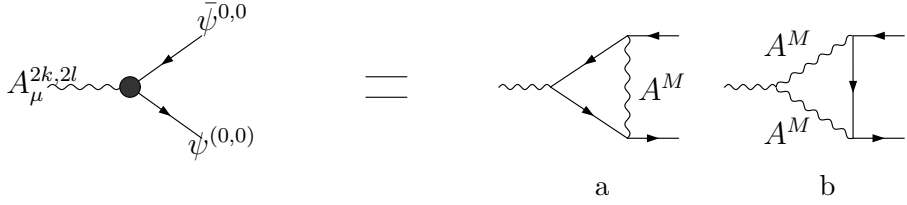

\begin{center}
\begin{feynartspicture}(12,5)(4,1)

\FADiagram{}
\FAProp(-0.,10.)(8.,10.)(0.,){/Sine}{0}
\FAProp(8.,10.)(15.,15.)(0.,){/Straight}{-1}
\FAProp(8.,10.)(15.,5.)(0.,){/Straight}{1}
\FAVert(8.,10.){-3}
\FALabel(17.,15.02)[b]{$\bar{\psi}^{0,0}$}
\FALabel(17.,6.18)[t]{$\psi^{(0,0)}$}
\FALabel(-0.5,12)[t]{$A_\mu^{2k,2l}$}

\FADiagram{}
\FAProp(9.,11.)(13.,11.)(0.,){/Straight}{0}
\FAProp(9,9.)(13.,9.)(0.,){/Straight}{0}

\FADiagram{a}
\FAProp(0.,10.)(5.,10.)(0.,){/Sine}{0}
\FAProp(5.,10.)(12.5,15.)(0.,){/Straight}{-1}
\FAProp(12.5,15.)(12.5,5.)(0.,){/Sine}{0}
\FALabel(13.52,10.)[l]{$A^{M}$}
\FAProp(12.5,5.)(5.,10.)(0.,){/Straight}{-1}
\FAProp(12.5,15.)(17.5,15.)(0.,){/Straight}{-1}
\FAProp(12.5,5.)(17.5,5.)(0.,){/Straight}{1}

\FADiagram{b}
\FAProp(0.,10.)(5.,10.)(0.,){/Sine}{0}
\FAProp(5.,10.)(12.5,15.)(0.,){/Sine}{0}
\FALabel(8.48,13.02)[br]{$A^{M}$}
\FAProp(12.5,15.)(12.5,5.)(0.,){/Straight}{1}
\FAProp(12.5,5.)(5.,10.)(0.,){/Sine}{0}
\FALabel(8.48,6.98)[tr]{$A^{M}$}
\FAProp(12.5,15.)(17.5,15.)(0.,){/Straight}{-1}
\FAProp(12.5,5.)(17.5,5.)(0.,){/Straight}{1}

\end{feynartspicture}
\end{center}
\caption{\footnotesize $A_\mu^{2k,2l}$ coupling to  two SM fermions.}
\label{fig:Affvertex2}
\end{figure}
The coupling can be expressed as
\begin{eqnarray}
i \Gamma_A^{a,\mu}=\frac{g^3}{16 \pi^4} \frac{i \pi^2}{\epsilon} \tau_A \gamma^\mu P_L t^a\,,
\end{eqnarray}
where $ t^a$ is the generator of the SU(N) gauge group.
The contributions from the diagrams in Figure~\ref{fig:Affvertex2} are listed in Table~\ref{tab:fermions3}.
\begin{table}[tb]
\begin{center} \begin{tabular}{|l|c|c|}
\hline
$\tau_{Aa}(A^\mu)$ & $\xi (C_2(r_f)-1/2C_2(G))$  \\
\hline
$\tau_{Aa} (A_\varphi + A_\pi)$ & $-2\ \frac{1}{2}(C_2(r_f)-1/2C_2(G))$ \\
\hline
$\tau_{Ab}(A^\mu)$ & $\frac{3}{4}(\xi+1)C_2(G) $ \\
\hline
$\tau_{Ab} (A_\varphi + A_\pi)$ & $ -2 \frac{1}{4}C_2(G)$ \\
\hline
\hline
$\tau_{A\, tot}$ & $\frac{1}{4}(4(\xi-1)C_2(r_f)+(\xi+3)C_2(G))$ \\
\hline
\end{tabular} \end{center}
\caption{\footnotesize contribution of non-Abelian gauge loops.} \label{tab:fermions3}
\end{table}
The total contribution is therefore:
\begin{eqnarray}
i \Gamma_A^{a, \mu}=\frac{g^3 }{16 \pi^4} \frac{i \pi^2}{\epsilon} \left( (\xi-1)C_2(r_f)+\frac{\xi+3}{4} C_2(G) \right)  \label{eq:fermions3}
\gamma^\mu P_L t^a\,.
\end{eqnarray}
For a U(1) gauge group, only diagrams of type ``a'' are present, and it would be enough to replace $C_2 (r_f) \to Y_f^2$, 
$C_2(G) \to 0$ and $t^a \to Y_f$.
Similarly, if the external gauge boson is different from the internal one, we would have (with the same replacements if one 
of the two gauge bosons is a U(1)):
\begin{eqnarray}
i \Gamma_{A'}^\mu=\frac{g {g'}^2 }{16 \pi^4} \frac{i \pi^2}{\epsilon} (\xi-1)C_2(r'_f) \gamma^\mu P_L t^a\,.
\end{eqnarray}

\subsubsection*{Yukawa couplings}

For massive SM fermions, Yukawa interactions can be sizeable and will give significant contributions to the vertex corrections.
Here we will quote the results in the case of a coupling in the form
\beq
y_f \, \bar{\psi}_D \phi \psi_S\,,
\eeq
where $D$ and $S$ label a singlet and doublet fields.
In this case, $Y_D = Y_H + Y_S$. 
Note that for the case $y_f \, \bar{\psi}_D \phi^\dagger \psi_S\,$, which applies to the top Yukawa, it would be enough to 
replace $Y_H \to  - Y_H$.








The coupling can be expressed as
\begin{eqnarray}
i \Gamma_A^\mu=\frac{g y_f^2}{16 \pi^4} \frac{i \pi^2}{\epsilon} f_x \gamma^\mu P_R t^a\,,
\end{eqnarray}
and the diagrams are the same as in Figure~\ref{fig:Affvertex2} with the internal gauge propagator replaced by a scalar one.
Because of the gauge transformations of the Higgs, we need to distinguish between the various gauge bosons (U(1), 
SU(2) and SU(3)), and the case of a singlet or doublet.
Note that for U(1) couplings the generator $t^a \to 1$.
The results are listed in table~\ref{tab:fermions4}.
In the case of a singlet, the diagram $(b)$ picks up a minus sign because it is $\phi^\dagger$ that runs in the loop, 
therefore the sign of the Yukawa coupling changes.

\begin{table}[tb]
\begin{center} \begin{tabular}{|l|c|c|c|c|c|}
\hline
   & U(1) to singlet & U(1) to doublet & SU(2) to singlet & SU(2) to doublet & SU(3) \\
\hline
$f_{a}$ & $Y_D$ & $1/2 Y_S$ & $0$ & $0$ & $1/2$   \\
\hline
$f_{b}$ & $- Y_H$ & $1/2 Y_H$ & $0$ & $1/2$ & $0$  \\
\hline
\hline
$f_{tot}$ & $Y_S$ & $1/2 Y_D$ & $0$ & $1/2$ & $1/2$  \\
\hline
\end{tabular} \end{center}
\caption{\footnotesize Contribution of Higgs loops. $Y_S$ and $Y_D$ are the hypercharges of the singlet and doublet 
respectively, while $Y_H$ is the hypercharge of the Higgs. We also used the fact that $Y_D = Y_S + Y_H$.} 
\label{tab:fermions4}
\end{table}

\subsubsection*{Heavy fermion $\psi_a^{2k,2l}$ coupling to a SM gauge boson and fermion.}
\begin{figure}[!tb]
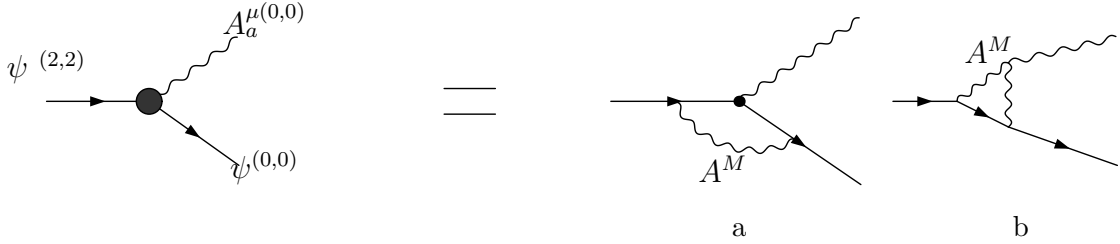

\begin{center}
\begin{feynartspicture}(15,4)(4,1)

\FADiagram{}
\FAProp(-0.,10.)(8.,10.)(0.,){/Straight}{1}
\FAProp(8.,10.)(15.,15.)(0.,){/Sine}{0}
\FAProp(8.,10.)(15.,5.)(0.,){/Straight}{1}
\FAVert(8.,10.){-3}
\FALabel(17.,15.02)[b]{$A_a^{\mu(0,0)}$}
\FALabel(17.,6.18)[t]{$\psi^{(0,0)}$}
\FALabel(0,14)[t]{$\psi^{\ (2,2)}$}

\FADiagram{}
\FAProp(9.,11.)(13.,11.)(0.,){/Straight}{0}
\FAProp(9,9.)(13.,9.)(0.,){/Straight}{0}

\FADiagram{a}
\FAProp(-0.,10.)(10.,10.)(0.,){/Straight}{1}
\FAProp(10.,10.)(19.5,16.5)(0.,){/Sine}{0}
\FAProp(10.,10.)(19.5,3.5)(0.,){/Straight}{1}
\FAProp(5.,10.)(14.5,7.)(0.4222,){/Sine}{0}
\FALabel(8.725,5.7344)[t]{$A^{M}$}
\FAVert(10.,10.){0}

\FADiagram{b}
\FAProp(0.,10.)(5.,10.)(0.,){/Straight}{1}
\FAProp(5.,10.)(9,13.)(0.,){/Sine}{0}
\FALabel(7.48,13.02)[b]{$A^{M}$}
\FAProp(9,13.)(9,8.)(0.,){/Sine}{0}
\FAProp(9,8.)(5.,10.)(0.,){/Straight}{-1}
\FAProp(9,13)(17.5,15.)(0.,){/Sine}{0}
\FAProp(9,8)(17.5,5.)(0.,){/Straight}{1}
\end{feynartspicture}
\end{center}
\caption{\footnotesize $\psi^{2k,2l}$ decay in SM scalar and gauge boson}
\label{fig:Affvertex}
\end{figure}

The diagrams in Figure~\ref{fig:Affvertex} generate the coupling of a heavy fermion $\psi_a^{2k,2l}$ with SM fermion and gauge 
vector: they are the same as the diagrams in Figure~\ref{fig:Affvertex2}, the only difference being a change in the momenta and 
masses of the external legs.
However, it can be shown that respective diagrams will lead to the same results, so that the same formulae as above apply.

\subsubsection*{Gauge scalars $A_{\varphi,\pi}^{2k,2l}$ couplings to SM fermions.}

\setlength{\unitlength}{1cm}
\begin{figure}[tb]
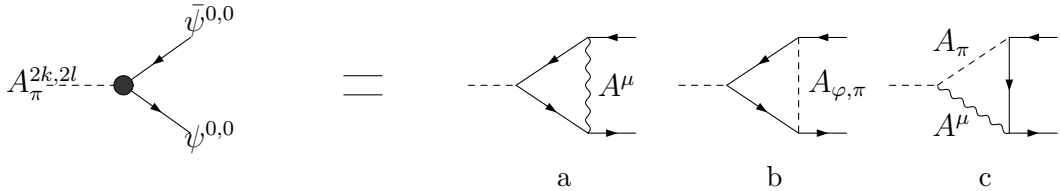

\begin{center}
\begin{feynartspicture}(14,5)(5,1)

\FADiagram{}
\FAProp(-0.,10.)(8.,10.)(0.,){/ScalarDash}{0}
\FAProp(8.,10.)(15.,15.)(0.,){/Straight}{-1}
\FAProp(8.,10.)(15.,5.)(0.,){/Straight}{1}
\FAVert(8.,10.){-3}
\FALabel(17.,15.02)[b]{$\bar{\psi}^{0,0}$}
\FALabel(17.,6.18)[t]{$\psi^{0,0}$}
\FALabel(-0.5,12)[t]{$A_\pi^{2k,2l}$}

\FADiagram{}
\FAProp(9.,11.)(13.,11.)(0.,){/Straight}{0}
\FAProp(9,9.)(13.,9.)(0.,){/Straight}{0}

\FADiagram{a}
\FAProp(0.,10.)(5.,10.)(0.,){/ScalarDash}{0}
\FAProp(5.,10.)(12.5,15.)(0.,){/Straight}{-1}
\FAProp(12.5,15.)(12.5,5.)(0.,){/Sine}{0}
\FALabel(13.52,10.)[l]{$A^{\mu}$}
\FAProp(12.5,5.)(5.,10.)(0.,){/Straight}{-1}
\FAProp(12.5,15.)(17.5,15.)(0.,){/Straight}{-1}
\FAProp(12.5,5.)(17.5,5.)(0.,){/Straight}{1}

\FADiagram{b}
\FAProp(0.,10.)(5.,10.)(0.,){/ScalarDash}{0}
\FAProp(5.,10.)(12.5,15.)(0.,){/Straight}{-1}
\FAProp(12.5,15.)(12.5,5.)(0.,){/ScalarDash}{0}
\FALabel(13.52,10.)[l]{$A_{\varphi,\pi}$}
\FAProp(12.5,5.)(5.,10.)(0.,){/Straight}{-1}
\FAProp(12.5,15.)(17.5,15.)(0.,){/Straight}{-1}
\FAProp(12.5,5.)(17.5,5.)(0.,){/Straight}{1}

\FADiagram{c}
\FAProp(0.,10.)(5.,10.)(0.,){/ScalarDash}{0}
\FAProp(5.,10.)(12.5,15.)(0.,){/ScalarDash}{0}
\FALabel(8.48,13.02)[br]{$A_\pi$}
\FAProp(12.5,15.)(12.5,5.)(0.,){/Straight}{1}
\FAProp(12.5,5.)(5.,10.)(0.,){/Sine}{0}
\FALabel(8.48,6.98)[tr]{$A^\mu$}
\FAProp(12.5,15.)(17.5,15.)(0.,){/Straight}{-1}
\FAProp(12.5,5.)(17.5,5.)(0.,){/Straight}{1}

\end{feynartspicture}
\end{center}
\caption{\footnotesize $A_\pi^{2k,2l} $ coupling to two SM fermions} \label{fig:Apiffvertex}
\end{figure}

One can show that $A_\varphi^{2k,2l}$ cannot couple to SM fermions because the loops cannot be closed, contrary to the 
Goldstone $A_\pi^{2k,2l}$.
Nevertheless, all the possible loops, listed in Figure~\ref{fig:Apiffvertex}, are finite. 
The power of the momentum flowing into the loops (a) and (b) is too small, so the loop is finite. The divergent contribution of the 
loop (c)  is vanishing because of a cancellation coming from symmetric diagrams. The proof is analogue to the scalar one.

\subsection{Yukawa couplings}
\subsubsection*{Heavy fermion $\psi_a^{2k,2l}$ coupling to a SM Goldstone $\phi$ and fermion.}

\begin{figure}[!tb]
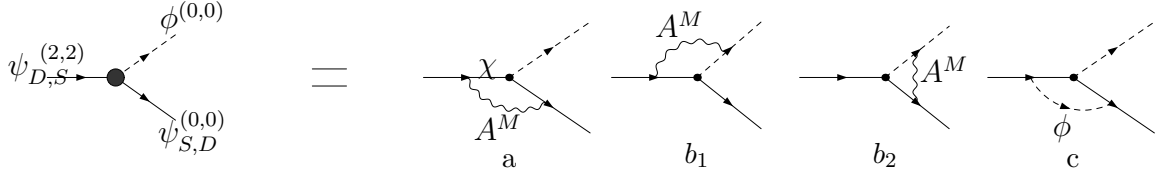

\begin{center}
\begin{feynartspicture}(15,4)(6,1)

\FADiagram{}
\FAProp(-0.,10.)(8.,10.)(0.,){/Straight}{1}
\FAProp(8.,10.)(15.,15.)(0.,){/ScalarDash}{1}
\FAProp(8.,10.)(15.,5.)(0.,){/Straight}{1}
\FAVert(8.,10.){-3}
\FALabel(17.,15.02)[b]{$\phi^{(0,0)}$}
\FALabel(17.,6.18)[t]{$\psi_{S,D}^{(0,0)}$}
\FALabel(0,14)[t]{$\psi_{D,S}^{\ (2,2)}$}

\FADiagram{}
\FAProp(9.,11.)(13.,11.)(0.,){/Straight}{0}
\FAProp(9,9.)(13.,9.)(0.,){/Straight}{0}

\FADiagram{a}
\FAProp(-0.,10.)(10.,10.)(0.,){/Straight}{1}
 \FALabel(7.5,12.18)[t]{$\chi$}
\FAProp(10.,10.)(19.5,16.5)(0.,){/ScalarDash}{1}
\FAProp(10.,10.)(19.5,3.5)(0.,){/Straight}{1}
\FAProp(5.,10.)(14.5,7.)(0.4222,){/Sine}{0}
\FALabel(8.725,5.7344)[t]{$A^{M}$}
\FAVert(10.,10.){0}

\FADiagram{$b_1$}
\FAProp(0.,10.)(10.,10.)(0.,){/Straight}{1}
\FAProp(10.,10.)(17.5,16.5)(0.,){/ScalarDash}{1}
\FAProp(10.,10.)(17.5,4.)(0.,){/Straight}{1}
\FAProp(5.,10.)(13.5,13.)(-0.5786,){/Sine}{0}
\FALabel(7.9494,14.7049)[b]{$A^{M}$}
\FAVert(10.,10.){0}

\FADiagram{$b_2$}
\FAProp(0.,10.)(10.,10.)(0.,){/Straight}{1}
\FAProp(10.,10.)(17.5,16.)(0.,){/ScalarDash}{1}
\FALabel(14.0821,12.4648)[tl]{$A^{M}$}
\FAProp(10.,10.)(17.5,4.)(0.,){/Straight}{1}
\FAProp(13.5,13.)(13.5,7.)(0.,){/Sine}{0}
\FAVert(10.,10.){0}

\FADiagram{c}
\FAProp(-0.,10.)(10.,10.)(0.,){/Straight}{1}
\FAProp(10.,10.)(19.5,16.5)(0.,){/ScalarDash}{1}
\FAProp(10.,10.)(19.5,3.5)(0.,){/Straight}{1}
\FAProp(5.,10.)(14.5,7.)(0.4222,){/ScalarDash}{1}
\FALabel(8.725,5.7344)[t]{$\phi$}
\FAVert(10.,10.){0} 
\end{feynartspicture}
\end{center}
\caption{\footnotesize $\psi^{2k,2l}$ decay into massless scalar $\phi$ and gauge boson}
\label{fig:fphifvertex}
\end{figure}

For definiteness, let's consider a Yukawa coupling between two fermions $\psi_D$ and $\psi_S$ in the form:
\beq
y_f\,  \bar{\psi}_D \phi \psi_S\,;
\eeq
were the representations of the fermions and scalar fields match in order for the coupling to be gauge invariant.
For instance, $D$ may label a doublet of SU(2), while $S$ is a singlet.
The loop-induced coupling can be expressed in general as
\begin{eqnarray}
i \Gamma_\phi=-\frac{y_f}{16 \pi^4} \frac{i \pi^2}{\epsilon}  (g^2\tau_\phi+y_f^2\tau_\phi')  P_{L(R)} \,,
\end{eqnarray}
were we emphasise that the representations of the two fermions and the scalar under the gauge group may be different.
For a generic SU(N) group, the contributions from the diagrams in Figure~\ref{fig:fphifvertex} are listed in Table~\ref{tab:fermYuk}.
\begin{table}[tb]
\begin{center} \begin{tabular}{|l|c|c|}
\hline
$\tau_{\phi a}(A^\mu)$ & $(\xi+3)\ t^a(r_D).t^a(r_S) $  \\
\hline
$\tau_{\phi a} (A_\varphi + A_\pi)$ & $0$ \\
\hline
$\tau_{\phi b}(A^\mu)$ & $\xi \  ( - t^a(r_S).t^a(r_\phi) +t^a(r_D).t^a(r_\phi))  $ \\
\hline
$\tau_{\phi b} (A_\varphi + A_\pi)$ & $ 0 $ \\
\hline
\hline
$\tau_{\phi\, tot}$ & $\xi \left( t^a(r_D).t^a(r_S)- t^a(r_S)).t^a(r_\phi) +t^a(r_D)).t^a(r_\phi) \right) + 3 t^a(r_D).t^a(r_S)$ \\
\hline
\hline
$\tau'_{\phi c}(\phi)$ & $ +1 $ \\
\hline
\end{tabular} \end{center}
\caption{\footnotesize Contribution of gauge loops and of scalar $\phi$; the generators $t^a (r_X)$ are in different representations 
of the gauge group, while the repeated index implies a sum over the generators.} \label{tab:fermYuk}
\end{table}
Note that the generators $t^a (r_D)$, $t^a (r_S)$ and $t^a (r_h)$ may be in different representations: only when they belong to 
the same representation we can use the relation $t^a (r) t^a (r) = C_2 (r) \cdot 1$.
In the case of a U(1), it is enough to replace the generators by the hypercharge.
Note that for the Yukawa coupling to exist, $Y_D = Y_S + Y_h$.

Note also that for a Yukawa coupling in the form
\beq
y_f\,  \bar{\psi}_D \phi^\dagger \psi_S\,,
\eeq
the same formula applies, with the only change, in the case of U(1) loops, $Y_h \to - Y_h$.

\subsection{Fermion counter-terms}

For fermions, many counter-terms are in principle allowed: in fact, we can consider the 4 chiral components of the 8-component 
6D field as independent and construct all possible 4D Lorentz invariant operators, without caring about the Lorentz structure in 6D and 
the 6D chiralities.
On the other hand, loops will respect the 6D Lorentz structure, even in the case of the divergent terms that, as we have seen, do 
correspond to 4D loops. We can therefore safely limit ourselves to operators in terms of the 6D gamma matrices:
\begin{multline}
\mathcal{L}_{\rm c-t} = i z_{1\pm}\; \bar{\Psi}_\pm \Gamma_\mu D^\mu \Psi_\pm - z_{5\pm}\; \left(i \bar{\Psi}_\pm 
\Gamma_5 D_5 \Psi_\pm + h.c. \right) - z_{6\pm}\; \left(i \bar{\Psi}_\pm \Gamma_6 D_6 \Psi_\pm + h.c. \right) \\
- z'_{5\pm}\; \left(i \bar{\Psi}_\pm \Gamma_6 D_5 \Psi_\pm + h.c. \right) - z'_{6\pm}\; \left(i \bar{\Psi}_\pm \Gamma_5 
D_6 \Psi_\pm + h.c. \right) - \delta m\; \left(\bar{\Psi}_+ \Psi_- + \bar{\Psi}_- \Psi_+  \right) \,,
\end{multline}
where the subscript $\pm$ refers to the two 6D chiral components.
We can focus in the following to the case of a left-handed zero mode, in which case all the right handed components of the 
fermion vanish on the fixed points (while their derivatives do not); moreover, we can write the counter-term Lagrangian in terms of 
4D chiral fields
\beq
\Psi_+ = \left( \begin{array}{c} \psi_{L+} \\ \psi_{R+} \end{array} \right)\,, \qquad \Psi_- = \left( \begin{array}{c} \psi_{R-} \\ 
\psi_{L-} \end{array} \right)\,.
\eeq
In this notation,
\begin{multline}
\mathcal{L}_{\rm c-t} = i z_{1\pm}\; \bar{\psi}_{L\pm} \gamma_\mu D^\mu \psi_{L\pm} + z_{5\pm}\; \left(\bar{\psi}_{L\pm} 
\gamma_5 \partial_5 \psi_{R\pm} + h.c. \right) + z_{6\pm}\; \left( \mp i \bar{\psi}_{L\pm} \gamma_5 \partial_6 \psi_{R\pm} + h.c. 
\right) \\
+ z'_{5\pm}\; \left(\mp i \bar{\psi}_{L\pm} \gamma_5 \partial_5 \psi_{R\pm} + h.c. \right) + z'_{6\pm}\; \left(\bar{\psi}_{L\pm} 
\gamma_5 \partial_6 \psi_{R\pm} + h.c. \right) \,.
\end{multline}
Note that gauge couplings only involve the massive vector, and their coefficient is the same as the correction to the kinetic term proportional to momentum: this is confirmed by Eqs.~\ref{eq:fermionmixing} and \ref{eq:fermions3} in $\xi=-3$ gauge!
For a zero mode, the wave functions, evaluated at any of the four fixed points, are given by
\beq
\psi_{L+} = p_g \psi_{L-} = \frac{1}{2 \sqrt{2} \pi R} \psi_L^{0,0}\,, \qquad \psi_{R+} = \psi_{R-} = 0\,,
\eeq
where $\psi^{0,0}$ is the 4D field and $p_g$ is the parity of the fermion under the glide symmetry.
The counter-term Lagrangian, therefore, reads for the zero mode
\beq
\delta \mathcal{L} = \frac{z_{1+}+z_{1-}}{4 \pi^2 \Lambda^2 R^2}\; i \bar{\psi}_L^{0,0} \gamma_\mu D^\mu_{0,0} \psi_L^{0,0} =  
\frac{z_{1}}{2 \pi^2 \Lambda^2 R^2}\; i \bar{\psi}_L^{0,0} \gamma_\mu D^\mu_{0,0} \psi_L^{0,0}\,,
\eeq
where, for convenience, we have defined $z_1 = \frac{z_{1+}+z_{1-}}{2}$, and $D^\mu_{0,0}$ is the covariant derivative that only 
contains the zero mode gauge vector.
Note that the coupling of $A_\mu^{2k,2l}$ with zero mode fermions will have the same coefficient up to a factor of 2 from the wave 
function normalisation.

Now, let us consider the terms with one fermion from level $(2k,2l)$ and zero modes: for the state $a$, the 
wave functions calculated at the fixed points are
\beq
\psi_{L+}^a = p_g \psi_{L-}^a &=& \frac{1}{\sqrt{2} \pi R} \psi_{a L}^{2k,2l}\,, \\
\partial_5 \psi_{R+}^a = p_g \partial_5 \psi_{R-}^a &=& - \frac{(2k)^2}{R \sqrt{(2k)^2 + (2l)^2}} \frac{1}{\sqrt{2} \pi R} 
\psi_{a R}^{2k,2l}\,, \\
\partial_6 \psi_{R+}^a = - p_g \partial_6 \psi_{R-}^a &=&  i \frac{(2l)^2}{R \sqrt{(2k)^2 + (2l)^2}} \frac{1}{\sqrt{2} \pi R} 
\psi_{a R}^{2k,2l}\,.
\eeq
Plugging those solutions in the counter-term Lagrangian, we obtain
\begin{multline}
\left. \delta \mathcal{L} \right|_{a} = \frac{z_{1}}{\pi^2 \Lambda^2 R^2}\; \left( i \bar{\psi}_L^{0,0} \gamma_\mu D^\mu_{0,0} 
\psi_{aL}^{2k,2l} + h.c. \right) - \left[ \left( \frac{z_{5+} +  z_{5-}}{2\pi^2 \Lambda^2 R^2} - i  \frac{z'_{5+} -  z'_{5-}}{2\pi^2 \Lambda^2 
R^2}  \right) \frac{2 k^2}{R \sqrt{k^2+l^2}} + \right. \\
\left. \left( \frac{z_{6+} +  z_{6-}}{2\pi^2 \Lambda^2 R^2} + i  \frac{z'_{6+} -  z'_{6-}}{2\pi^2 \Lambda^2 R^2}  \right) \frac{2 l^2}{R 
\sqrt{k^2+l^2}} \right] \left( \bar{\psi}_L^{0,0} \psi_{aR}^{2k,2l} + h.c. \right)\,.
\end{multline}
For a fermion $b$
\beq
\psi_{L+}^b = \psi_{L-}^b &=& 0\,, \\
\partial_5 \psi_{R+}^b = - p_g \partial_5 \psi_{R-}^b &=& i \frac{(2k) (2l)}{R \sqrt{(2k)^2 + (2l)^2}} \frac{1}{\sqrt{2} \pi R} 
\psi_{b R}^{2k,2l}\,, \\
\partial_6 \psi_{R+}^b = p_g \partial_6 \psi_{R-}^b &=& \frac{(2l) (2k)}{R \sqrt{(2k)^2 + (2l)^2}} \frac{1}{\sqrt{2} \pi R} 
\psi_{b R}^{2k,2l}\,;
\eeq
and therefore
\beq
\left. \delta \mathcal{L} \right|_{b} = \left( i \frac{z_{5+} - z_{5-}- z_{6+} + z_{6-}}{2 \pi^2 \Lambda^2 R^2}  + \frac{z'_{5+} 
+ z'_{5-} + z'_{6+} + z'_{6-}}{2 \pi^2 \Lambda^2 R^2} \right) \frac{2 k l}{R \sqrt{k^2+l^2}}   \left( \bar{\psi}_L^{0,0} 
\psi_{bR}^{2k,2l} + h.c. \right)\,.
\eeq

From the kinetic term, we can easily check that in the magic gauge the loop calculation agrees with the presence of a 
unique counter-term, and
\beq
\frac{z_1}{\pi^2 \Lambda^2 R^2} = \left( - C_2(r_f) \frac{\alpha}{\pi} + \frac{1}{8} \frac{y_f^2}{4 \pi^2} \right) \log \Lambda R\,.
\eeq
Note however that we cannot determine if $z_{1+} = z_{1-}$!
Regarding the mass terms, from the fact that the fermion $b$ does not receive divergent corrections and that the correction to
the fermion $a$ is proportional to $\sqrt{k^2+l^2}$, we can deduce that $z_{5\pm} = z_{6\pm}$ and that 
$z'_{5+} = z'_{5-}=- z'_{6+} = -z'_{6-} = z'$.
Defining $z_2 = \frac{z_{5+} + z_{5-}}{2} =  \frac{z_{6+} + z_{6-}}{2}$ and matching with the loop calculation, we have
\beq
\frac{z_2}{\pi^2 \Lambda^2 R^2} =\frac{1}{8} \frac{y_f^2}{4 \pi^2} \log \Lambda R\,.
\eeq
As before, we cannot tell if $z_{5+} = z_{5-}$ or if $z'=0$.
In Table~\ref{tab:fermCT} we list the coefficient for the counter-terms of the SM matter content.
\begin{table}[ht]
\begin{center} \begin{tabular}{|l|c|c|}
\hline
 & $\frac{z_1}{\pi^2 \Lambda^2 R^2}$ &   $\frac{z_2}{\pi^2 \Lambda^2 R^2}$  \\
\hline
\hline
$E$  & $-\frac{\alpha_1}{\pi} + \frac{1}{8} \frac{y_e^2}{4 \pi^2}$  &  $\frac{1}{8} \frac{y_e^2}{4 \pi^2}$ \\
\hline
$L$  & $-\frac{1}{4} \frac{\alpha_1}{\pi} -\frac{3}{4} \frac{\alpha_2}{\pi} + \frac{1}{8} \frac{y_e^2}{4 \pi^2}$  &  
$\frac{1}{8} \frac{y_e^2}{4 \pi^2}$ \\
\hline
$U$  & $-\frac{4}{9} \frac{\alpha_1}{\pi} - \frac{4}{3} \frac{\alpha_3}{\pi} + \frac{1}{8} \frac{y_u^2}{4 \pi^2}$  &  
$\frac{1}{8} \frac{y_u^2}{4 \pi^2}$ \\
\hline
$D$  & $-\frac{1}{9} \frac{\alpha_1}{\pi} - \frac{4}{3} \frac{\alpha_3}{\pi} + \frac{1}{8} \frac{y_d^2}{4 \pi^2}$  &  
$\frac{1}{8} \frac{y_d^2}{4 \pi^2}$ \\
\hline
$Q$  & $-\frac{1}{36} \frac{\alpha_1}{\pi} -\frac{3}{4} \frac{\alpha_2}{\pi} - \frac{4}{3} \frac{\alpha_3}{\pi} + 
\frac{1}{8} \frac{y_u^2 + y_d^2}{4 \pi^2}$  &  $\frac{1}{8} \frac{y_u^2 + y_d^2}{4 \pi^2}$ \\
\hline
\end{tabular} \end{center}
\caption{\footnotesize Coefficients of the fermion counter-terms in the SM.} \label{tab:fermCT}
\end{table}

\section{Physical observables: effective couplings and decays}
\label{sec:physobs}
\setcounter{equation}{0}
\setcounter{footnote}{0}

In this section we will focus on the effective couplings between one state in the level $(2k, 2l)$ and SM states (zero modes): 
those are phenomenologically important because they control the decays and single production cross sections of the 
heavy states.
They are generated by diagrams with the insertion of the loops (or counter-terms) we computed in the previous sections.
We will show that, as expected, they are gauge independent (in the limit of on-shell external particles): this fact has an 
important consequence. In fact, we can use the simple structure of the counter-terms in the ``magic'' gauge $\xi=-3$ to predict 
the decay rates and single production cross sections of all KK tiers!
This is possible because in all the relevant cases we need the coupling with on-shell particles.
The contributions of off-shell particles will be sub-leading compared to tree-level bulk couplings, therefore gauge 
dependent contributions can be safely neglected.

In the following we will analyse all the possible decays and explicitly check the gauge invariance of the effective coupling.
In the limit of massless SM particles, the decay into heavy gauge bosons (the $W$ and $Z$) is given, thanks to the equivalence 
principle, by the decay rate in Higgs bosons.

\subsection{Heavy gauge boson decays}
\label{sec:AAAdecay}

\subsubsection*{Heavy vector to vectors, $A^{2k,2l}_\mu \longrightarrow A^{0,0}_\nu A^{0,0}_\rho $: case of massless SM fields}

\begin{figure}[!tb]
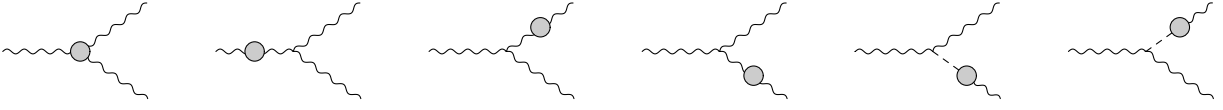

\begin{center}
\begin{feynartspicture}(17,3)(6,1)

\FADiagram{}
\FAProp(-0.,10.)(8.,10.)(0.,){/Sine}{0}
\FAProp(8.,10.)(15.,15.)(0.,){/Sine}{0}
\FAProp(8.,10.)(15.,5.)(0.,){/Sine}{0}
\FAVert(8.,10.){-1}

\FADiagram{}
\FAProp(-0.,10.)(8.,10.)(0.,){/Sine}{0}
\FAProp(8.,10.)(15.,15.)(0.,){/Sine}{0}
\FAProp(8.,10.)(15.,5.)(0.,){/Sine}{0}
\FAVert(4.,10.){-1}

\FADiagram{}
\FAProp(-0.,10.)(8.,10.)(0.,){/Sine}{0}
\FAProp(8.,10.)(15.,15.)(0.,){/Sine}{0}
\FAProp(8.,10.)(15.,5.)(0.,){/Sine}{0}
\FAVert(11.5,12.5){-1}

\FADiagram{}
\FAProp(-0.,10.)(8.,10.)(0.,){/Sine}{0}
\FAProp(8.,10.)(15.,15.)(0.,){/Sine}{0}
\FAProp(8.,10.)(15.,5.)(0.,){/Sine}{0}
\FAVert(11.5,7.5){-1}

\FADiagram{}
\FAProp(-0.,10.)(8.,10.)(0.,){/Sine}{0}
\FAProp(8.,10.)(11.5,7.5)(0.,){/ScalarDash}{0}
\FAProp(11.5,7.5)(15.,5.)(0.,){/Sine}{0}
\FAProp(8.,10.)(15.,15.)(0.,){/Sine}{0}
\FAVert(11.5,7.5){-1}

\FADiagram{}
\FAProp(-0.,10.)(8.,10.)(0.,){/Sine}{0}
\FAProp(8.,10.)(11.5,12.5)(0.,){/ScalarDash}{0}
\FAProp(11.5,12.5)(15.,15.)(0.,){/Sine}{0}
\FAProp(8.,10.)(15.,5.)(0.,){/Sine}{0}
\FAVert(11.5,12.5){-1}
\end{feynartspicture}
\end{center}
\caption{\footnotesize Contribution to the decay from bilinear and trilinear counter-terms (grey dots).}
\label{fig:AAA}
\end{figure}

The effective vertex of one heavy vector to SM vectors is given by the diagrams in Figure \ref{fig:AAA}, where a grey dot stands for a 
counter-term insertion (either trilinear coupling or mixing term). The mixing terms contribute by converting a massive state in the 
external leg into a zero mode (while the bulk couplings do preserve KK number).
Summing all the contributions for an SU(N) gauge boson, we obtain:
\beq
i V^{\mu\nu\rho}=\frac{ig^3 f^{abc}C_2(G)}{16 \pi^4}\frac{i \pi^2}{\epsilon} \  \mathcal{T}^{\mu\nu\rho}\,,
\eeq
where the tensor 
\begin{align}
\mathcal{T}^{\mu \nu \rho} = &\frac{4 M^2 ((\xi -11) \xi +6) (q_1^\nu -q_2^\rho ) g^{\mu  \rho}+ ((\xi-12) \xi+3) 
(q_2^\mu-q_1^\mu)q_1^\nu q_2^\rho }{16   M^2   \xi} \nonumber\\
&+\frac{  (\xi -1) (\xi +1) (\xi +3) (q_1^\mu +q_2^\mu ) (q_1^\nu q_1^\rho - q_2^\nu q_2^\rho)  }{16   M^2  \xi}
\end{align}
is gauge dependent.
In the limit of massless zero modes, from gauge invariance we would expect this decay to vanish: in fact
\beq
\mathcal{T}^{\mu\nu\rho} \epsilon^\rho(q_2)\epsilon^\nu(q_1) =0 \,,
\eeq
where $\epsilon^\nu(q_1)$ and $\epsilon^\rho(q_2)$ are the polarisation vectors of the two (massless) vectors.
Note that the other two contractions
\beq
\mathcal{T}^{\mu\nu\rho} \epsilon^\mu(q)\epsilon^\nu(q_1) &\neq& 0  \nonumber\\
\mathcal{T}^{\mu\nu\rho} \epsilon^\rho(q_2)\epsilon^\mu(q)&\neq & 0  \nonumber
\label{eq:AAAmassless}
\eeq
do not vanish because $A^{2k,2l}$ is massive.
This means, for instance, that a heavy gluon cannot decay into two SM gluons.
On the other hand, the decay into massive gauge bosons, the $W$ and $Z$, in this limit, is given by the decay into the 
Higgs components, as discussed below.

\subsubsection*{Heavy vector to fermions, $A^{2k,2l}_\mu \longrightarrow \bar{f} f $}

\begin{figure}[tb]
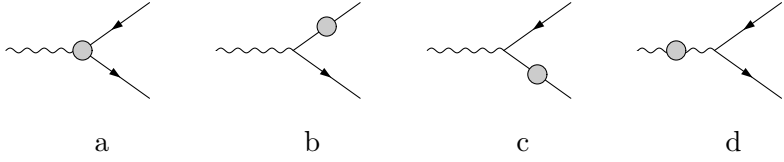

\begin{center}
\begin{feynartspicture}(14,4)(5,1)

\FADiagram{a}
\FAProp(-0.,10.)(8.,10.)(0.,){/Sine}{0}
\FAProp(8.,10.)(15.,15.)(0.,){/Straight}{-1}
\FAProp(8.,10.)(15.,5.)(0.,){/Straight}{1}
\FAVert(8.,10.){-1}

\FADiagram{b}
\FAProp(-0.,10.)(8.,10.)(0.,){/Sine}{0}
\FAProp(8.,10.)(15.,15.)(0.,){/Straight}{-1}
\FAProp(8.,10.)(15.,5.)(0.,){/Straight}{1}
\FAVert(11.5,12.5){-1}

\FADiagram{c}
\FAProp(-0.,10.)(8.,10.)(0.,){/Sine}{0}
\FAProp(8.,10.)(15.,15.)(0.,){/Straight}{-1}
\FAProp(8.,10.)(15.,5.)(0.,){/Straight}{1}
\FAVert(11.5,7.5){-1}

\FADiagram{d}
\FAProp(-0.,10.)(8.,10.)(0.,){/Sine}{0}
\FAProp(8.,10.)(15.,15.)(0.,){/Straight}{-1}
\FAProp(8.,10.)(15.,5.)(0.,){/Straight}{1}
\FAVert(4.,10.){-1}
\end{feynartspicture}
\end{center}
\caption{\footnotesize Contribution to the decay from bilinear and trilinear counter-terms.}
\label{fig:Aff}
\end{figure}

Fermion decays of the heavy gauge vector are given by the diagrams in Figure~\ref{fig:Aff}: like in the previous case, we 
will consider the external particle on-shell. In this limit, for a SU(N) the effective vertex can be written as:
\begin{eqnarray}
i\Gamma^\mu_A=\frac{g^3}{16 \pi^4} \frac{i \pi^2}{\epsilon} \tau_A \gamma^\mu P_L t^a\,, \label{eq:Affgen}
\end{eqnarray}
where the contribution of the single diagrams are listed in Table~\ref{tab:Aff}.

\begin{table}[tb]
\begin{center} \begin{tabular}{|l|c|c||c|}
\hline
$\tau_{x}$ & $R_\xi$ gauge & $\xi=1$ \scriptsize{Feynman gauge} & Higgs\\
\hline
$\tau_{a}$ & $\frac{1}{4}(4(\xi-1)C_2(r_f)+(\xi+3)C_2(G))$ &$C_2(G)$ & 0\\
\hline
$\tau_{b+c}$ & $-2\ \frac{1}{2}(\xi+3)C_2(r_f)$ &$-4 C_2(r_f)$ & 0 \\
\hline
$\tau_{d}$ & $  \frac{1}{4}(13-\xi)C_2(G)$& $ 3 C_2(G)$  & $-\frac{1}{3} C (r_h)$\\
\hline
\hline
$\tau_{tot}$ & $4(C_2(G)-C_2(r_f))$ & $4C_2(G)-4C_2(r_f) $ & $-\frac{1}{3} C (r_h)$ \\
\hline
\end{tabular} \end{center}
\caption{\footnotesize Contribution to the effective vertex with two massless fermions. The results in Feynman gauge $\xi=1$ are 
also listed to be compared with the direct calculation. } \label{tab:Aff}
\end{table}
The total result is, as expected, independent on $\xi$ and given by:
\begin{eqnarray}
i\Gamma^\mu=-\frac{g^3 }{16 \pi^4} \frac{i \pi^2}{\epsilon}\ 4 (C_2(r_f)-C_2(G)) \gamma^\mu P_L t^a\,.
\end{eqnarray}

For a U(1) gauge boson, it is enough to replace $C_2(G) \to 0$, $C_2(r_f) \to Y_f^2$ and $t^a \to Y_f$.
If we consider the corrections by a gauge group SU(N') to the decays of an SU(N) gauge boson, in the formula we replace 
$g^3 \to g {g'}^2$,  $C_2(G) \to 0$ and $C_2(r_f) \to C_2 {r'_f}$.

Finally we quote the effective couplings in the case of the Standard Model on the Real Projective plane. 
As more gauge groups contribute to the same effective coupling, we parameterise a generic coupling as
\beq
i \Gamma^\mu_X = - i \frac{g}{\pi \epsilon}\ C_X \gamma^\mu P_L t^a\,, \label{eq:Aff}
\eeq
where the various coefficients $C_X$ are listed in Table~\ref{tab:AffSM}.

\begin{table}[ht]
\begin{center} \begin{tabular}{|l|c|c|c|}
\hline
U(1)   & gauge & Higgs & Yukawa  \\
 \hline
 $B \to E$ & $\alpha_1$  & $\frac{1}{24} \alpha_1$ & $\frac{1}{2} \frac{y_l^2}{4 \pi}$ \\
\hline 
$B \to L$ & $\frac{1}{4} \alpha_1 + \frac{3}{4} \alpha_2$& $\frac{1}{24} \alpha_1$ & $\frac{1}{4} \frac{y_l^2}{4 \pi}$\\
\hline 
$B \to U $ & $ \frac{4}{9} \alpha_1 + \frac{4}{3} \alpha_3$& $\frac{1}{24} \alpha_1$ & $\frac{1}{2} \frac{y_u^2}{4 \pi}$\\
 \hline
$B \to D$ & $\frac{1}{9} \alpha_1 + \frac{4}{3} \alpha_3$& $\frac{1}{24} \alpha_1$ & $\frac{1}{2} \frac{y_d^2}{4 \pi}$ \\
\hline
$B\to Q$ & $\frac{1}{36} \alpha_1 + \frac{3}{4} \alpha_2 + \frac{4}{3} \alpha_1$ & $\frac{1}{24} \alpha_1$ & $\frac{1}{4} 
\frac{y_u^2+y_d^2}{4 \pi}$\\
\hline
\hline
SU(2)  &  gauge & Higgs & Yukawa \\
\hline
$W \to L$ & $\frac{1}{4} \alpha_1 - \frac{5}{4} \alpha_2$ & $\frac{1}{24} \alpha_2$ & $\frac{1}{4} \frac{y_l^2}{4 \pi}$ \\
\hline
$W \to Q$ & $\frac{1}{36} \alpha_1 - \frac{5}{4} \alpha_2 + \frac{4}{3} \alpha_3$ & $\frac{1}{24} \alpha_2$ & $ \frac{1}{4} 
\frac{y_u^2+y_d^2}{4 \pi}$ \\
\hline
\hline
SU(3)  &  gauge & Higgs & Yukawa \\
\hline
$G \to U$ & $\frac{4}{9} \alpha_1 - \frac{5}{3} \alpha_3$ & $0$ & $\frac{1}{2} \frac{y_u^2}{4 \pi}$\\
\hline
$G \to D$ & $\frac{1}{9} \alpha_1 - \frac{5}{3} \alpha_3$ & $0$ & $\frac{1}{2} \frac{y_d^2}{4 \pi}$\\
\hline
$G \to Q$ & $\frac{1}{36} \alpha_1 + \frac{3}{4} \alpha_2 - \frac{5}{3} \alpha_3$ & $0$ & $\frac{1}{4} \frac{y_u^2+y_d^2}{4 \pi}$\\
\hline
 \end{tabular} \end{center}
\caption{\footnotesize Coefficients $C$ as defined in Eq.~\ref{eq:Aff} of the effective couplings in the Standard Model in the gauge basis.} \label{tab:AffSM}
\end{table}

\subsubsection*{Yukawa couplings}

For heavy SM fermions, we need to take into account the contribution of the Yukawa couplings.
They will generate counter-terms for the vertex and fermion propagators, therefore they will contribute to diagrams (a), (b) 
and (c) in Figure~\ref{fig:Aff}.
As singlets and doublets couple differently to the Higgs, we need to distinguish the two cases: the results are summarised in 
Table~\ref{tab:Aff2}, where the coefficients refer to Eq.~\ref{eq:Affgen} with $g^3 \to g\, y_f^2 $. 
\begin{table}[tb]
\begin{center} \begin{tabular}{|l|c|c|c|c|c|}
\hline
   & U(1) to singlet & U(1) to doublet & SU(2) to singlet & SU(2) to doublet & SU(3) \\
\hline
(a) & $Y_S$ & $1/2 Y_D$ & $0$ & $1/2$ & $1/2$   \\
\hline
(b)+(c) & $Y_S$ & $1/2 Y_D$ & $0$ & $1/2$ & $1/2$  \\
\hline
\hline
tot & $2 Y_S$ & $Y_D$ & $0$ & $1$ & $1$  \\
\hline
\end{tabular} \end{center}
\caption{\footnotesize contribution of Higgs loops. $Y_S$ and $Y_D$ are the hypercharges of the singlet and 
doublet respectively, while $Y_H$ is the hypercharge of the Higgs.} \label{tab:Aff2}
\end{table}
Note that the contribution to the U(1) gauge boson coupling are always proportional to the fermion hypercharge, as expected.
The total contribution to SM fields is summarised in Table~\ref{tab:AffSM}.

\subsubsection*{Heavy vector to scalars : $A^{2k,2l}_\mu \longrightarrow \phi^\dagger \phi $}
\label{sec:Aphiphi}

\begin{figure}[tb]
\begin{center}
\begin{feynartspicture}(14,3)(5,1)

\FADiagram{}
\FAProp(-0.,10.)(8.,10.)(0.,){/Sine}{0}
\FAProp(8.,10.)(15.,15.)(0.,){/ScalarDash}{-1}
\FAProp(8.,10.)(15.,5.)(0.,){/ScalarDash}{1}
\FAVert(8.,10.){-1}

\FADiagram{}
\FAProp(-0.,10.)(8.,10.)(0.,){/Sine}{0}
\FAProp(8.,10.)(15.,15.)(0.,){/ScalarDash}{-1}
\FAProp(8.,10.)(15.,5.)(0.,){/ScalarDash}{1}
\FAVert(4.,10.){-1}

\FADiagram{}
\FAProp(-0.,10.)(8.,10.)(0.,){/Sine}{0}
\FAProp(8.,10.)(15.,15.)(0.,){/ScalarDash}{-1}
\FAProp(8.,10.)(15.,5.)(0.,){/ScalarDash}{1}
\FAVert(11.5,12.5){-1}

\FADiagram{}
\FAProp(-0.,10.)(8.,10.)(0.,){/Sine}{1}
\FAProp(8.,10.)(15.,15.)(0.,){/ScalarDash}{-1}
\FAProp(8.,10.)(15.,5.)(0.,){/ScalarDash}{1}
\FAVert(11.5,7.5){-1}
\end{feynartspicture}
\end{center}
\caption{\footnotesize Contribution to the decay from bilinear and trilinear counter-terms.}
\label{fig:Aphiphi}
\end{figure}
In this section we will focus on the decay into scalars. 
The contribution of the diagrams in Figure~\ref{fig:Aphiphi} are listed in Table~\ref{tab:Aphiphi}. For an SU(N) gauge boson, we obtain the following result: 
\begin{table}[tb]
\begin{center} \begin{tabular}{|l|c|c|}
\hline
$\tau_{x}$ & gauge & Higgs \\
\hline
$\tau_{a}$ & $- (\xi-3) C_2 (r_h) - \frac{1}{4} (\xi+3) C_2 (G)$ &$0$ \\
\hline
$\tau_{b}$ & $\frac{1}{4} (\xi-13) C_2 (G)$ &$\frac{1}{3} C (r_h)$ \\
\hline
$\tau_{c+d}$ & $  \frac{1}{2} (2\xi+1) C_2(r_h)$& $0$ \\
\hline
\hline
$\tau_{tot}$ & $\frac{7}{2} C_2(r_h)-4C_2(G)$ & $ \frac{1}{3} C(r_h)$ \\
\hline
\end{tabular} \end{center}
\caption{\footnotesize Contribution to the effective vertex with two massless scalars from the diagrams in figure~\ref{fig:Aphiphi}.} \label{tab:Aphiphi}
\end{table}

\begin{eqnarray}
i V^\mu=\frac{g^3}{16 \pi^4}\left\{\frac{i \pi^2}{\epsilon}\ 4 \left(\frac{7}{8} C_2(r_h)- C_2(G) + \frac{1}{12} C (r_h) \right) -
\frac{2 i \pi^3}{\eta M^2} C_2(r_h) \right\} \  (q_1^\mu-q_2^\mu) t^a\,.
\end{eqnarray}
Note that in this case the result is gauge independent as expected; in the following we will neglect the quadratically divergent 
contribution that, as explained before, should be of the same order of the Higgs mass.
As in the fermion case, the previous calculation made for generic SU(N) can be generalised: for a U(1) gauge boson, it is 
enough to replace $C_2 (G) \to 0$, $C_2 (r_h) \to Y_h^2$,  $C (r_h) \to 2 Y_h^2$ (the factor of 2 takes into account the fact that 
the scalar is a doublet) and $t^a \to Y_h$; for corrections of a different gauge group, $C_2 (G)$ and $C(r_h) \to 0$, $C_2 (r_h) \to 
C_2 (r'_h)$ and $g^3 \to g {g'}^2$. 

The only scalar in the SM is the Higgs, which is a doublet of SU(2) with hypercharge $Y_h = 1/2$. 
Only SU(2) and U(1) gauge fields, therefore, can decay into Higgs fields.
Note also that, in the limit of unbroken SU(2)$\times$ U(1) that we are considering here, the Higgs doublet contains both the 
physical Higgs boson $h$ and the Goldstone bosons which are the longitudinal polarisations of the massive zero mode vectors, 
$W$ and $Z$. Therefore, we can use the equivalence principle to calculate the decay to Goldstone bosons and identify these 
partial widths with the ones in massive vectors.

The effective coupling of a heavy vector to scalars can be parameterised as
\begin{eqnarray}
i V^\mu=i \frac{g}{\pi \epsilon}\ C_X\,  (q_1^\mu-q_2^\mu) t^a\,;
\label{eqn:Aphiphi}
\end{eqnarray}
with
\beq
C_{U(1)} =  \alpha_1 \frac{25}{96} + \alpha_2 \frac{21}{32}\,, \qquad C_{SU(2)} = \alpha_1 \frac{7}{32} - \alpha_2 \frac{125}{96}\,.
\eeq

\subsubsection*{Gauge scalar decays: $A_\phi^{2k,2l} \longrightarrow \bar{f} f, \ \phi^\dagger \phi$}

\begin{figure}[tb]
\begin{center}
\begin{feynartspicture}(14,4)(5,1)

\FADiagram{a}
\FAProp(-0.,10.)(8.,10.)(0.,){/ScalarDash}{0}
\FAProp(8.,10.)(15.,15.)(0.,){/Straight}{-1}
\FAProp(8.,10.)(15.,5.)(0.,){/Straight}{1}
\FAVert(11.5,12.5){-1}

\FADiagram{b}
\FAProp(-0.,10.)(8.,10.)(0.,){/ScalarDash}{0}
\FAProp(8.,10.)(15.,15.)(0.,){/Straight}{-1}
\FAProp(8.,10.)(15.,5.)(0.,){/Straight}{1}
\FAVert(11.5,7.5){-1}

\FADiagram{}

\FADiagram{a'}
\FAProp(-0.,10.)(8.,10.)(0.,){/ScalarDash}{0}
\FAProp(8.,10.)(15.,15.)(0.,){/ScalarDash}{-1}
\FAProp(8.,10.)(15.,5.)(0.,){/ScalarDash}{1}
\FAVert(11.5,12.5){-1}

\FADiagram{b'}
\FAProp(-0.,10.)(8.,10.)(0.,){/ScalarDash}{0}
\FAProp(8.,10.)(15.,15.)(0.,){/ScalarDash}{-1}
\FAProp(8.,10.)(15.,5.)(0.,){/ScalarDash}{1}
\FAVert(11.5,7.5){-1}
\end{feynartspicture}
\end{center}
\caption{\footnotesize Contribution to the decay from bilinear and trilinear counter-terms.}
\label{fig:Aphiff}
\end{figure}

The gauge scalar can also, in principle, decay to a pair of fermions or scalars via the diagrams in figure~\ref{fig:Aphiff}: note that 
counter-terms of the trilinear coupling and of the mixing of the gauge scalar are absent.
However, the sum of the two diagrams vanish once we take on-shell fermions: the reason is that the part of the fermion mixing 
counter-term proportional to the momentum vanishes on shell, while the term proportional to the heavy mass $M$ cancels out 
between the two diagrams. All in all, therefore, this effective vertex is absent.

\subsection{Heavy fermion decays}

\subsubsection*{Into a (massless) gauge boson, $\psi^{2k,2l} \to f A_\mu$}

Heavy fermions can couple to a light fermion plus a zero mode gauge boson.
In this section, we will limit ourselves to the case of exactly massless gauge bosons: the decay to massive ones, like the $W$ and 
$Z$, is in this approximation given by the decay to Higgs bosons.
The effective vertex is given by the diagrams in Figure~\ref{fig:FAf}: we will always consider on-shell external particles.

\begin{figure}[tb]
\begin{center}
\begin{feynartspicture}(15,4)(5,1)

\FADiagram{a}
\FAProp(-0.,10.)(8.,10.)(0.,){/Straight}{1}
\FAProp(8.,10.)(15.,15.)(0.,){/Sine}{0}
\FAProp(8.,10.)(15.,5.)(0.,){/Straight}{1}
\FAVert(8.,10.){-1}

\FADiagram{$b_1$}
\FAProp(-0.,10.)(8.,10.)(0.,){/Straight}{0}
\FAProp(8.,10.)(15.,15.)(0.,){/Sine}{1}
\FAProp(8.,10.)(15.,5.)(0.,){/Straight}{1}
\FAVert(11.5,12.5){-1}

\FADiagram{$b_2$}
\FAProp(-0.,10.)(8.,10.)(0.,){/Straight}{1}
\FAProp(11.5,12.5)(15.,15.)(0.,){/Sine}{0}
\FAProp(8.,10.)(11.5,12.5)(0.,){/ScalarDash}{0}
\FAProp(8.,10.)(15.,5.)(0.,){/Straight}{1}
\FAVert(11.5,12.5){-1}

\FADiagram{c}
\FAProp(-0.,10.)(8.,10.)(0.,){/Straight}{1}
\FAProp(8.,10.)(15.,15.)(0.,){/Sine}{0}
\FAProp(8.,10.)(15.,5.)(0.,){/Straight}{1}
\FAVert(11.5,7.5){-1}

\FADiagram{d}
\FAProp(-0.,10.)(8.,10.)(0.,){/Straight}{1}
\FAProp(8.,10.)(15.,15.)(0.,){/Sine}{0}
\FAProp(8.,10.)(15.,5.)(0.,){/Straight}{1}
\FAVert(4.,10.){-1}

\end{feynartspicture}
\end{center}
\caption{\footnotesize Contribution to the decay from bilinear and trilinear counter-terms.}
\label{fig:FAf}
\end{figure}
In general, the effective vertex is given by the following formula:
\begin{eqnarray}
i\Gamma^\mu_X=\frac{g^3}{16 \pi^4} \frac{i \pi^2}{\epsilon} (\tau_{x} \gamma^\mu + \sigma_x q_1^\mu)P_L t^a\,;
\end{eqnarray}
the $\tau$-term is a correction to the usual gauge coupling and, from gauge invariance, we would expect it to vanish.
In fact, if present, it would correspond to a non-diagonal kinetic term.
The second term ($q_1$ is the momentum of the gauge boson) is potentially non-zero: however, when calculating the decay 
width, is must be contracted with the polarisation vector of the gauge boson:
\beq
 i\mathcal{M}=i\Gamma^\mu \epsilon_\mu(q1) \propto q_1^\mu . \epsilon_\mu(q_1) = 0\,,
\eeq
where the product vanishes because of the masslessness of the gauge boson.
Therefore, we expect that the fermion decay width is zero in the massless gauge boson limit.

\begin{table}[tb]
\begin{center} \begin{tabular}{|l|c|c|}
\hline
diagram & $\tau_{x}$&$\sigma _{x}$ 
 \\
\hline
$a$ & $\frac{1}{4}(4(\xi-1)C_2(r_f)+(\xi+3)C_2(G))$ &$0$\\
\hline
$b_1+b_2$ & $-\frac{1}{4}(\xi+3)C_2(G)$ &$\frac{1}{8M}(13-\xi)C_2(G)$ \\
\hline
$c$ & $  -\frac{1}{2} (\xi+3)C_2(r_f)$& $ 0$ \\
\hline
$d$ & $  \ \frac{1}{2}(-\xi+5) C_2(r_f)$& $0)$ \\
\hline
\hline
$tot$ & $0$ & $\frac{1}{8M}(13-\xi)C_2(G)$\\
\hline
\end{tabular} \end{center}
\caption{\footnotesize Contribution to the decay of a heavy fermion into a gauge boson from the diagrams in Figure~\ref{fig:FAf}. As expected, the coefficient $\tau$ vanishes due to gauge invariance.} \label{tab:FAmuf}
\end{table}

The contributions of each diagram are listed in Table~\ref{tab:FAmuf}: all in all we have
\begin{eqnarray}
i\Gamma^\mu=\frac{g^3 }{16 \pi^4} \frac{i \pi^2}{8 M \epsilon}(13-\xi)C_2(G) q_1^\mu P_L t^a\,.
\label{eq:vertexFfA}
\end{eqnarray}
One can see that this vertex is gauge dependent, however it does not contribute to the decay width.

\subsubsection*{Into a (massless) Higgs boson, $\psi^{2k,2l} \to f \phi$}

\begin{figure}[tb]
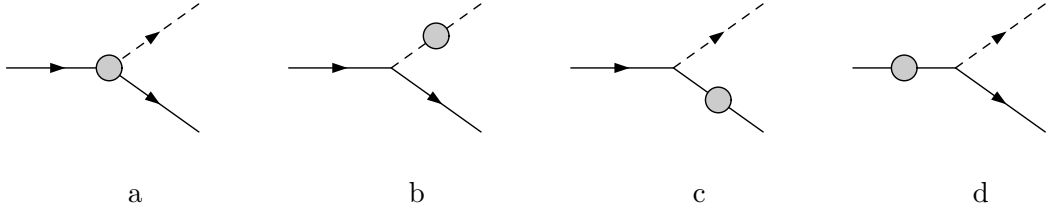

\begin{center}
\begin{feynartspicture}(15,4)(4,1)

\FADiagram{a}
\FAProp(-0.,10.)(8.,10.)(0.,){/Straight}{1}
\FAProp(8.,10.)(15.,15.)(0.,){/ScalarDash}{1}
\FAProp(8.,10.)(15.,5.)(0.,){/Straight}{1}
\FAVert(8.,10.){-1}

\FADiagram{b}
\FAProp(-0.,10.)(8.,10.)(0.,){/Straight}{1}
\FAProp(8.,10.)(15.,15.)(0.,){/ScalarDash}{1}
\FAProp(8.,10.)(15.,5.)(0.,){/Straight}{1}
\FAVert(11.5,12.5){-1}
 
\FADiagram{c}
\FAProp(-0.,10.)(8.,10.)(0.,){/Straight}{1}
\FAProp(8.,10.)(15.,15.)(0.,){/ScalarDash}{1}
\FAProp(8.,10.)(15.,5.)(0.,){/Straight}{1}
\FAVert(11.5,7.5){-1}

\FADiagram{d}
\FAProp(-0.,10.)(8.,10.)(0.,){/Straight}{1}
\FAProp(8.,10.)(15.,15.)(0.,){/ScalarDash}{1}
\FAProp(8.,10.)(15.,5.)(0.,){/Straight}{1}
\FAVert(4.,10.){-1}

\end{feynartspicture}
\end{center}
\caption{\footnotesize Contribution to the decay from bilinear and trilinear counter-terms.}
\label{fig:Fphif}
\end{figure}

Heavy fermions can also couple to a light fermion plus a zero mode scalar via Yukawa couplings: this effective vertex will describe 
both the coupling to the Higgs boson, and the decay into longitudinal $W$ and $Z$ bosons.
The effective vertex is given by the diagrams in Figure~\ref{fig:Fphif}: we will always consider on-shell external particles while the 
outgoing particles are massless.

In general, the effective vertex can be parameterised as
\begin{eqnarray}
i \Gamma =-\frac{y_\psi}{16 \pi^4} \frac{i \pi^2}{\epsilon}  (g^2\tau_\phi+y_\psi^2\tau_\phi')  P_{L(R)} \,;
\end{eqnarray}
here we will consider the case $\psi_D \to \psi_S + \phi$, for which $Y_D = Y_S + Y_h$.
The contributions to $\tau$ and $\tau'$ are given in Table \ref{tab:Fphif}. 
\begin{table}[ht]
\begin{center} \begin{tabular}{|l|c|c|}
\hline
diagram & $\tau_{x}$ & $\tau'_{x}$ 
 \\
\hline
$a $ & $(\xi+3)t^a(r_D).t^a(r_S)+\xi (- t^a(r_S)+t^a(r_D)).t^a(r_h)$&  $1$ \\
\hline
$ b$ & $- \frac{1}{4} (2\xi+1)C_2(r_h) $ &  $0$ \\
\hline
$ c$ & $  - \frac{1}{2}(\xi+3) C_2(r_S) $ &$ - \frac{1}{2}$ \\
\hline
$ d$ & $   \frac{1}{2} (-\xi+5) C_2(r_D)$& $0$  \\
\hline
\hline
$ tot  $ &  $\begin{array}{l} \xi (t^a(r_D).t^a(r_S) -  t^a(r_S).t^a(r_h) +  t^a(r_D).t^a(r_h)) + \\ - \xi/2 ( C_2 (r_h) + C_2 (r_D) + C_2 (r_S)) +\\ 3 t^a(r_D).t^a(r_S) - 1/4 C_2 (r_h) - 3/2 C_2 (r_S) + 5/2 C_2 (r_D) \end{array} $ & $\frac{1}{2}$ \\
\hline
\end{tabular} \end{center}
\caption{\footnotesize Contribution to the decay of a heavy fermion into scalars. By the equivalence principle, these couplings also give the decay rates into massive SM gauge bosons.} \label{tab:Fphif}
\end{table}
While $\tau'_x$ is gauge invariant, $\tau_x$ may in general depend on the $\xi$ parameter: therefore we need to check case by case 
that the gauge dependence is cancelled. For a U(1) gauge boson, $t^a (r_X) \to Y_X$ and $C_2 (r_X) \to Y_X^2$:
\begin{multline}
\tau_{U(1)}^D = \frac{\xi}{2} (Y_D - Y_S - Y_h)^2 + 3 Y_D Y_S - 1/4 Y^2_h - 3/2 Y^2_S + 5/2 Y^2_D = \\  3 Y_D Y_S - 1/4 Y^2_h - 3/2 Y^2_S + 5/2 Y^2_D \,.
\end{multline}
The latter result has a quite interesting property: using the fact that $Y_S = Y_D - Y_h$, it can be rewritten as:
\beq
\tau_{U(1)}^D =  4 Y_D^2 - 7/4 Y^2_h\,;
\eeq
this means that the result will be the same for the two components of the doublet, because it does 
not depend on the sign of $Y_h$~\footnote{This is a consequence of custodial symmetry: in fact, neglecting Yukawa couplings and the masses of the light fermions, the two right-handed SM fermions form a doublet and therefore the decays of up and down component of the doublet D should be the same.}. In the case of SU(2), when $D$ and the scalar are doublets and $S$ is a singlet, 
we have $ t^a(r_D).t^a(r_h) = C_2(r_h)$ while $t^a (r_S) = 0$ and $C_2 (r_S) = 0$; therefore 
\begin{multline}
\tau_{SU(2)}^D = \frac{\xi}{2} (2 C_2(r_h) - C_2(r_h) - C_2(r_h)) + \left(- \frac{1}{4} + \frac{5}{2} \right) 
C_2(r_h)= \frac{9}{4} C_2(r_h) = \frac{27}{16} \,.
\end{multline}
Finally, for SU(3), it is the Higgs that does not couple to gluons, therefore $ t^a(r_D).t^a(r_S) = C_2(r_f)$ 
while $t^a (r_h) = 0$ and $C_2 (r_h) = 0$.
The total result is:
\begin{multline}
\tau_{SU(3)}^D = \frac{\xi}{2} (2 C_2(r_f) - C_2(r_f) - C_2(r_f)) + \left(3 - \frac{3}{2} + \frac{5}{2} \right) 
C_2(r_f)= 4 C_2(r_f) = \frac{16}{3}\,.
\end{multline}

For the decay of heavy singlets into doublets, the formulae are a bit different: in Table~\ref{tab:Fphif}, the only difference 
is that diagram (c) is proportional to $C(r_S)$, while diagram (d) to $C(r_D)$.
The above results are therefore modified to
\beq
\tau_{U(1)}^S &=& 3 Y_D Y_S - 1/4 Y^2_h - 3/2 Y^2_D + 5/2 Y^2_S\,, \\
\tau_{SU(2)}^S & = & \left(- \frac{1}{4} - \frac{3}{2} \right) C_2(r_h)= - \frac{7}{4} C_2(r_h) = -\frac{21}{16} \,,\\
\tau_{SU(3)}^S &=& \tau_{SU(3)}^D \,.
\eeq
Note that in this case, the decays of up and down singlets will be different: this is a consequence of the breaking of the custodial symmetry by gauge interactions~\footnote{Already, the masses of the two heavy fermions will be different because they have different hypercharges.}.
Furthermore, the contribution of diagram (c) to $\tau'_X$ picks a factor of $1/2$ because it applies to the correction to 
a doublet fermion, and it is proportional to both Yukawa couplings of the doublet (this is relevant in the case of the quarks).

To summarise the results in the case of SM fermions, if we parameterise the effective coupling as
\beq
i \Gamma = - i \frac{y_f}{\pi \epsilon} C_X P_{L(R)}\,,
\eeq
the coefficients are given by
\beq
C_{L\to E \phi} &=& \frac{9}{64} \alpha_1 + \frac{29}{64} \alpha_2 + \frac{1}{8} \frac{y_l^2}{4 \pi}\,, \\
C_{E\to L \phi^\dagger} &=& \frac{57}{64} \alpha_1 - \frac{21}{64} \alpha_2 + \frac{3}{16} \frac{y_l^2}{4 \pi}\,, \\
C_{Q\to U \phi^\dagger} &=& - \frac{47}{576} \alpha_1 + \frac{29}{64} \alpha_2 + \frac{4}{3} \alpha_3 + \frac{1}{8} 
\frac{y_u^2}{4 \pi}\,, \\
C_{Q\to D \phi} &=& -\frac{47}{576} \alpha_1 + \frac{29}{64} \alpha_2 + \frac{4}{3} \alpha_3 + \frac{1}{8} 
\frac{y_d^2}{4 \pi}\,,\\
C_{U\to Q \phi} &=& \frac{193}{576} \alpha_1 - \frac{21}{64} \alpha_2 + \frac{4}{3} \alpha_3 + \frac{1}{16} 
\frac{3y_u^2 + 2 y_d^2}{4 \pi}\,, \\
C_{D\to Q \phi^\dagger} &=& \frac{1}{576} \alpha_1 - \frac{21}{64} \alpha_2 + \frac{4}{3} \alpha_3 + \frac{1}{16} 
\frac{3y_d^2+ 2y_u^2}{4 \pi}\,.
\eeq

\section{A ``genuine'' 6D loop calculation: the level $(2 n, 0)$}
\label{sec:6Dloop}
\setcounter{equation}{0}
\setcounter{footnote}{0}

In the previous sections we used the case of the mixing of states in the tier $(2n, 2m)$ to show that with a 4D calculation we 
can predict the divergent contribution of all one loop corrections to masses and couplings.
The argument is that the log divergences are regularised by localised counter-terms and, in the magic $\xi = -3$ gauge, we 
only need a handful of operators in gauge invariant form.
The only limitation of this method is that it does not allow to calculate finite contributions, that may be relevant for light tiers and 
for states that vanish on the singular points (and that therefore only receive finite contributions from one loop corrections).
In such a case, a detailed 6D calculation is necessary.

\setlength{\unitlength}{1cm}
\begin{figure}[tb]
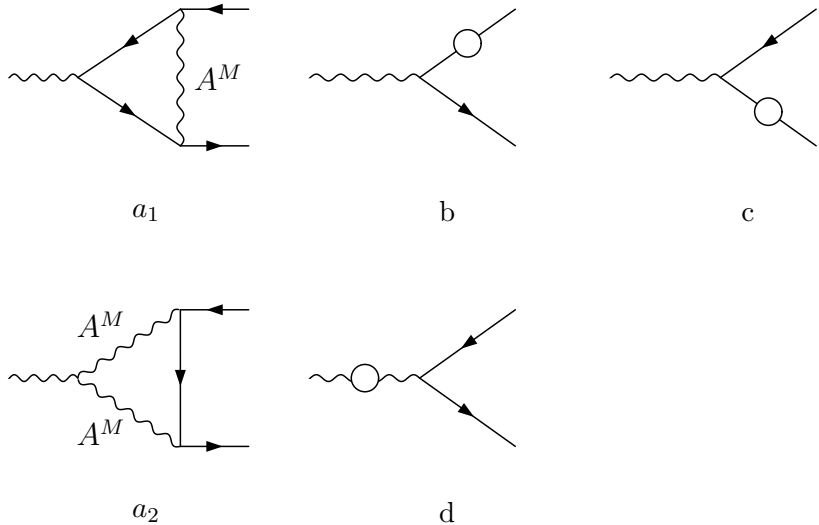

\begin{center}
\begin{feynartspicture}(12,8)(3,2)

\FADiagram{$a_1$}
\FAProp(0.,10.)(5.,10.)(0.,){/Sine}{0}
\FAProp(5.,10.)(12.5,15.)(0.,){/Straight}{-1}
\FAProp(12.5,15.)(12.5,5.)(0.,){/Sine}{0}
\FALabel(13.52,10.)[l]{$A^{M}$}
\FAProp(12.5,5.)(5.,10.)(0.,){/Straight}{-1}
\FAProp(12.5,15.)(17.5,15.)(0.,){/Straight}{-1}
\FAProp(12.5,5.)(17.5,5.)(0.,){/Straight}{1}

\FADiagram{b}
\FAProp(-0.,10.)(8.,10.)(0.,){/Sine}{0}
\FAProp(8.,10.)(15.,15.)(0.,){/Straight}{-1}
\FAProp(8.,10.)(15.,5.)(0.,){/Straight}{1}
\FAVert(11.5,12.5){-0.1}

\FADiagram{c}
\FAProp(-0.,10.)(8.,10.)(0.,){/Sine}{0}
\FAProp(8.,10.)(15.,15.)(0.,){/Straight}{-1}
\FAProp(8.,10.)(15.,5.)(0.,){/Straight}{1}
\FAVert(11.5,7.5){-0.1}

\FADiagram{$a_2$}
\FAProp(0.,10.)(5.,10.)(0.,){/Sine}{0}
\FAProp(5.,10.)(12.5,15.)(0.,){/Sine}{0}
\FALabel(8.48,13.02)[br]{$A^{M}$}
\FAProp(12.5,15.)(12.5,5.)(0.,){/Straight}{1}
\FAProp(12.5,5.)(5.,10.)(0.,){/Sine}{0}
\FALabel(8.48,6.98)[tr]{$A^{M}$}
\FAProp(12.5,15.)(17.5,15.)(0.,){/Straight}{-1}
\FAProp(12.5,5.)(17.5,5.)(0.,){/Straight}{1}

\FADiagram{d}
\FAProp(-0.,10.)(8.,10.)(0.,){/Sine}{0}
\FAProp(8.,10.)(15.,15.)(0.,){/Straight}{-1}
\FAProp(8.,10.)(15.,5.)(0.,){/Straight}{1}
\FAVert(4.,10.){-0.1}
\end{feynartspicture}
\end{center}
\caption{\footnotesize decays of $A^{2n,0}_\mu$ into two SM fermions}
\label{fig:loopstriangle}
\end{figure}

In this section, we want to check our results by calculating loop corrections to the level  $(2n,0)$: in this case the loops will 
contain a sum over the KK modes. This tier is also more interesting phenomenologically because, for $n=1$, it contains the 
lightest even tier that can decay into SM states at loop level, without missing energy.
We used the method consisting in expanding in KK modes only along one direction and using the re-summed 5D 
propagator~\cite{Puchwein:2003jq} in the sum, and we checked the result, when possible, using the complete expansions 
in 6D KK modes~\cite{Cheng:2002iz}. Moreover, we will work in the Feynman gauge $\xi = 1$, where the loops are greatly simplified.
In general, the loop contribution (that we generically label $V$) can be divided in 4 pieces
\beq
V = V_T + p_g V_G + p_g p_r V_{G'} + p_r V_R\,,
\eeq
where $p_i$ are the parities under glide and rotation of the particles running in the loop.
The first term, $V_T$, is the contribution one would get on a torus compactification: after renormalisation of the bulk kinetic terms, 
it leaves a finite contribution.
Note that in this section we will be interested in a coupling between one state in the tier $(2n,0)$ and two SM particles $(0,0)$.
This vertex is absent in the bulk because it violates the bulk sum rules, therefore we would expect that the torus contribution is 
finite or vanishing. The other three terms correspond to the two glides and rotation, in the sense that their sign depends on 
the parities $(p_r, p_g)$ of the fields running in the loop. The contribution of the two glides is finite because the glides do not have 
any fixed points where a counter-term could be localised. 
On the other hand, the rotation does generate divergences which will be cut-off by counter-terms localised on the four points left 
fixed by the rotation, i.e. the two singular points of the orbifold.
The singularities will be equally spread on the two points, because of the extended global symmetries of the bulk interactions. 
In the calculation of the 6D loops, we will regularise the divergences by cutting off the 4D momentum integral at a cut-off 
$\Lambda$ after re-summing over the infinite KK states.
For simplicity we will focus on a single coupling: that of a massive vector boson $A_\mu^{2n,0}$ with massless fermions.

\begin{table}[tb]
\begin{center}
\begin{tabular}{|c||c|c|c|c|}
\hline
$\alpha$& torus & $\times p_g$ & $\times p_g p_r$ & $\times p_r$ \\
 \hline
 $a_1$ &  $0$ & $0 $ & {\small $(-1)^n (\Phi_1(n)+\Phi_2(n)+\Phi_3(n)-\Phi_4(n) )$} &  $0$ \\
 $b$ & $0$ & $0$ &{\small  $(-1)^{n}  (-12\Phi_2(n))$} &  $-8 n^2 \pi^2 \log \Lambda^2 R^2$ \\
 $c$& $0$ & $0$ & {\small $(-1)^{n}  (-12\Phi_2(n))$} &  $-8 n^2 \pi^2 \log \Lambda^2 R^2$ \\
\hline
Total & $0$ & $0$ & {\small $(-1)^n (\Phi_1(n) - 23 \Phi_2(n)+\Phi_3(n)-\Phi_4(n) )$} & $-16 n^2 \pi^2 \log \Lambda^2 R^2$\\
 \hline
 \hline
$\beta$& torus  & $\times p_g$ & $\times p_g p_r$ & $\times p_r$ \\
 \hline
$ a_{2}\ (A_\mu - A_\nu)$ & $0$ & $0$ &{\small  $(-1)^{n}\ \left(-3\Phi_1(n)+8\Phi_2(n)-3\Phi_3(n)\right)$ }&  $4 \pi^2 n^2  
\log \Lambda^2 R^2$ \\
$a_{2}\ (A_\mu - A_5)$ & $0$ & $0$ & {\small $(-1)^{n}\ 2 \left(-\Phi_1(n)-\Phi_2(n)-\Phi_3(n)-\Phi_4(n)\right)$} &  $0$ \\
$a_{2}\ (A_\mu - A_6)$ & $0$ & $0$ & {\small $(-1)^{n}\ 6 \left(-\Phi_2(n)-\Phi_4(n)\right)$} &  $0$ \\
$d$ & $0$ & $0$ & {\small $(-1)^{n} \left(2\Phi_1(n)-12\Phi_2(n)+\Phi_3(n)\right) $} &  $12 n^2 \pi^2  \log \Lambda^2 R^2 $ \\
 \hline
 Total & $0$ & $0$ & {\small $(-1)^{n} \left(-3\Phi_1(n)-12\Phi_2(n)-4 \Phi_3(n)- 8\Phi_4 (n) \right) $} &  $ 16 n^2 \pi^2 
 \log \Lambda^2 R^2 $ \\
 \hline
  \end{tabular} \end{center} 
\caption{\footnotesize Contributions of the single diagrams to the parameters $\alpha$ and $\beta$ in Feynman gauge. The three contributions to 
graph $a_2$ come from two vectors, one vector and one $A_5$ and one vector and one $A_6$ running in the loop respectively.}
\label{tab:loopstriangle}
\end{table}

The diagrams contributing to the effective vertex are depicted in Figure~\ref{fig:loopstriangle}: the white blob represents a 
one loop mixing between the state in tier $(2n,0)$ and a SM state in $(0,0)$.
Note that these diagrams are in one-to-one correspondence with the diagrams in Figure~\ref{fig:Aff}: $a_1+a_2$ represents 
the correction to the vertex and the others corrections to the propagators.
We will see that, with the 6D calculation, we can recover the previous results.
For a generic SU(N) group, we can parameterise the effective vertex as:
\beq
i V^\mu = i \ \frac{1}{\sqrt{2}} \frac{g^3}{16 \pi^4} \frac{1}{8 n^2}\  \left(\alpha C_2 (r) + \beta C_2 (G) \right) \  
t^a \gamma^\mu P_{L(R)}\,, 
\eeq
where $P_{L(R)}$ refer to the coupling to left- or right-handed SM fermions.
The results of the individual diagrams is given in Table~\ref{tab:loopstriangle}: the finite functions $\Phi_i (n)$ are 
given in Appendix~\ref{app:formulae}.
They typically give a very small contribution to the total coupling.
Note also that, as expected, the torus contribution vanishes.

Putting all the results together, the effective vertex is given by
\beq
i V^\mu = i \ \frac{1}{\sqrt{2}} \frac{g^3}{16 \pi^2} \log (\Lambda R)\  \left(-4 C_2 (r) + 4 C_2 (G) \right) \  t^a \gamma^\mu 
P_{L(R)} + \mbox{finite}\,.
\eeq
This result reproduces exactly the vertex in the previous section summarised in Table~\ref{tab:Aff} (recall that $1/\epsilon = \log \Lambda R$), up to a factor 
of $1/\sqrt{2}$.
This is however not a surprise: if we think in terms of counter-terms, the only difference between a vector in the tier $(2n,2m)$ 
and one in the tier $(2n,0)$ is a normalisation factor coming from the fact that the wave function of $A_\mu^{2n,0}$ does 
not depend on $x_6$.
This normalisation factor accounts for the $\sqrt{2}$ factor, therefore the result of the 6D calculation matches with the 
counter-term prediction.

\section{LHC phenomenology of tier $(2,0)$ (and $(0,2)$)}
\label{sec:tier2}
\setcounter{equation}{0}
\setcounter{footnote}{0}

In this section we will discuss an important application of the results we obtained in the previous sections to the 
phenomenology of the tier $(2,0)$ and $(0,2)$. The peculiarity of these two tiers is that they are even under the KK parity, and particles in these levels can decay directly to a pair of SM states without missing energy. 
Potentially, this could lead to very clean signatures with resonances.
The level $(2,0)$ is not the lightest even level: the tier $(1,1)$ contains lighter states, however it cannot decay into SM 
particles via loops but via interactions localised on the singular points which violate maximally the symmetries of the bulk.
Therefore in the $(1,1)$ case the decay rates cannot be predicted and crucially depend on the UV completion of the theory.
The operators responsible for the decays into SM states are higher order operators, suppressed by a cut-off scale.
One can assume that they are small compared to one-loop effects; under this assumption all the states in the tier will decay 
into the lightest particle plus soft SM particles.
The fate of the lightest state in the $(1,1)$ is UV dependent: it may be stable, if protected by a symmetry in the UV, or long lived 
and be seen as missing energy in the experiments, or it may decay into a pair of SM states.
In~\cite{4tops}, we studied the case of decay into two tops, which leads to a very clean signature with 4 tops in the final state: 
the fact that the whole tier contributes to the production of the lightest state gives rise to very large cross section 
and potentially severe bounds on the mass and on the branching ratio into top pair.

\begin{figure}[tb]
\begin{center}
\includegraphics[width=10cm]{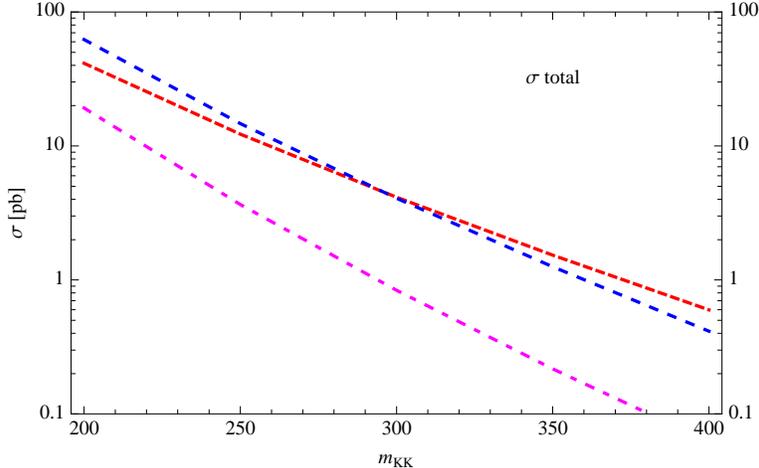}
\end{center}
\caption{\footnotesize Cross sections at the LHC at 7 TeV for the pair productions of heavy states: the red line represents the channel with a pair of heavy quarks, in blue an heavy gluon in 
association with an heavy quark, in magenta a pair of heavy gluons. All cross sections are calculated with calcHEP, using the PDF set {\tt cteq6m}.} \label{fig:xsects}
\end{figure}

The results of our calculation allows us to predict the divergent contributions to the masses of the states in the level $(2,0)$.
Here we will consider the non-symmetric case, where the radii of the two extra dimensions are different and the difference 
is larger than the loop contribution to the masses, therefore the mixing between the two levels can be neglected.
It is very similar to the spectrum of the tiers $(1,0)$ and $(0,1)$ with the exception that the loop corrections are larger compared 
to the Higgs VEV contribution due to the fact that the former are proportional to the mass of the tier.
At the LHC, the most important states to be produced are quarks and gluons, because they have strong interactions and also 
because the proton is made of quarks.
The tier contains two massive quarks for each SM quark: one corresponding to the SM singlet, $q_R^{2,0}$, and one 
corresponding to the SM doublet, $q_L^{2,0}$.
The mixing between the two is induced by the Yukawa couplings and it is negligible except for tops.
Here we will be mostly interested in partners of the light quarks, because of the larger cross sections.
Note also that the labels $L/R$ refer to the chirality of the SM fermion, while the heavy fermions are vector-like.
The main production cross sections as a function of the KK mass are plotted in Figure~\ref{fig:xsects}: we calculated the 
inclusive production cross sections for a pair of heavy quarks (and antiquarks), a single heavy quark in association with an 
heavy gluon and a pair of heavy gluons.  All other cross sections are sub-dominant.
The channels with heavy quarks are larger due to the dominant contribution coming from the production of heavy ups, which 
are enhanced due to the presence of a valence up quark in the initial state.
Nevertheless, all the other channels involving down and second generation quarks do contribute significantly and cannot be neglected.
In the plots, we limit the range of $m_{KK}$ between $200$ and $400$ GeV: this is in fact the preferred range by 
WMAP data~\cite{Komatsu:2010fb} on the relic Dark Matter abundance.
For $m_{KK} < 200$ GeV, the lightest particle is a charged singlet lepton, therefore this region is not viable.
A precise relic abundance calculation~\cite{relicDM} shows that the preferred masses are $m_{KK} \sim 220$ GeV 
in the degenerate case ($R_5 = R_6$), and $m_{KK} \sim 300$ GeV in the asymmetric case.  
Notwithstanding the precise measurements of WMAP, those predictions depend on the Cosmological model and 
shouldn't be taken too seriously.
Nevertheless, we would consider the model disfavoured if the LHC could exclude masses below, say, $400$ GeV.
At a 7 TeV centre of mass energy, the cross sections are quite large due to the many channels contributing, and they can 
sum up to several pb (even without taking into account the probable enhancement due to QCD corrections).

To determine the phenomenology of the tier, it is important to know the loop-induced effective vertices which may mediate direct 
decays into a pair of SM states: those are given by the formulas we computed for the level $(2k,2l)$ up to an extra factor of 
$1/\sqrt{2}$ due to the different normalisation of the wave function, as we discussed in Section~\ref{sec:6Dloop}. 
Therefore, one can directly use our results to calculate the branching ratios.
For quarks and leptons, the decay into a SM quark (lepton) plus a gauge boson is suppressed by the mass of the gauge boson, 
while the decay into Higgs and longitudinal polarisation of the gauge bosons (Goldstone bosons) is suppressed by the mass of the SM
fermions, and it is therefore negligible except for the top and left-handed bottom partners.
This means that the heavy fermions will mainly decay into a SM fermion plus an electroweak gauge boson in the same tier (the 
gluon is too heavy), or into a pair of fermion and electroweak gauge scalar in the level $(1,0)$.
Noticeably, there is no phase space for decays into another fermion in tier $(2,0)$ because such decay would also contain a 
massive gauge boson.
The heavy gauge bosons prefer to decay into fermions: in this case, the dominant modes are in a pair of fermions in tier $(1,0)$, 
or a fermion in the same tier with a SM fermion; moreover, the gluon will decay uniquely into heavy quarks, while the $W$ and 
$Z$ will only decay into heavy leptons.
The decays into SM fermions are also present and competitive: moreover, due to the largest corrections from 
QCD couplings, the quark final states are always preferred to leptonic ones.
Finally, decay modes into pair of heavy SM vectors and Higgs are also present and usually smaller.
The lightest particle in the tier is a special case: due to its light mass, it can only decay into a pair of SM fermions: because of the 
QCD loops, it will decay roughly 80\% into light jets and 17\% into a pair of tops, while decays into leptons are at the level of a 
percent. 
A more detailed discussion of masses and branching ratios can be found in~\cite{jeremie}.
For the purpose of this work, we focused on a benchmark point with $m_{KK} = 300$ GeV: this point corresponds to the mass preferred by present WMAP data.
In Table~\ref{tab:benchmark}, we show the masses of all states and their branching ratios.

\begin{table}[tb]
\begin{center}
\begin{tabular}{|c||c|c||l|}
\hline
& mass & width & Branching Ratios \\
& GeV & MeV & \\
 \hline
 $A_\mu^{2,0}$ & $600.9$ & $19$ & $\begin{array}{l} jj: 80\%\,, \quad t\bar{t}: 17\%\,,  \quad \nu \bar{\nu}: 1.6\%\,, 
 \quad l^+ l^-: 0.22\%\,, \quad Z H: 0.26\%\,. \end{array}$ \\
  \hline
 $Z_\mu^{2,0}$ & $620.4$ & $95$ & $\begin{array}{l} (1,0): 47\%\,, \quad \nu^{2,0} \nu: 18\%\,, \quad l l_L^{2,0}: 13\%\,, 
 \quad l l_R^{2,0}: 0.7\% \\ jj: 16\%\,, \quad t\bar{t}: 1.5\%\,, \quad \nu \bar{\nu}: 1.8\%\,, \quad l^+ l^-: 0.5\%\,, \\
 W^+ W^-: 0.3\%\,, \quad Z H:0.2\%\,. \end{array}$ \\
  \hline
   $W_\mu^{2,0}$ & $619.6$ & $150$ & $\begin{array}{l} (1,0): 29\%\,, \quad l \nu^{2,0}: 9.5\%\,, \quad \nu l_L^{2,0}: 9.5\%\,, \\ 
   jj: 34\%\,, \quad t b: 11\%\,,  \nu l: 2.\%\,,  \quad W Z: 0.17\%\,, \quad W H: 0.14\%\,. \end{array}$ \\
  \hline
   $G_\mu^{2,0}$ & $653.0$ & $1900$ & $\begin{array}{l} (1,0): 44\%\,, \quad u u_L^{2,0}: 7\%\,, \quad d d_L^{2,0}: 10\%\,, \\ 
   u u_R^{2,0}: 10\%\,, \quad d d_R^{2,0}: 16\%\,,\quad jj: 10\%\,, \quad t \bar{t}: 0.24\% \end{array}$ \\
  \hline
  \hline
   $l_R^{2,0}$ & $602.3$ & $0.05$ & $\begin{array}{l} l A_\mu^{2,0}: 98\%\,, \quad l_R^{1,0} A_\phi^{1,0}: 2\% \end{array}$ \\
  \hline
   $l_L^{2,0}$ & $606.2$ & $0.39$ & $\begin{array}{l} l A_\mu^{2,0}: 73\%\,, \quad l_L^{1,0} A_\phi^{1,0}: 27\% \end{array}$ \\
    \hline
   $\nu^{2,0}$ & $606.2$ & $0.1$ & $\begin{array}{l} \nu A_\mu^{2,0}: 94\%\,, \quad \nu^{1,0} A_\phi^{1,0}: 6\% \end{array}$ \\
  \hline
  \hline   
  $u_R^{2,0}$ & $625.7$ & $7.4$ & $\begin{array}{l} u A_\mu^{2,0}: 87\%\,, \quad u_R^{1,0} A_\phi^{1,0}: 13\%\end{array}$ \\
  \hline
   $d_R^{2,0}$ & $625.0$ & $1.7$ & $\begin{array}{l} d A_\mu^{2,0}: 87\%\,, \quad d_R^{1,0} A_\phi^{1,0}: 13\%\end{array}$ \\
  \hline
   $u_L^{2,0}$ & $630.4$ & $9.2$ & $\begin{array}{l} d W_\mu^{2,0}: 51\%\,, \quad u Z_\mu^{2,0}: 22\%\,,\quad 
   u A_\mu^{2,0}: 21\%\,, \quad  \quad u_L^{1,0} A_\phi^{1,0}: 6.8\%\end{array}$ \\
  \hline
   $d_L^{2,0}$ & $630.4$ & $7.0$ & $\begin{array}{l} u W_\mu^{2,0}: 67\%\,, \quad d Z_\mu^{2,0}: 32\%\,,
   \quad d A_\mu^{2,0}: 0.2\%\,, \quad  \quad d_L^{1,0} A^{1,0}: 0.8\%\end{array}$ \\
  \hline
  \end{tabular} \end{center} 
  \caption{\footnotesize Branching ratios for the states in tier $(2,0)$ for the benchmark point $m_{KK} = 300$ GeV; $(1,0)$ 
  labels the inclusive decay into a pair of states in tier $(1,0)$. For the Higgs, we considered $m_H = 120$ GeV.}
\label{tab:benchmark}
\end{table}

From the table we can see that most of the events will contain a chain decay within level $(2,0)$ to the lightest state, 
$A_\mu^{2,0}$, or a decay into states of tier $(1,0)$ that will chain decay to the stable Dark Matter candidate.
In both cases, the SM particles emitted in the chain will have very little energy due to the small splitting between the masses.
Moreover, the states in level $(1,0)$ will give rise to missing energy: however we expect the particle in level $(2,0)$ that initiate 
the chain to have very little transverse momentum.
The consequence is that the states in tier $(1,0)$ at the end of the decay chain will also have little transverse momentum, and 
therefore the event will not contain significant missing energy after all.
This simple picture is affected by the fact that some events will have sufficient transverse momentum at production and 
that initial state 
radiation may boost the whole event. To study this effect, we need a detailed Monte Carlo simulation that we leave for further 
studies, therefore we will not consider these cases here.

\begin{figure}[tb]
\begin{center}
\includegraphics[width=16cm]{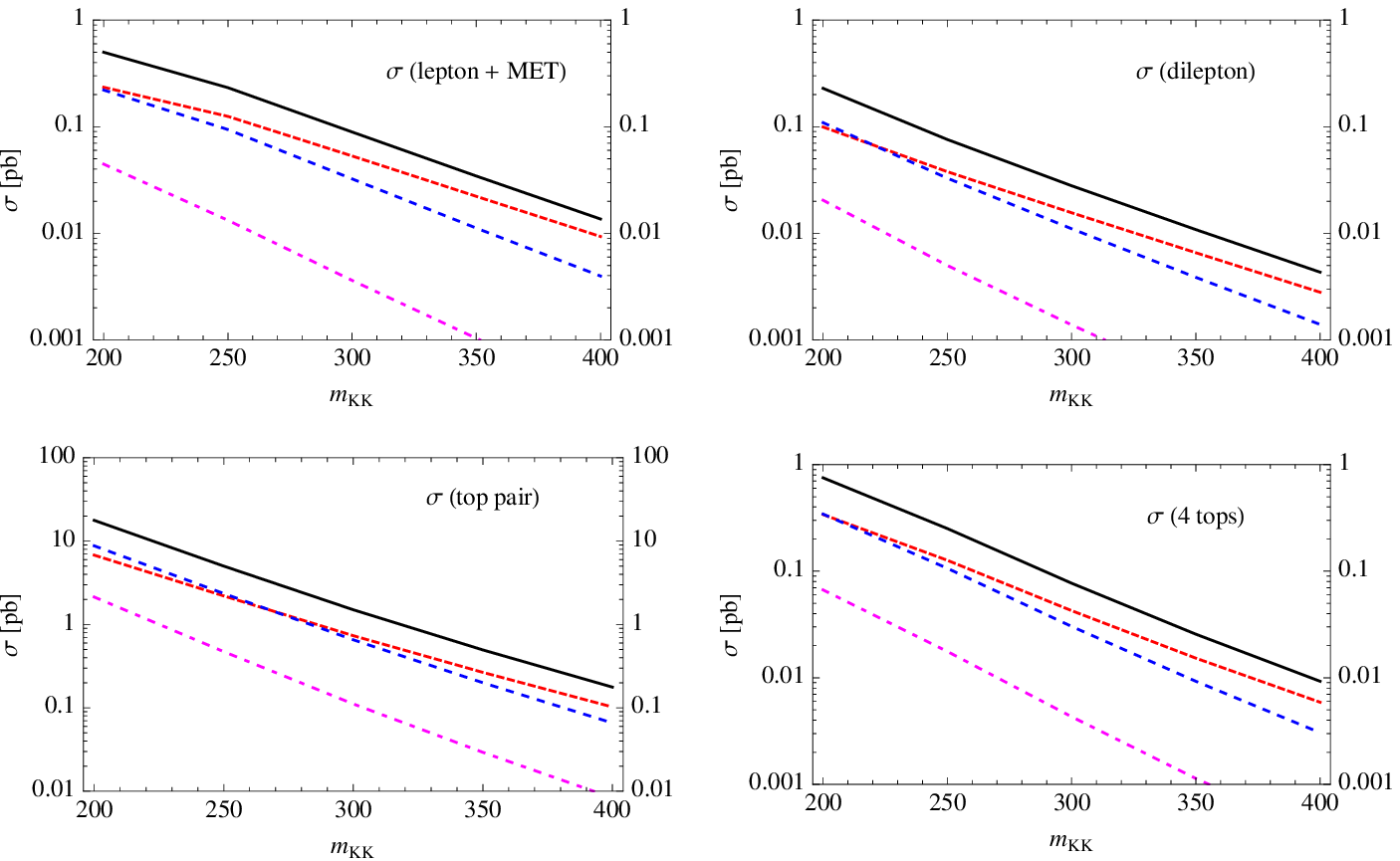}
\end{center}
\caption{\footnotesize Effective and inclusive cross 
sections for a final state with a single lepton plus neutrino (top-left), di-lepton $e^+ e^-$ or $\mu^+ \mu^-$ (top-right), top pair $t\bar{t}$ (bottom-left) and 4 tops $t \bar{t} 
t \bar{t}$ (bottom-right). In all figures, the red line represents the channel with a pair of heavy quarks, in blue an heavy gluon in 
association with an heavy quark, in magenta a pair of heavy gluons, and in black the total cross sections.} \label{fig:xsects2}
\end{figure}

The heavy gauge bosons, on the other hand, can decay directly into a pair of energetic SM fermions at any stage in the 
decay chain.
Even though the branching ratios are rather small, the signal can be very clean and easy to be detected by the experiments.
The main mode will be into light jets and will be affected by a large QCD background; the cleanest signatures will therefore be 
given by energetic leptons and tops.
We identify 4 interesting final states: di-lepton ($e^+ e^-$ and $\mu^+ \mu^-$), top pair and 4 top, coming from the decays of the 
$A_\mu^{2,0}$ and $Z_\mu^{2,0}$, and a single energetic lepton plus a neutrino from decays of the charged $W^{\pm\; 2,0}_\mu$.
In Figure~\ref{fig:xsects2} we show a plot of the effective inclusive cross sections for the above mentioned final states: in all cases, we considered 
all possible decay chains, and the final state is always accompanied by softer SM states from the decay chains.
The di-lepton and single lepton cross sections in the plot include only one lepton flavour (the cross sections are the same).
These channels are very clean because the lepton and top pairs resonate with the mass of the tier, while the 4 top channel has 
negligible background from the SM.
In Table~\ref{tab:benchmark2} we show in detail the cross sections for the case $m_{KK} = 300$ GeV.

\begin{table}[tb]
\begin{center}
\begin{tabular}{|c||c|c|c|c|c|}
\hline
 channel & inclusive & di-lepton & lepton & top pair & 4 tops \\
  \hline
heavy quark pair & $4.14$ pb & $15.5$ fb & $53$ fb & $733$ fb & $42.5$ fb \\
heavy quark + gluon & $4.1$ pb & $11$ fb & $32$ fb & $654$ fb & $30$ fb \\
heavy gluon pair & $0.8$ pb & $1.4$ fb & $3.5$ fb & $112$ fb & $4.3$ fb \\
\hline
total & $9$ pb &  $28$ fb & $89$ fb & $1.5$ pb & $77$ fb \\
\hline
  \end{tabular} \end{center} 
  \caption{\footnotesize Cross sections for the final states di-lepton, single lepton plus neutrino, top pair and 4 tops for the benchmark point 
  $m_{KK} = 300$ GeV.}
\label{tab:benchmark2}
\end{table}

\begin{figure}[tb]
\begin{center}
\includegraphics[width=16cm]{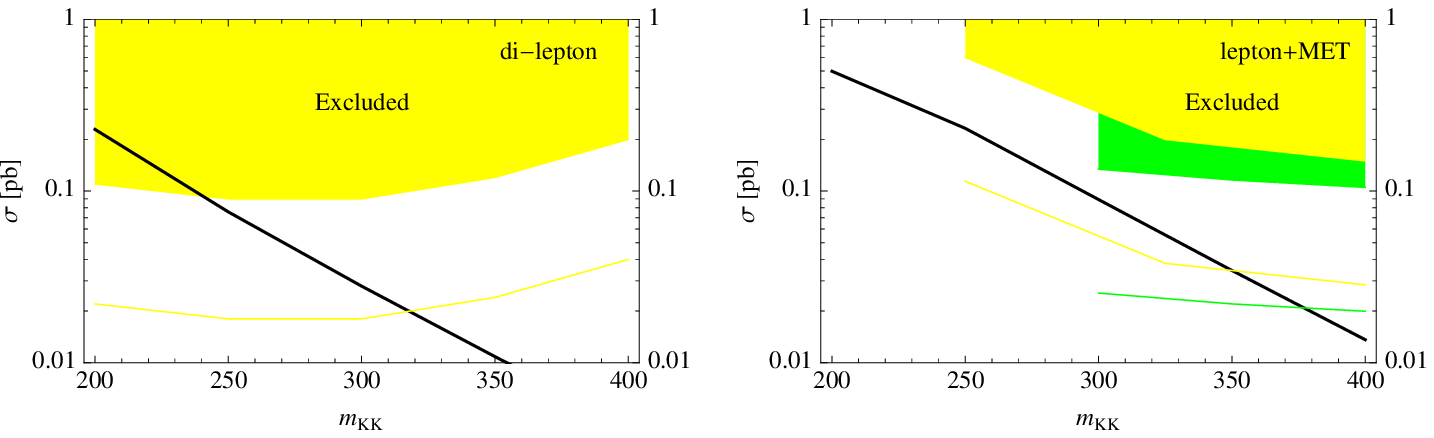}
\end{center}
\caption{\footnotesize ATLAS bounds on di-lepton (left) and single lepton (right) $\sigma \times$ Br, compared to our inclusive cross sections (black line). 
The yellow regions are already excluded, while the solid yellow lines indicate the projected reach for an integrated luminosity of 
1 fb$^{-1}$.
In green, the CMS exclusion and projected reach in the single lepton channel.} 
\label{fig:xsects3}
\end{figure}

The LHC experiments have already analysed the data from the first run, which collected approximately $45$ pb$^{-1}$ luminosity.
One of the first channels they looked at is resonances in di-lepton channels: both ATLAS~\cite{Aad:2011xp} and CMS~\cite{CMS} 
published their results for $40$ pb$^{-1}$ of integrated luminosity.
Here we will consider the bound on cross section times branching ratio published by ATLAS: we will compare our estimated 
inclusive cross section directly with their result, without going through a detailed simulation of the signal.
For $M_{l^+ l^-} $ between $400$ and $500$ GeV (that would correspond, in our case, to $200 < m_{KK} < 250$ GeV), the bound 
on the effective cross section from the combined analysis of electron and muon channels is around $100$ fb.
For $M_{l^+ l^-} > 500$ GeV, the bound is much looser due to an excess in the data; furthermore, the effective cross section in 
our case drops well below $100$ fb.
In Figure~\ref{fig:xsects3}, we compare the bound with our predicted cross section (left panel): we can see that low values of the masses are already excluded or disfavoured, and a bound of roughly $m_{KK} < 240$ GeV can be inferred. 
This comparison is very rough and just indicative: in fact, we haven't taken into account the full event that may contain extra leptons of SM jets, and experimental resolutions.
In the ATLAS study, the trigger efficiency on the $Z$ pole is very close to 100\%, while the total acceptance for $Z' \to l^+ l^-$ 
events at an di-lepton mass of $1$ TeV is 60\% for electrons and 40\% for muons.
However, it can still be used to estimate the strongest bound that present LHC data pose on the model.
We also rescaled the bound to an integrated luminosity of 1 fb$^{-1}$ that should be collected in 2011, assuming that the 
uncertainty in the bound is dominated by statistics.
This estimate gives the solid yellow line in the plot: this shows that the LHC has the potential to probe masses up to 
$m_{KK} \sim 320$ GeV.  A more detailed study is nevertheless necessary to confirm the estimates.
In the right panel of Figure~\ref{fig:xsects3} we also compared the single lepton channel with bounds on leptonic decays of $W'$s from ATLAS~\cite{ATLAS2} and CMS~\cite{CMS2}.
The ATLAS study only focuses on $W'$s with masses above $500$ GeV ($600$ GeV for CMS), corresponding to $m_{KK} > 250$ GeV ($300$ GeV for CMS). 
At lower masses, the sensitivity to new physics decreases.
From the plot we can see that the present bound, calculated for an integrated luminosity of $36$ pb$^{-1}$, is a factor of few above our prediction; however the projection for 1 fb$^{-1}$ seems to be sensitive to KK masses up to $350$ GeV.
This estimate may be affected by the full event, that may contain extra missing energy from the decays into $(1,0)$ states, and by experimental resolutions and acceptances, nevertheless it confirms that the early LHC has the potentiality to probe the interesting mass range of this model.

\section{Conclusion}
\label{sec:concl}
\setcounter{equation}{0}
\setcounter{footnote}{0}

We have discussed the one loop structure of a 6D implementation of universal extra dimensions on the real projective plane, 
motivated by the absence of fixed points and the presence of a stable dark matter candidate due to a relic of the 
Lorentz invariance on the orbifold, with no additional discrete symmetries imposed by hand. The typical radius of the extra 
dimensions is in the range of few hundred GeV and it is bounded on both sides by the dark matter relic abundance, therefore 
the full spectrum at tree level is fixed by this requirement.  
Loop effects, both on the spectrum and the couplings, play a crucial role in understanding the phenomenology 
of the model. We computed the counter-terms and showed that in the special gauge $\xi = -3$ the loop corrections require 
gauge invariant counter-terms. This allows to correlate mass corrections to loop induced couplings and obtain a very 
predictive framework.

Concerning phenomenology we discussed  in more detail the $(2,0)$ and $(0,2)$ tiers. These states mainly chain decay 
within the same tier to the lightest state of this tier or decay in pairs of $(1,0)$ or $(0,1)$ modes. Due to the fact that the 
mass of the $(2,0)$ and $(0,2)$ tiers is equal to twice the mass of the $(1,0)$ or $(0,1)$ tier at three 
level, the possibility that such decays are kinematically open really depends on the loop and Higgs induced splitting, and, 
in general, those decays will be suppressed by the small phase space. However the loops also induce decays directly into 
SM particles and single production of the heavy states, therefore cross sections and branching fractions for this level to SM 
particles can be calculated and detected or excluded at the LHC. The decays into SM particles make this level easy to observe 
in final states without missing energy and with many clear resonances. Few sample decay modes were analysed: di-lepton 
($e^+ e^-$ and $\mu^+ \mu^-$), single lepton plus neutrino, top pair and 4 tops, coming from the decays of the gauge vectors $A_\mu^{2,0}$, $Z_\mu^{2,0}$ and $W^{2,0}_\mu$. 
Based on production cross-sections to these clean modes, we showed that existing LHC data already place limits on the model in the interesting mass range and that an integrated luminosity of 1 fb$^{-1}$, that will probably be collected by the end of 2011, has the potentiality to cover the whole interesting mass range.

\section*{Acknowledgements}
We thank B.Kubik and L.Panizzi for useful discussions and comments. 
G.C. also thanks J.Ellis for useful discussions on the possible phenomenology of this model.

\newpage
\appendix

\section{Appendix: loop integrals}
\label{app:formulae}
\setcounter{equation}{0}

The functions of $n$ appearing in the loop corrections can be expressed in terms of the three following integrals:
\beq
\Phi_1 (n) &=& 2 \pi^3 \int_0^\infty dk\; \frac{k^3}{\sqrt{k^2+n^2} \sinh \pi \sqrt{k^2+n^2}}\,, \\
\Phi_2 (n) &=& 2 \pi^3 \int_0^\infty dk\; \frac{k n (\sqrt{k^2+n^2}-n)}{\sqrt{k^2+n^2} \sinh \pi \sqrt{k^2+n^2}}\,, \\
\Phi_3 (n) &=& 2 \pi^3 \int_0^\infty dk\; \frac{k^3 (\sqrt{k^2+n^2}-n)}{n \sqrt{k^2+n^2} \sinh \pi \sqrt{k^2+n^2}}\,
\\
\Phi_4 (n) &=& 2 \pi^3 \int_0^\infty dk\; \frac{k n^2 }{ \sqrt{k^2+n^2} \sinh \pi \sqrt{k^2+n^2}}\,.
\eeq
Numerically the integrals are suppressed for large $n$:
\begin{center} \begin{tabular}{|l|c|c|c|}
\hline
   & $n=1$ & $n=2$  & $n=3$ \\ 
\hline
$\Phi_1$ & $1.43$ & $0.109$ & $0.0067$ \\
\hline
$\Phi_2$ & $0.54$ &  $0.047$ & $0.0030$\\
\hline
$\Phi_3 $ & $1.02$ & $0.037$ &  $0.0015$ \\
\hline
$\Phi_4 $ & $1.71$ & $0.294$ &  $0.0286$ \\
\hline
\end{tabular} \end{center}


\end{document}